\documentclass{aa}
\usepackage{txfonts}
\usepackage{epsfig}
\usepackage{subfigure}

\usepackage{times}
\usepackage{natbib}
\usepackage{rotating}
\usepackage{lscape}
\bibpunct{(}{)}{;}{a}{}{,}

\begin{document}

\title{The Great Observatories Origins Deep Survey}
\subtitle{VLT/VIMOS Spectroscopy in the GOODS-South Field: Part II}

\author{I. Balestra\inst{1} 
      \and
      V. Mainieri\inst{1}
      \and
      P. Popesso\inst{2}
      \and
      M. Dickinson\inst{3}
      \and 
      M. Nonino\inst{4}
      \and
      P. Rosati\inst{1}
      \and 
      H. Teimoorinia\inst{1,5}
      \and
      E. Vanzella\inst{4}
      \and
      S. Cristiani\inst{4}
      \and
      C. Cesarsky\inst{1}
      \and
      R.A.E. Fosbury\inst{1}
      \and
      H. Kuntschner\inst{1}
      \and
      A. Rettura\inst{6}
      the GOODS team}

\offprints{I. Balestra, \email{ibalestr@eso.org}}

\institute{European Southern Observatory, Karl-Schwarzschild-Strasse 2, D-85748 Garching, Germany
    \and
    Max-Planck-Institut fur extraterrestrische Physik, Giessenbachstrasse 2, 85748 Garching, Germany
    \and
    National Optical Astronomy Observatory, P.O. Box 26732, Tucson, AZ 85726
    \and
    INAF - Osservatorio Astronomico di Trieste, Via G.B. Tiepolo 11, 40131 Trieste, Italy
    \and
    Institute for Advanced Studies in Basic Sciences, P.O. Box 45159-1159, Zanjan 45159, Iran
    \and
    Department of Physics and Astronomy, University of California, Riverside, CA 92521, USA
\thanks{Based on observations made at the European Southern
Observatory, Paranal, Chile (ESO program 171.A-3045 {\it The Great
Observatories Origins Deep Survey: ESO Public Observations of the
SIRTF Legacy/HST Treasury/Chandra Deep Field South.})}
\thanks{Catalogs and data products are available in electronic form at 
\texttt{http://archive.eso.org/cms/eso-data/
data-packages}}
}

\date{Received 2009/ Accepted 2009}

\authorrunning{I. Balestra et al.}

\abstract {We present the full data set of the VIsible Multi-Object Spectrograph (VIMOS) 
spectroscopic campaign of the ESO/GOODS program in the Chandra Deep Field South (CDFS), 
which complements the FORS2 ESO/GOODS spectroscopic campaign.} 
{The ESO/GOODS spectroscopic programs are aimed at reaching
signal-to-noise ratios adequate to measure redshifts for galaxies with 
AB magnitudes in the range $\sim24-25$ in the $B$ and $R$ band using VIMOS, 
and in the $z$ band using FORS2.}
{The GOODS/VIMOS spectroscopic campaign is structured in two separate surveys using two 
different VIMOS grisms. The VIMOS Low Resolution Blue (LR-Blue) and Medium Resolution 
(MR) orange grisms have been used to cover 
different redshift ranges. The LR-Blue campaign is aimed at observing 
galaxies mainly at $1.8 < z < 3.5$, while the MR campaign mainly aims at galaxies at $z < 1$ 
and Lyman Break Galaxies (LBGs) at $z > 3.5$.}
{The full GOODS/VIMOS spectroscopic campaign consists of 20 VIMOS masks.
This release adds 8 new masks to the previous release (12 masks, Popesso et al. 2009). 
In total we obtained 5052 spectra, 3634 from the 10 LR-Blue masks and 
1418 from the 10 MR masks. A significant fraction of the extracted spectra 
comes from serendipitously observed sources: $\sim21\%$ in the LR-Blue and 
$\sim16\%$ in the MR masks. We obtained 2242 redshifts in the LR-Blue campaign 
and 976 in the MR campaign for a total success rate of $62\%$ and $69\%$ 
respectively, which increases to $66\%$ and $73\%$ if only primary targets 
are considered. The typical redshift uncertainty is estimated to be 
$\sigma_{z}\simeq0.0012$ ($\sim255\,\rm{km\,s}^{-1}$) for the LR-Blue grism 
and $\sigma_{z}\simeq0.00040$ ($\sim120\,\rm{km\,s}^{-1}$) for 
the MR grism. By complementing our VIMOS spectroscopic catalog with all existing 
spectroscopic redshifts publicly available in the CDFS, we compiled a redshift 
master catalog with 7332 entries, which we used
to investigate large scale structures out to $z\simeq3.7$. We produced stacked spectra 
of LBGs in a few bins of equivalent width (EW) of the Ly-$\alpha$ and found evidence for
a lack of bright LBGs with high EW of the Ly-$\alpha$. 
Finally, we obtained new redshifts for 12 X-ray sources of the CDFS and extended-CDFS.} 
{After the completion of the two complementary ESO/GOODS spectroscopic campaigns 
with VIMOS and FORS2 at VLT, the number of spectroscopic redshifts in 
CDFS/GOODS field increased dramatically, in particular at $z\ga2$. These data 
provide the redshift information indispensable to achieve the scientific goals 
of GOODS, such as tracing the evolution of galaxy masses, morphologies, clustering, 
and star formation.}


\maketitle

%________________________________________________________________

\section{Introduction}

The Great Observatories Origins Deep Survey (GOODS) is a public,
multi-facility project aimed at gathering the best and deepest 
multi-wavelenght data to investigate some of the most profound cosmological 
issues, such as the formation and evolution of galaxies and active galactic nuclei, 
the distribution of luminous and dark matter at high redshift, the 
cosmological parameters from distant supernovae, and the extragalctic 
background light
\citep[for an overview of GOODS, see][]{dic03, ren03, gia04}.
The program has targeted
two carefully selected fields, the Hubble Deep Field North (HDFN) and
the Chandra Deep Field South (CDFS), with three NASA Great
Observatories (HST, Spitzer and Chandra), ESA's XMM-Newton, and a wide
variety of ground-based facilities. The area common to
all the observing programs is 320 arcmin$^2$, equally divided between
the North and South fields.

Spectroscopy is crucial to reach the scientific goals of GOODS.
In order to reconstruct the evolutionary history
of galaxy masses, morphologies, clustering, or star formation,
reliable redshifts are essential.
Consequently, the CDFS has been target of several spectroscopic campaigns 
over the last decade \citep{cri00, cro01, cim02, bun03, sta04, str04, van04, dic04, 
szo04, lef05, mig05, van05, van06, rav07, van08, pop09}.

The ESO/GOODS spectroscopic program was designed to observe galaxies for which VLT
optical spectroscopy was likely to obtain useful data. The program was
organized into two large campaigns, carried out with the FOcal Reducer and low 
dispersion Spectrograph (FORS2) at VLT/UT1 and with the
VIsible Multi-Object Spectrograph (VIMOS) at VLT/UT3. The ESO/GOODS spectroscopic
program made full use of the VLT instrument capabilities, matching
targets to instrument and disperser combinations in order to maximize
the effectiveness of the observations.

The FORS2 campaign is completed \citep{van05, van06, van08}.
As a result, 1715 spectra of 1225 individual targets were observed and
887 redshifts were determined with a typical uncertainty of 
$\sigma_z \simeq 0.001$. Galaxies were selected adopting three 
different color criteria and using photometric redshifts. The 
resulting redshift distribution spans two redshift domains:
from $z=0.5$ to 2 and from $z=3$ to 6.5. The reduced spectra and the derived 
redshifts were released to the community through the ESO web pages 
\texttt{http://archive.eso.org/cms/eso-data/
data-packages}.

The GOODS/VIMOS spectroscopic survey complements the FORS2 survey 
in terms of completeness and sky coverage. The FORS2 campaign was designed 
to take advantage of the instrument's very high throughput at red wavelengths, 
which allows detecting rest-frame optical and near-ultraviolet spectral features 
(such as the [OII]3727~$\AA$ emission line) 
out to $z \simeq 1.6$, and rest-frame UV emission and absorption lines at 
$z>4$. On the other hand, the VIMOS campaign takes advantage of VIMOS's very large 
field of view, multiplexing capability, and good instrumental throughput 
at roughly 3600-9000~$\AA$, which enable measuring large numbers of redshifts 
at $z<1.4$ from the [OII]3727~$\AA$ emission line and other optical and near-UV 
features, as well as redshifts between 1.5 and 3.5 from Lyman-$\alpha$ emission
and rest-frame UV absorption lines. The cumulative source counts on the CDFS 
taken from the deep public FORS1 data \citep{szo04}, show that down to
$V_{AB}=25$ mag there are $\sim6000$ objects over the 160~arcmin$^2$ 
of the GOODS-S field. Only the high multiplexing
capabilities of VIMOS at VLT could ensure to reach the required
completeness in a reasonable amount of time. 

The GOODS/VIMOS program used two different observational configurations, 
with different object selection criteria. Observations with the Medium 
Resolution (MR) orange grism target galaxies in the redshift ranges 
$0.5<z<1.3$ (primarily from [OII]) and $z>3.5$ (from Ly-$\alpha$).
Observations with the Low Resolution Blue (LR-Blue) grism cover the 
wavelengths of Ly-$\alpha$ and UV rest-frame absorption lines at $1.8<z<3.5$, 
a range not covered by the FORS2 spectroscopy. On average, $\sim$330 objects 
per mask have been observed with the low resolution ($R\simeq 180$) blue grism 
and $\sim$140 with the medium resolution ($R\simeq 580)$ orange grism. 
The overall goal of the GOODS spectroscopic campaign was to reach 
signal-to-noise (S/N) ratios adequate to measure redshifts for galaxies with 
AB magnitudes in the range $\sim24-25$, in the $B$ band for objects observed 
with the VIMOS LR-Blue grism, in the $R$ band for objects observed with the 
VIMOS MR grism, and in the $z$ band for objects observed with FORS2. 

The first part of the GOODS/VIMOS spectroscopic survey has been recently released
\citep[][hereafter P09]{pop09}. The first release includes 2344 spectra from 
6 LR-Blue masks and 968 from 6 MR masks. The number of redshifts obtained is 
1481 and 656 for the LR-Blue and the MR campaign, respectively. 

In this paper we report on the full data set of the VIMOS spectroscopic
follow-up campaign in the CDFS, carried out with the VIMOS instrument at 
the VLT from ESO observing periods P74 through P78 
(mid 2004 through early 2007). This final release includes a total of 10 masks for 
the LR-Blue grism and 10 masks for the MR grism. 

The plan of the paper is as follows. In Sect.~2 we describe the GOODS/VIMOS survey 
and the target selection criteria used. In Sect.~3 we describe the observations and 
the data reduction. The redshift determination and the full data set is presented 
in Sect.~4 and discussed in Sect.~5. 
In Sect.~6 we summarize the results of the GOODS/VIMOS spectroscopic campaign and 
our conclusions. 

Throughout this paper the
magnitudes are given in the AB system (AB~$\equiv 31.4 -
2.5\log\langle f_\nu / \mathrm{nJy} \rangle$), and the ACS F435W,
F606W, F775W, and F850LP filters are denoted hereafter as $B_{435}$,
$V_{606}$, $i_{775}$ and $z_{850}$, respectively. We assume a
cosmology with $\Omega_{\rm tot}, \Omega_M, \Omega_\Lambda = 1.0, 0.3,
0.7$ and $H_0 = 70$~km~s$^{-1}$~Mpc$^{-1}$.

%__________________________________________________________________

\section{Target selection}

\begin{figure}
\centering
\includegraphics[width=9 cm, angle=0]{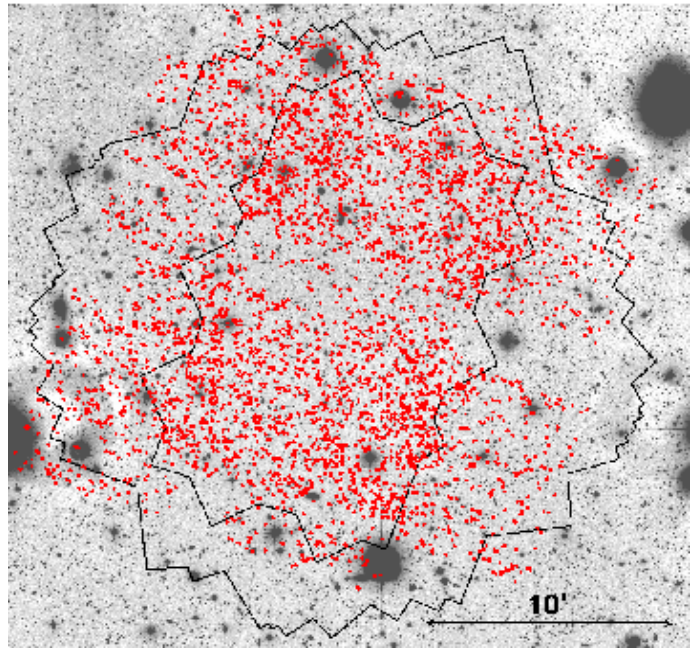}
\caption{Spatial distribution of objects from the whole (LR-Blue plus MR grism) 
GOODS/VIMOS spectroscopic campaign (\textit{red circles}) over the ESO-WFI $R$-band image. 
The contours outline the area of the 2~Ms exposure of the CDFS and the GOODS-S field.}
\label{lay}
\end{figure}

VIMOS \citep{lef03} is a 4-channel imaging spectrograph, each channel (``quandrant'')
covering $\sim7\times8$~arcmin$^2$ for a total field of view (``pointing'') of
$\sim218$~~arcmin$^2$. Because of its geometry (a $16'\times18'$ field of view, 
with a cross gap of $2'$ between the quadrants) only $\sim50\%$ of VIMOS field of view
overlaps with the $10'\times16'$ area roughly defining the GOODS-S field.
The spectroscopic campaign was designed to cover the whole GOODS area. 
At least 3 VIMOS pointing are required to cover the GOODS area, filling the gaps
between the quadrants, with some fraction of the VIMOS coverage extending outside
the GOODS-S field into the CDFS and the ``Extended'' CDFS (ECDFS).
Figure \ref{lay} shows the spatial distribution of the sources observed in the two 
VIMOS surveys (both LR-Blue and MR objects), which outlines the VIMOS coverage of the 
GOODS/CDFS field. 

The survey strategy and target-selection criteria are the same as in P09,
where a detailed description can be found. Here, we recall
the imaging data, the source catalogs, and the main criteria used 
for target selection in the GOODS/VIMOS spectroscopic campaign.
The imaging data and source catalogs used are:
\begin{itemize}
\item CTIO $4$~m MOSAIC $U$-band imaging and ESO $2.2$~m WFI $B$- and $R$-band 
imaging, covering the $30^{\prime}\times 30^{\prime}$ ECDFS, with 
AB magnitude 5$\sigma$ depths 26, 26.2 and 25.8 mag, respectively \citep{gia04}, 
used for Lyman break U-dropout and ``sub-U-dropout'' color selection, both 
inside and outside the nominal GOODS-S area; 
\item HST-ACS $B_{435}$ and $z_{850}$ imaging, covering the GOODS-South 
field ($\sim165$~arcmin$^2$) with depth 27.8 and 27.4~mag \citep{gia04}, 
used for the $BzK$ color-selection technique within the GOODS-S field;
\item VLT-ISAAC $K_s$-band imaging covering the GOODS-S field with depth 
25.1~mag (Retzlaff et al. in preparation), for applying the $BzK$ selection 
technique in GOODS-S field; 
\item CDFS X-ray catalog \citep{gia02, leh05}, covering 
an area larger than GOODS-S, down to an on-axis flux limit of
approximately $5\times10^{-17}$~erg~cm$^2$~s$^{-1}$ in the $0.5-2$~keV band.
\end{itemize}

The wavelength range covered by the VIMOS LR-Blue grism (3500-6900~$\AA$)
is suitable for the detection of ultraviolet absorption and emission features 
of objects in the redshift range $1.8<z<3.8$. 
Targets for the VIMOS LR-Blue grism were selected using the following criteria:
\begin{itemize}
\item U-dropouts, i.e. Lyman-break color-selection of galaxies, using 
the CTIO $U$ and WFI $B$ and $R$ photometry, designed to select blue, star-forming 
galaxies at $z\approx3$;
\item so-called ``sub-U-dropouts'', i.e. $UBR$ color-selected objects with $U-B$ 
colors somewhat bluer than those of the `normal' $z\sim3$ U-dropout Lyman break 
selection criteria, designed to select star-forming galaxies 
at somewhat lower redshifts than those of the regular U-dropouts, nominally 
$z\approx1.8$ to 2.5;
\item $BzK$ color-selection \citep{dad04}, designed to select galaxies at 
$1.4<z<2.5$. Late in the VIMOS campaign, additional Spitzer/IRAC color criteria 
were applied to try to refine the $BzK$ method (see P09);
\item X-ray sources from the CDFS and the ECDFS X-ray catalogs \citep{gia02, leh05}.
\end{itemize}

No low redshift galaxies were intentionally targeted for the LR-Blue masks, 
although as discussed in P09, some foreground interlopers do ``contaminate'' 
the color-selected samples, particularly the sub-U-dropouts. A magnitude cut at 
$B<24.5$~mag was applied to all target catalogs listed above.

The wavelength range of the VIMOS MR grism is 4000-10000~$\AA$, similar to 
that of FORS2. However, because of the stronger fringing at red wavelength 
($\lambda \ge 7000$~$\AA$) and the lower red throughput of VIMOS compared to FORS2, 
optical rest-frame spectral features for galaxies at $z>1$ and ultraviolet rest-frame 
spectral features of Lyman break galaxies (LBGs) at $z\ga4.8$ are harder to detect 
with VIMOS than with FORS2. Therefore, the target selection for the VIMOS MR grism 
was limited to brighter galaxies (mainly expected to be at $z<1.2$), and to 
color-selected LBGs in the redshift range $2.8<z<4.8$. 
Target selection for the MR grism used the available imaging data and photometry 
catalogs according to the following criteria:
\begin{itemize}
\item galaxies with $R<24.5$, with no other color pre-selection, excluding 
VIMOS LR-Blue targets and objects already observed in other spectroscopic programs. 
In the later VIMOS campaigns, some preference was given to galaxies detected at 
24~$\mu$m from the GOODS Spitzer MIPS data (Dickinson et al. in preparation; 
Chary et al. in preparation), meeting the same $R<24.5$ mag limit. 
\item relatively bright Lyman break galaxies at $i_{775}<25$, selected as $B_{435}$, 
$V_{606}$ dropouts (nominally, redshifts $z\approx4$ and 5, respectively), according 
to the same color criteria described in \citet{van05, van06, van08}.
\end{itemize}

No photometric redshifts were used, nor a surface brightness selection was applied 
when selecting galaxies for observations. GOODS/VIMOS masks were designed to 
avoid as much as possible objects already observed in previous redshift surveys
\citep[e.g.][]{cim02, szo04, lef05, van05, van06, van08}.

%__________________________________________________________________

\section{Observations and data reduction}

The VLT/VIMOS spectroscopic observations were carried out in service 
mode during ESO observing periods P74-P78.
A log of all the GOODS/VIMOS observations is presented in Table~\ref{obs}.
The total exposure time per mask is 4 h. Each LR-Blue mask consists of 10 
exposures of 24 min, while each MR mask consists of 12 exposures of 20 min. 
This work adds 8 new masks to the previous release (12 masks, P09), 
for a total of 20 VIMOS masks. All the masks were designed with $1\arcsec$ slits. 
The spatial scale for VIMOS is $0.205\arcsec$/pixel.

In the LR-Blue campaign, the LR-Blue grism was used together with
the Order Sorting OS\_Blue cutoff filter. In this configuration the useful 
wavelength range is $3700-6700\,\AA$, the nominal resolution is 
$R=\lambda/\Delta\lambda=180$, corresponding to a spectral resolution of 
$\sim28\,\AA$, and the dispersion is $5.7\,\AA/$pixel.

The MR grism and the GG475 filter were used in the MR campaign.
In this configuration the useful wavelength range is $4800-10000\,\AA$, 
the nominal resolution is $R=\lambda/\Delta\lambda=580$, which corresponds 
to a spectral resolution of $\sim13\,\AA$, and the dispersion is $2.55\,\AA/$pixel.

For a detailed description of the preparation of VIMOS observations and 
the procedure used for data reduction we refer the reader to P09.

\begin{table}
\caption{Log of VIMOS observations for the entire GOODS/VIMOS spectroscopic campaigns.
Columns list the following information: (1) GOODS/VIMOS mask identification number, (2)
date of the observations, and (3) number of exposures per mask times 
duration of single exposure.}
\begin{center}
\begin{tabular}{l c c}
\hline
\hline
Mask ID & Date & Exp. time (s) \\
(1)     & (2)  & (3)           \\
\hline
\multicolumn{3}{l}{LR-Blue masks (P09)} \\
\hline
GOODS\_LRb\_001         & Sept.-Oct. 2004   & $10\times1440$ \\
GOODS\_LRb\_001\_1      &       Nov. 2004   & $10\times1440$ \\
GOODS\_LRb\_002         &  Oct.-Nov. 2004   & $10\times1440$ \\
GOODS\_LRb\_003\_new    &       Oct. 2005   & $10\times1440$ \\
GOODS\_LRb\_003\_new\_1 &  Oct.-Nov. 2005   & $10\times1440$ \\
GOODS\_LRb\_003\_new\_2 &  Nov.-Dec. 2005   & $10\times1440$ \\
\hline
\multicolumn{3}{l}{MR masks (P09)}  \\
\hline
GOODS\_MR\_001          &       Nov. 2004   & $12\times1200$ \\
GOODS\_MR\_002\_1       &       Jan. 2005   & $12\times1200$ \\
GOODS\_MR\_new\_1       &       Dec. 2005   & $12\times1200$ \\
GOODS\_MR\_new\_2       &       Jan. 2006   & $12\times1200$ \\
GOODS\_MR\_new\_2\_1    &       Jan. 2006   & $12\times1200$ \\
GOODS\_MR\_new\_3\_c    & Sept.-Oct. 2006   & $12\times1200$ \\
\hline
\multicolumn{3}{l}{LR-Blue masks (this work)}  \\
\hline
GOODS\_LRb\_002\_1      &       Dec. 2004   & $10\times1440$ \\
GOODS\_LRb\_dec06\_1    &       Nov. 2006   & $10\times1440$ \\
GOODS\_LRb\_dec06\_2    &       Nov. 2006   & $10\times1440$ \\
GOODS\_LRb\_dec06\_3    &  Nov.-Dec. 2006   & $10\times1440$ \\
\hline
\multicolumn{3}{l}{MR masks (this work)}  \\
\hline
GOODS\_MR\_dec06\_1     &       Oct. 2006   & $12\times1200$ \\
GOODS\_MR\_dec06\_2     &  Oct.-Nov. 2006   & $12\times1200$ \\
GOODS\_MR\_dec06\_3     &       Oct. 2006   & $12\times1200$ \\
GOODS\_MR\_dec06\_4     &       Jan. 2007   & $12\times1200$ \\
\hline
\end{tabular}
\end{center}
\label{obs}
\end{table}

\subsection{Target coordinates}\label{tar}

\begin{figure}
\centering
\includegraphics[width=8.5 cm, angle=0]{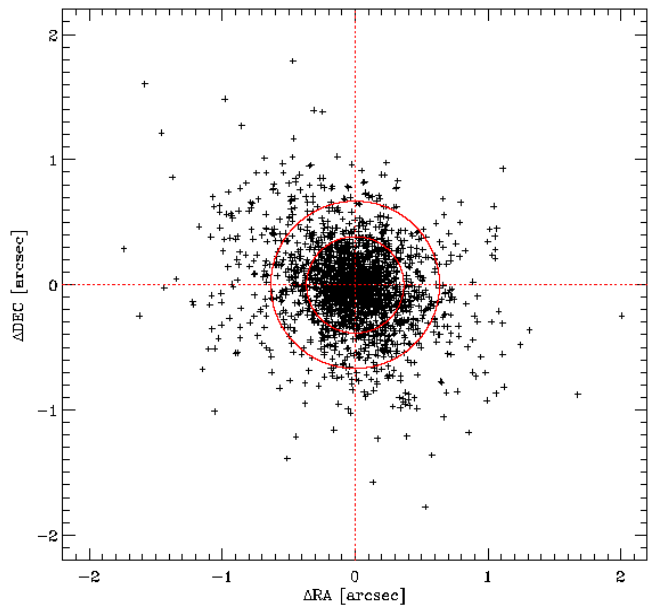}
\caption{$\Delta$RA--$\Delta$DEC distribution computed between VIMOS 
``reconstructed'' coordinates and WFI coordinates of matching WFI $R$-band object
for the VIMOS LR-Blue spectroscopic catalog. The rms dispersion is about $0.2\arcsec$
on both axes. We display two regions encircling 68\% and 90\% of the objects,
which have radii of $0.307''$ and $0.635''$, respectively.}
\label{coord}
\end{figure}

The pipeline processing of the GOODS/VIMOS data was carried out 
using the VIMOS Interactive Pipeline Graphical Interface 
\citep[VIPGI,][]{sco05}.
As it was pointed out in the previuos release (P09),
since the rotation angle of the GOODS/VIMOS pointings ($-20$~deg) 
is different from the default values accepted by VIPGI 
($0$ or $90$~deg), the software does not provide the
astrometry for the extracted spectra.
We followed the same procedure described in P09 to retrieve target 
coordinates. In addition, we cross-correlated the ``reconstructed'' coordinates, 
obtained as in P09, with the WFI $R$-band catalog. In our final released catalogs
we provide for each object both the ``reconstructed'' VIMOS coordinates and the coordinates
of the matching (within a positional tolerance of $1\arcsec$) WFI $R$-band object. If no 
match is found, we repeat the ``reconstructed'' VIMOS coordinates. 
Fig.~\ref{coord} shows the distribution of 
$\Delta$RA and $\Delta$DEC computed between VIMOS 
``reconstructed'' coordinates and the coordinates of the matching WFI $R$-band object
for the VIMOS LR-Blue spectroscopic catalog. The rms dispersion is smaller than 
$0.3\arcsec$ on both coordinates ensuring accurate target identification.

\begin{figure}
\centering
\includegraphics[width=8 cm, angle=0]{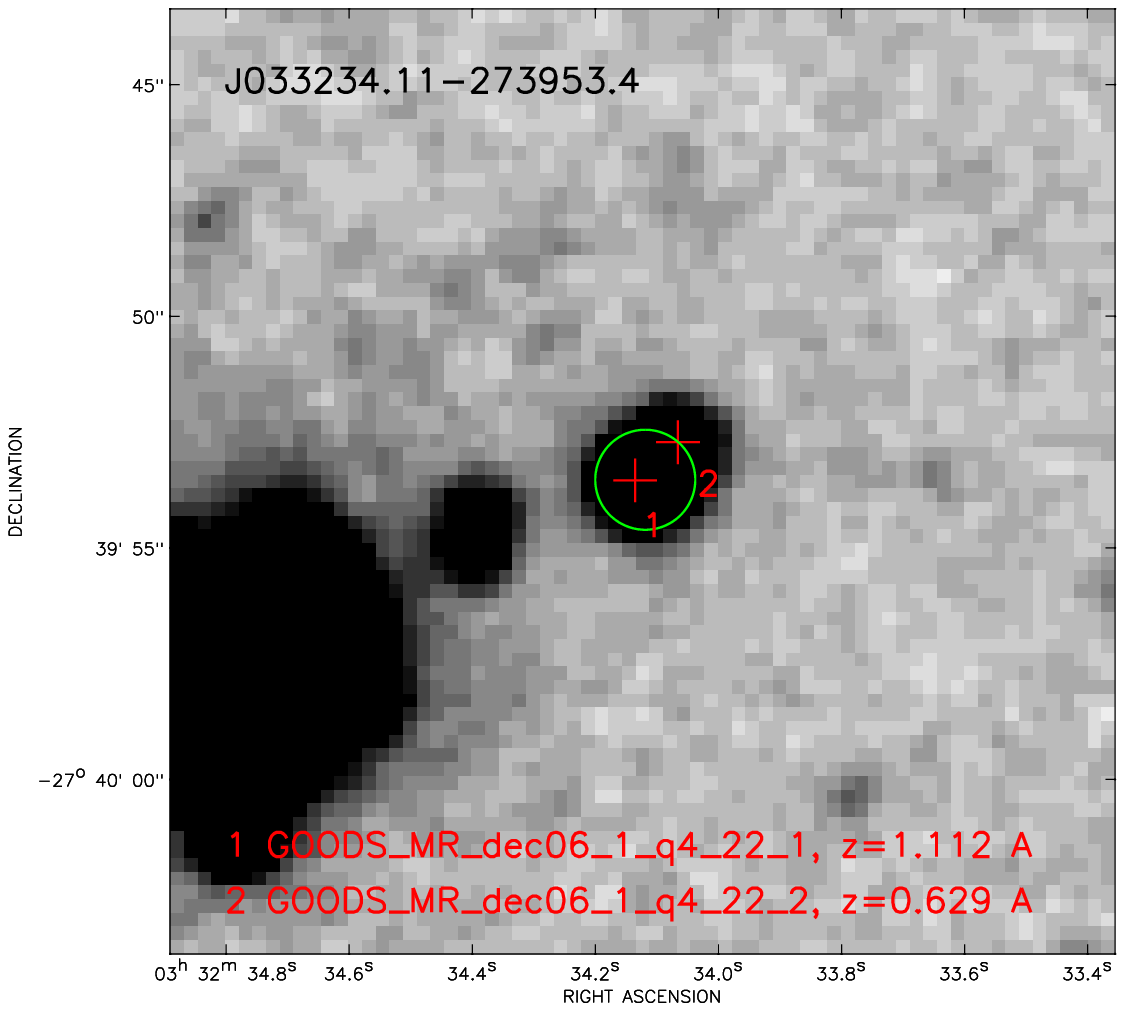}
\includegraphics[width=8 cm, angle=0]{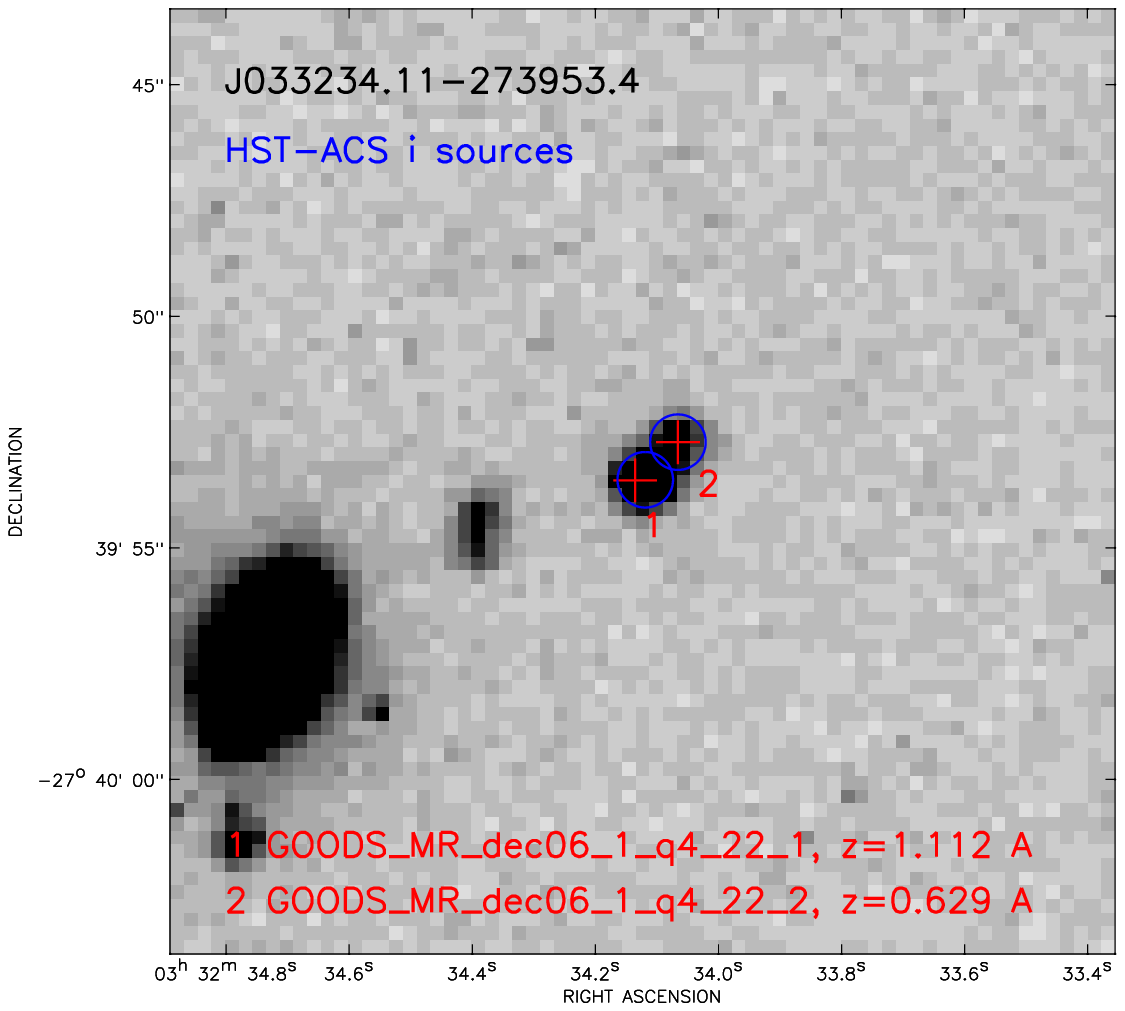}
\caption{WFI $R$-band (\textit{upper panel}) and ACS $z$-band (\textit{lower panel}) 
$20''\times20''$ cutout centered on J033234.11-273953.4. Red crosses display the 
reconstructed VIMOS coordinates of the two objects in slit. In the upper panel the green
circle indicates the position of the WFI-matched object. In the lower panel 
the blue circles correspond to the position of the matching ACS $i$-band sources.}
\label{excoo}
\end{figure}

It is worth noticing that 48 cases (39 in the LR-Blue and 9 in the MR grism) could be 
identified, where two closely separated ($\la1''$) spectra are extracted in the same 
slit for two objects that are blended in WFI images and, therefore, match a single 
WFI object. In all these cases, the WFI coordinates and the coordinate-based names 
assigned to each pair are the same. However, the information on the ``correct'' 
position may still be retrieved from their reconstructed VIMOS coordinates.
In Fig.~\ref{excoo} we show one such case. The comparison between the WFI $R$-band
and the ACS $z$-band images clearly reveals the close pair of sources and the good 
accuracy of their VIMOS reconstructed position.

As mentioned in P09, due to a bug, VIPGI assigns wrong focal plane coordinates 
to a small number of objects in slits with more than 2 spectra. In most cases 
the extracted spectra of these objects lie very close to the edge of the 2D spectrum.
Targets with uncertain coordinates were 81 in the LR-Blue campaign 
(out of which 80\% had no redshift determination) and 34 in the MR campaign 
(50\% without redshift), all of which are serendipitously observed objects. 
Among them, we selected objects having a redshift determination
(18 in the LR-Blue catalog and 16 in the MR catalog), in order to attempt to retrieve 
their coordinates manually.
From a visual inspection of the 2D spectrum and using the coordinates of 
primary targets and, if present, other objects in the same slit as a reference, 
we could unambigously identify and retrieve coordinates for all of the 34 objects 
selected. Their new coordinates are included in the final released catalogs.

The remaining 63 and 18 (from the LR-Blue and the MR catalog respectively) 
serendipitous objects having uncertain coordinates and no redshift estimate 
were removed from the final released catalogs.

\section{Redshift determination}

The total number of GOODS/VIMOS spectra extracted is 3634 in the LR-Blue
and 1418 in the MR campaign. We were able to determine 2242 and 976 redshifts in 
the LR-Blue and the MR campaign, respectively. Serendipitous objects constitute
$\sim21$\% of the LR-Blue and $\sim16$\% of the MR objects. 
We have identified 3305 single LR-Blue objects and 1297 single MR objects. 
Out of those, we were able to determine 2074 and 885 redshifts in 
the LR-Blue and the MR campaign, respectively.

The redshifts were estimated following a different procedure compared to P09.
We avoid the use of the \textit{rvsao} package in IRAF to cross-correlate individually 
observed spectra with template spectra.
Instead, we proceed via visual inspection first, assigning redshifts 
only to the most obvious cases, where more than 2 spectral features could be 
unambigously identified. VIPGI's software \textit{EZ} was subsequently 
used for the cross-correlation with template spectra (i.e. ordinary S0, Sa, Sb, Sc, 
and elliptical galaxies at low redshift, Lyman break galaxies and quasars 
at high redshift) in dubious cases only.

Redshift were determined through the identification of 
prominent features of galaxy spectra:
\begin{itemize}
\item at low redshift the absorption features: the $4000\,\AA$ break, Ca H and
K, H$\delta$ and H$\beta$ in absorption, G-band, MgII 2798
\item and the emission features: [O\,{\sc ii}]3727, [O\,{\sc
iii}]5007, H$\beta$, H$\alpha$
\item at high redshift: Ly$\alpha$, in emission and absorption,
ultraviolet absorption features such as [Si\,{\sc ii}]1260, [O\,{\sc
i}]1302, [C\,{\sc ii}]1335, [Si\,{\sc iv}]1393,1402, [S\,{\sc
ii}]1526, [C\,{\sc iv}]1548,1550, [Fe\,{\sc ii}]1608 and [Al\,{\sc
iii}]1670
\end{itemize} 

In analogy to the complementary GOODS/FORS2 redshift campaign 
\citep{van05,van06,van08}, we use
four quality flags to indicate the quality of a redshift estimate. 
Quality flags are assigned with the following criteria:
\begin{itemize}
\item flag A, high quality, i.e. several emission lines and strong absorption 
features are well identified;
\item flag B, intermediate quality, i.e. at least two spectral features are well 
identified, for instance one emission line plus few absorption features;
\item flag C, low quality, i.e. spectral features, either in emission or in absorption, 
are less clearly identified;
\item flag X, no redshift estimated. No features identified.
\end{itemize}

\begin{figure*}
\centering
\includegraphics[width=8.5 cm, angle=0]{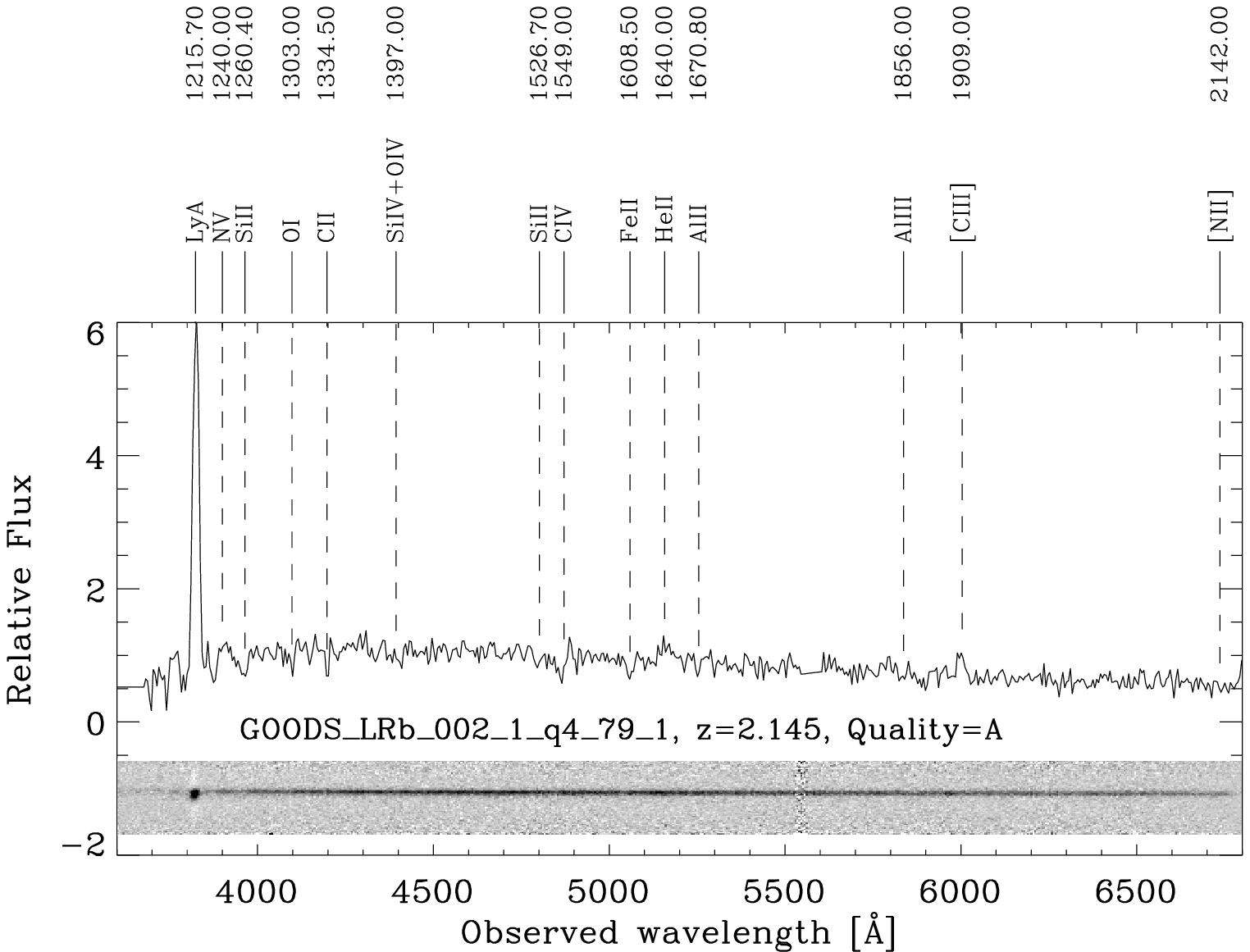}
\includegraphics[width=8.5 cm, angle=0]{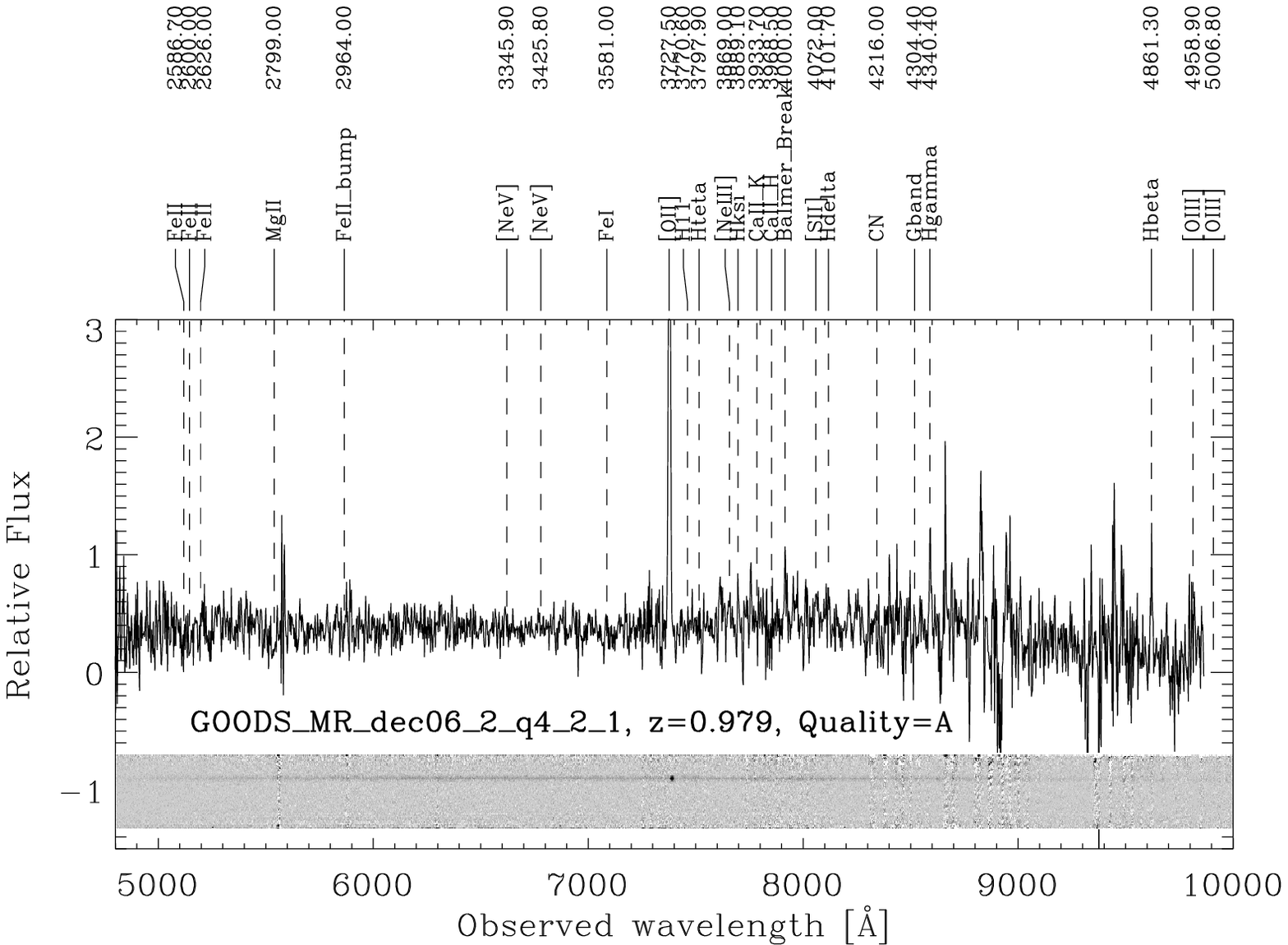}
\includegraphics[width=8.5 cm, angle=0]{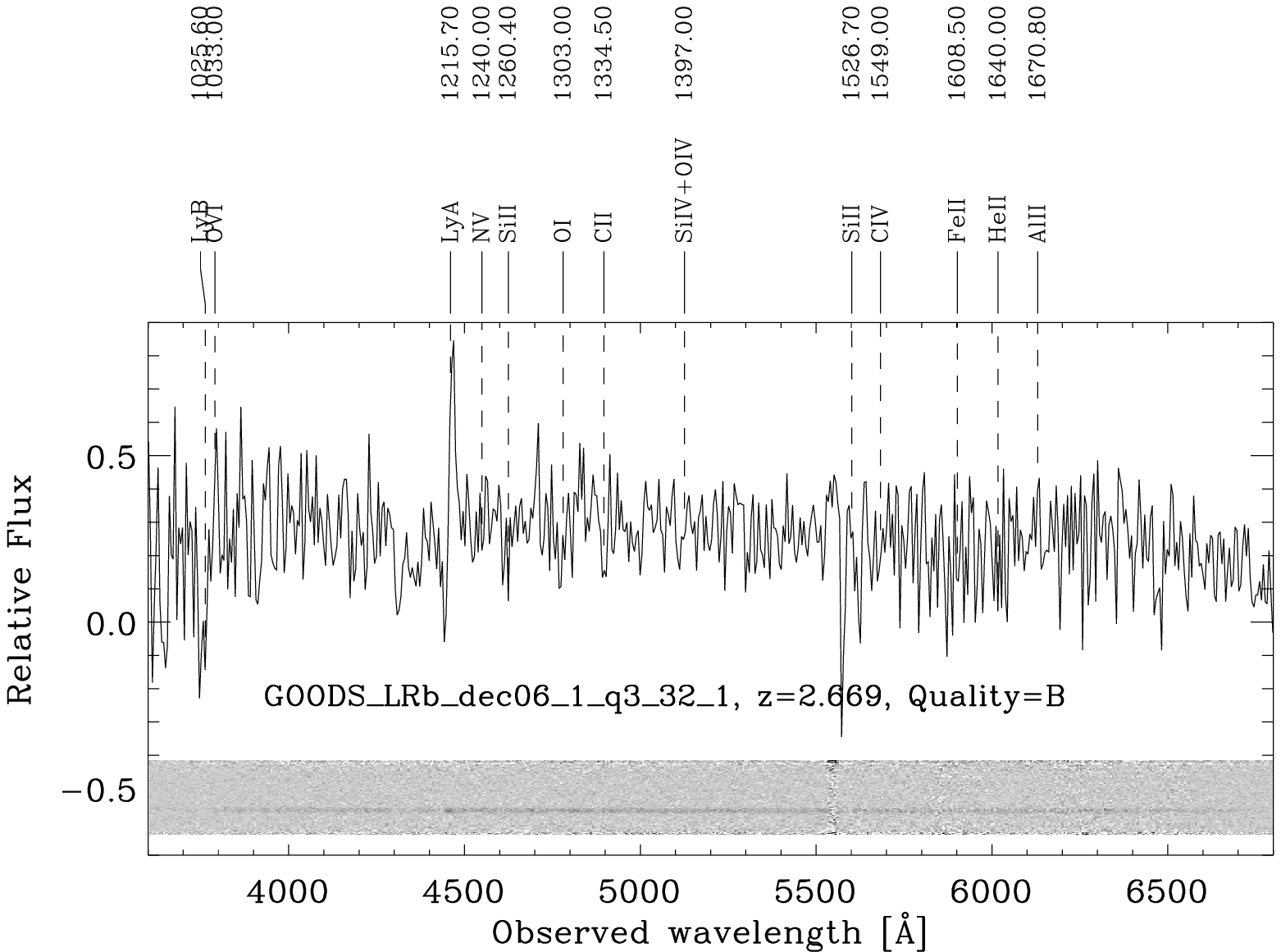}
\includegraphics[width=8.5 cm, angle=0]{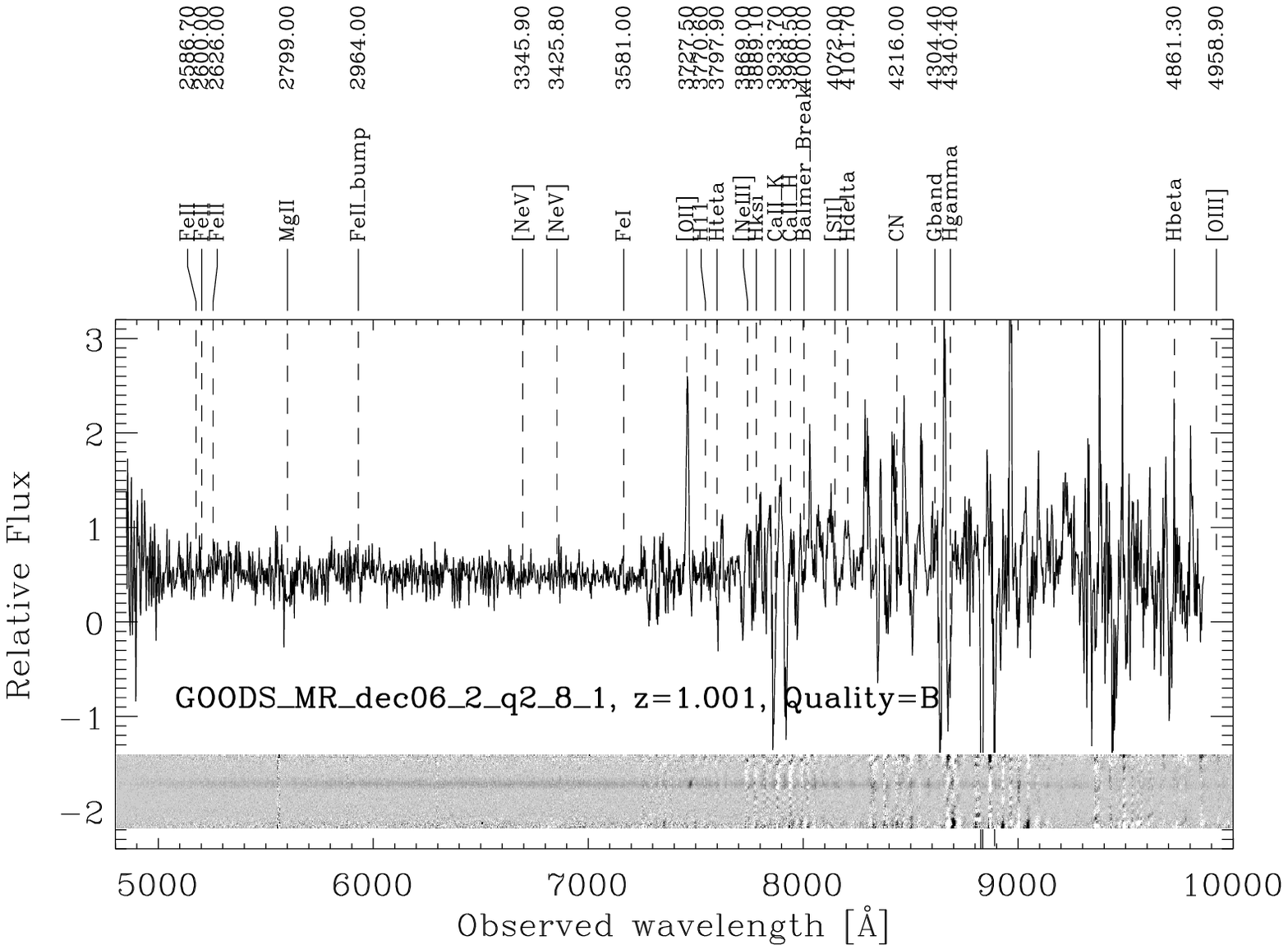}
\includegraphics[width=8.5 cm, angle=0]{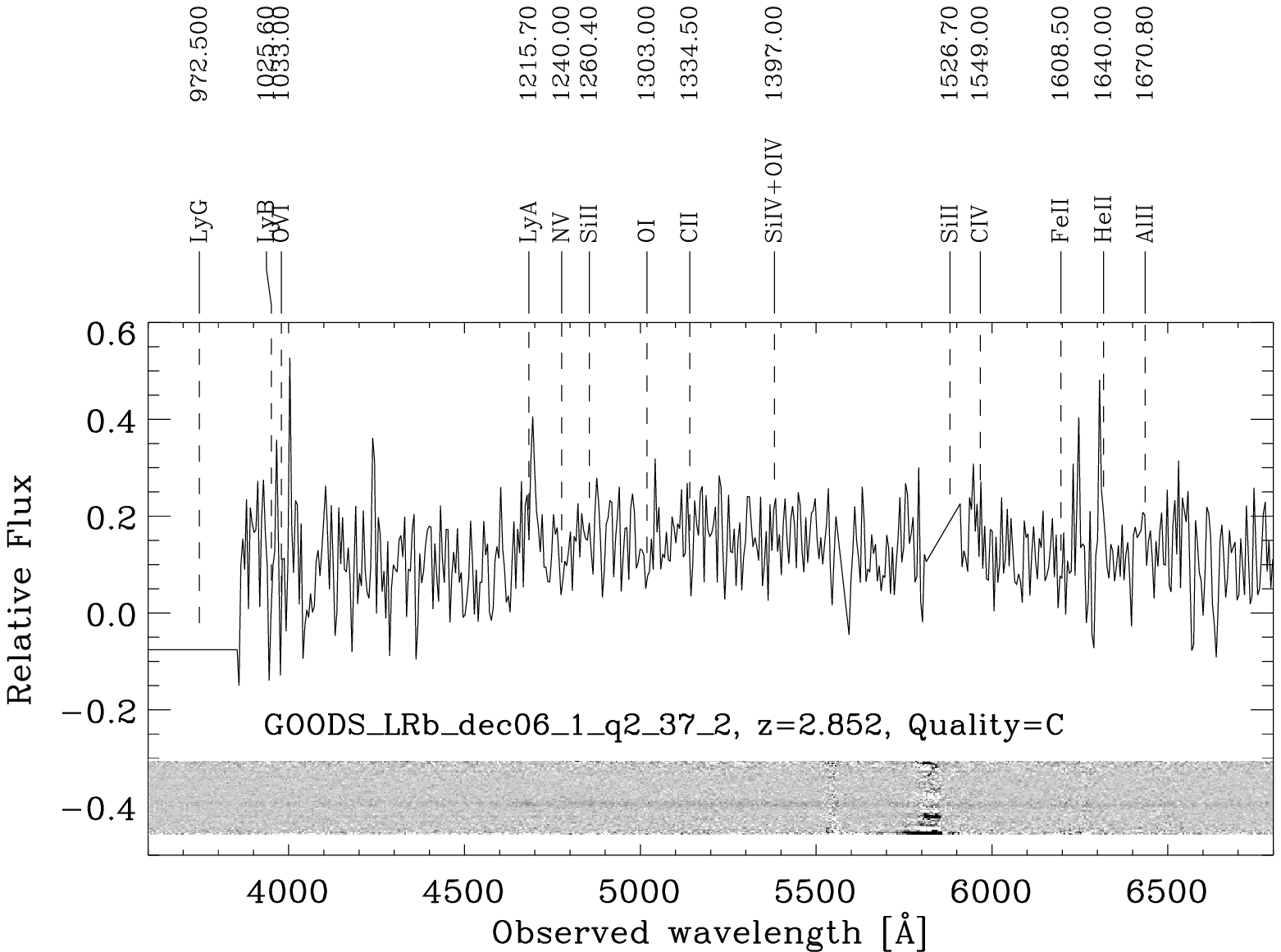}
\includegraphics[width=8.5 cm, angle=0]{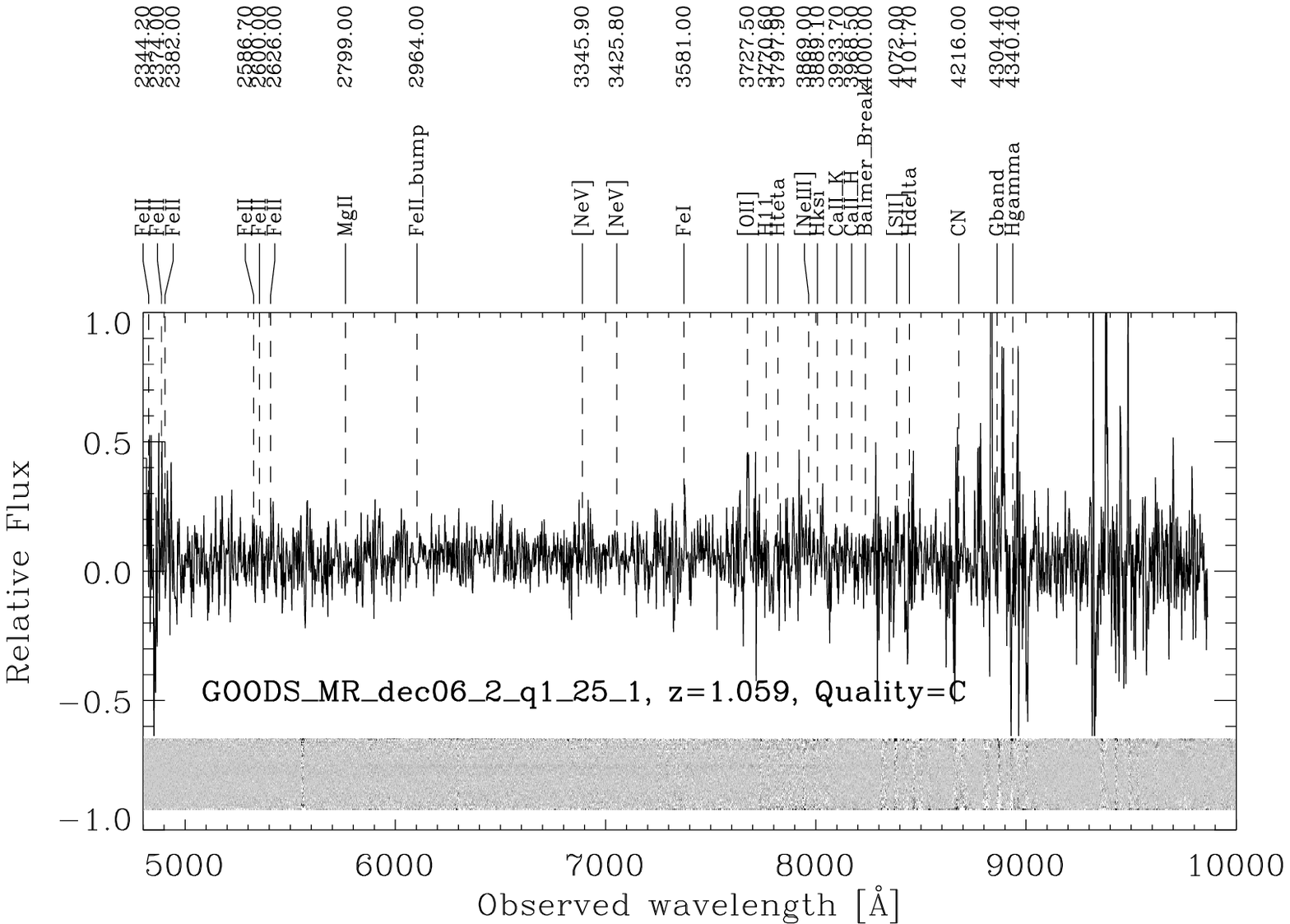}
\caption{Examples of VIMOS spectra (unsmoothed) ranked by quality of the redshift estimate. 
The \textit{left column} shows 1D and 2D spectra obtained with the LR-Blue grism for 
objects with quality-A, -B, and -C redshift. The \textit{right column} shows the same 
for the MR grism.}
\label{ex_spe}
\end{figure*}

Each spectrum with the superposed main spectral
features, is visually inspected independently by different people and the 
redshift determination and the quality flag assignment are eventually further refined. 
On average, each spectrum is checked more than three times.
Fig.~\ref{ex_spe} shows the quality of VIMOS spectra ranked by S/N: typical 
LR-Blue and MR spectra for quality-A, -B, and -C redshift etimates are shown for 
comparison.

In $\sim 15\%$ of the cases the redshift is based only on one emission
line, usually identified with [O\,{\sc ii}]3727 or Ly-$\alpha$.  In
these cases the continuum shape, the presence of breaks, the absence
of other spectral features in the observed spectral range and the
broad band photometry are particularly important in the evaluation.
In general these solo-emission line redshifts are classified as
``likely'' (B) or ``tentative'' (C) if no other information is
provided by the continuum.

In order to investigate possible differences in the redshift estimates that could be 
introduced by the different methods utilized in the two releases, we compared
the reshift resulting from the two different methods for two quadrants of one of the 
new LR-Blue masks. We verified that our procedure produced equivalent 
results to that used in the previous release, with the only noticeable exception
of the absence of an overabundance of low-quality estimates in the range
$1.8\lesssim z\lesssim2.2$, that was affecting the previous release (see P09; see also 
Sect.\ref{master} and Figure \ref{zmag_LR}) and which was probably due to the 
different method used in determining redshifts.

\begin{figure}
\centering
\includegraphics[width=8.5 cm, angle=0]{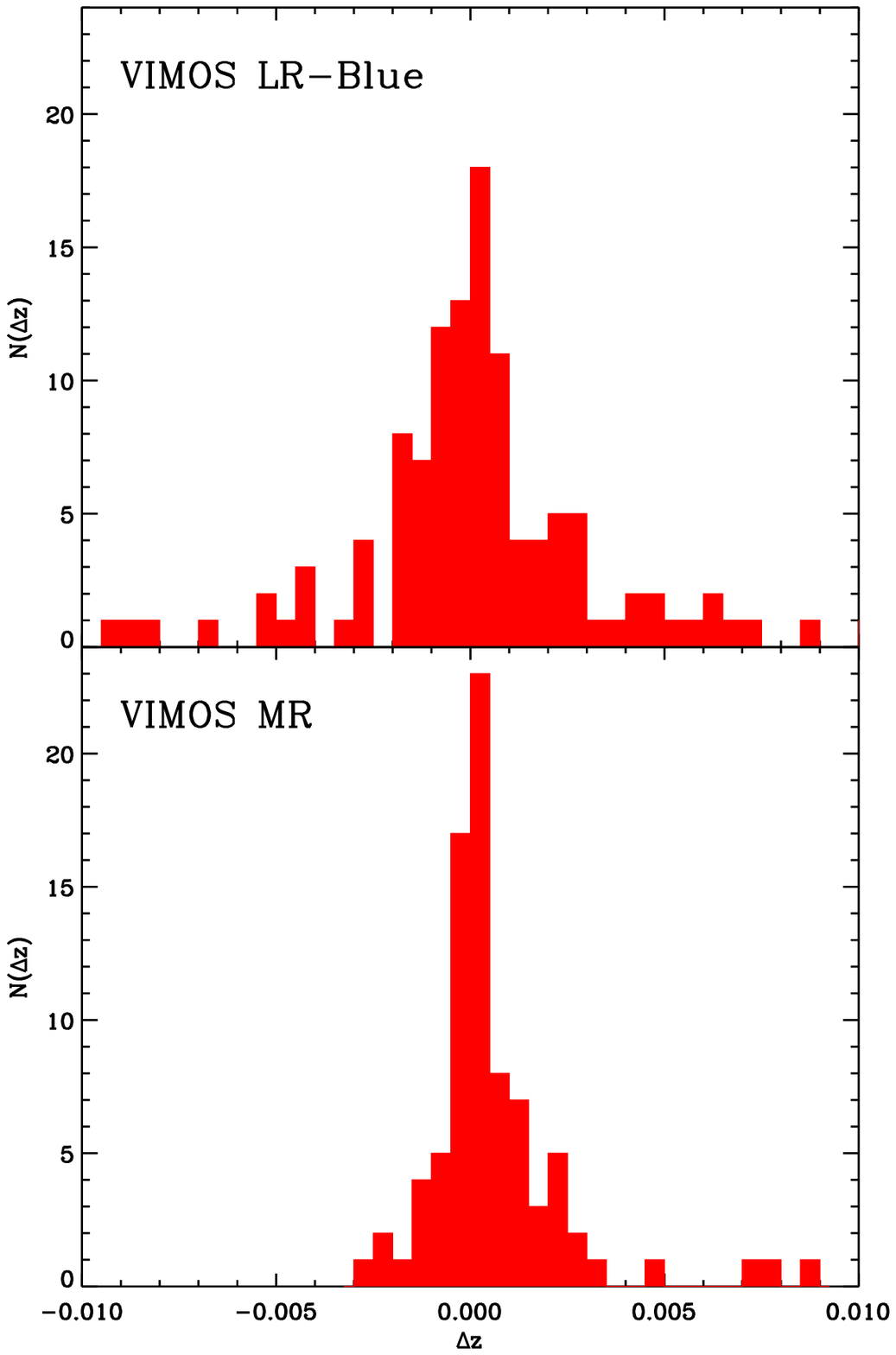}
\caption{Redshift differences between objects observed twice or more in independent 
VIMOS masks. The distribution of $\Delta z$ for the LR-Blue masks ({\em upper panel}) 
is Gaussian with $\sigma_{\Delta z}=0.00120$, therefore the accuracy on single redshift 
measurements is $\sigma_z=\sigma_{\Delta z}/\sqrt{2}=0.00084$ 
($\sim255\,\rm{km\,s}^{-1}$), while for the MR masks ({\em lower panel})
$\sigma_{\Delta z}=0.00056$ and hence $\sigma_z=0.00040$ 
($\sim120\,\rm{km\,s}^{-1}$).}
\label{hist_dz}
\end{figure}

We estimated the internal redshift accuracy by comparing the redshift 
measurements of all the objects observed twice in independent VIMOS masks.
We found 118 such objects in the LR-Blue masks and 83 in the
MR masks with redshift quality flag A or B. In Figure~\ref{hist_dz} we plot histograms 
of the difference in redshift measurements ($\Delta z=z_1-z_2$) separately for 
the LR-Blue and MR grisms. The difference of redshift measurements in the LR-Blue 
campaign has a Gaussian distribution with $<\Delta z>= 5.9\times10^{-5}$ and 
$\sigma_{\Delta z}=0.00120$, therefore the accuracy of single redshift 
measurements is $\sigma_z=\sigma_{\Delta z}/\sqrt{2}=0.00084$ ($\sim255\,\rm{km\,s}^{-1}$). 
For the MR redshifts the Gaussian distribution has a mean $<\Delta z>= 1.2\times10^{-4}$ 
and a dispersion $\sigma_{\Delta z}=0.00056$, therefore the accuracy on a single 
measurement is $\sigma_z=0.00040$ ($\sim120\,\rm{km\,s}^{-1}$).

\subsection{Reliabilty of redshifts}\label{master}

In order to assess the reliability of our redshift estimates we compared them with 
independent measurements from other publicly available spectroscopic surveys in the 
CDFS: the GOODS-FORS2 campaign \citep{van05, van06, van08}, the K20 survey \citep{mig05},
the \citet{szo04} survey, the VVDS survey \citep{lef05}, and the IMAGES survey 
\citep{rav07}. 
As in P09, we combined the redshift information from all these surveys into a 
``master catalog\footnote{This compilation of redshifts from all the public 
spectroscopic surveys is available in electronic form at: 
\textit{http://www.eso.org/sci/activities/projects/goods/}}'', 
cleaned from duplicate observations (in case of 
double or multiple observations we kept the mean value of the redshift estimates).
However, to refine the test on redshift reliability performed in P09, 
we created a ``secure'' redshift reference sample by selecting only 
the best-quality (all reliable at $>99\%$ c.l.) redshift determinations for each survey 
(i.e. GOODS-FORS2 quality A, K20 quality 1, VVDS quality 4, \citet{szo04} quality 3, 
and IMAGES quality 2 redshifts).

We found 95 VIMOS LR-Blue targets matching with objects of the high-quality reference 
sample within a spatial accuracy of 0.3 arcsec or better. Out of them, 21 have 
VIMOS quality flag A, 19 have flag B, 27 have flag C and 28 do not have a redshift
estimate (flag X). Most of the objects without redshift 
lie in a redshift range not accessible to the LR-Blue grism (i.e. $0.8 < z < 1.7$). 
``Catastrophic'' discrepancies ($|z_{VIMOS}-z_{Others}| > 0.015$) are found
only for VIMOS lower-quality redshifts (i.e. flag B and C), which are always 
less convincing than the corresponding estimates from other surveys. 
The resulting confidence level is $>99.99\%$ for quality-A 
(all of the 21 redshifts compared with the reference catalog are correct), 
$\sim63\%$ for quality-B (7 discrepancies out of 19 compared redshifts), and 
$\sim19\%$ for quality-C (22 discrepancies out of 27) redshift determinations.

In the VIMOS MR survey we found 78 objects in common with the high-quality 
reference sample, matching within a spatial accuracy of 0.3 arcsec. 
Out of them, 10 have VIMOS quality flag A, 35 have quality flag B, 13 have quality flag C, 
and 20 have no redshift estimate (flag X). We only find 7 catastrophic discrepancies 
($|z_{VIMOS}-z_{Others}| > 0.015$): 2 have flag B and 5 have flag C.
Hence, we estimate a confidence level of $>99.99\%$ for quality-A (all of the 10 
redshifts compared with the reference catalog are correct), $\sim94\%$ for quality B
(2 discrepancies out of 35), and $62\%$ for quality C (5 discrepancies out of 13) 
redshift determinations.

Our new estimates of confidence levels for the reliability of different quality flags 
are consistent with those obtained in the previous release (see P09) with the only 
exception of LR-Blue quality-C redshift, which are found to be less reliable then 
previously estimated.

\begin{figure}
\centering
\includegraphics[width=8 cm, angle=0]{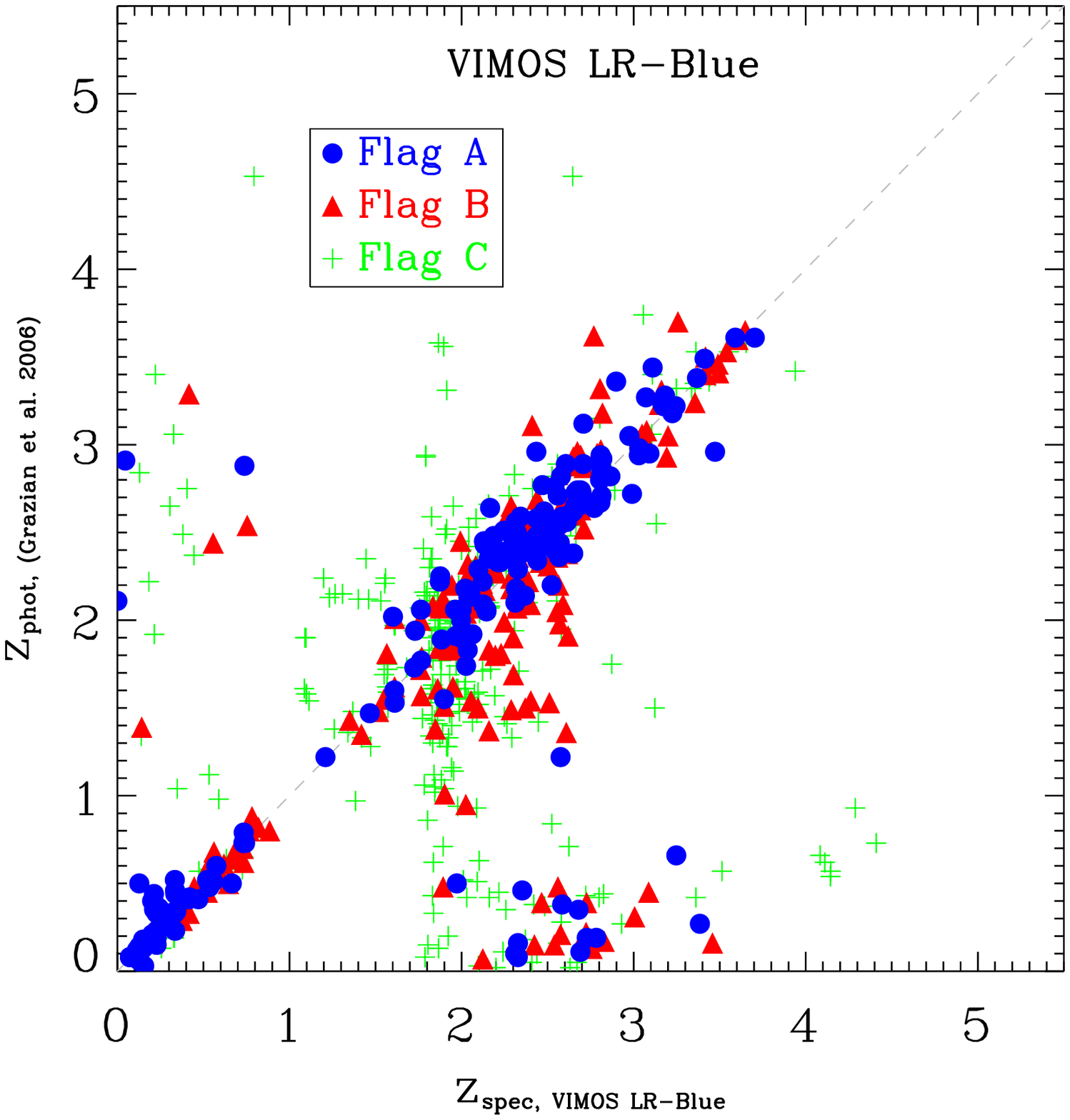}
\includegraphics[width=8 cm, angle=0]{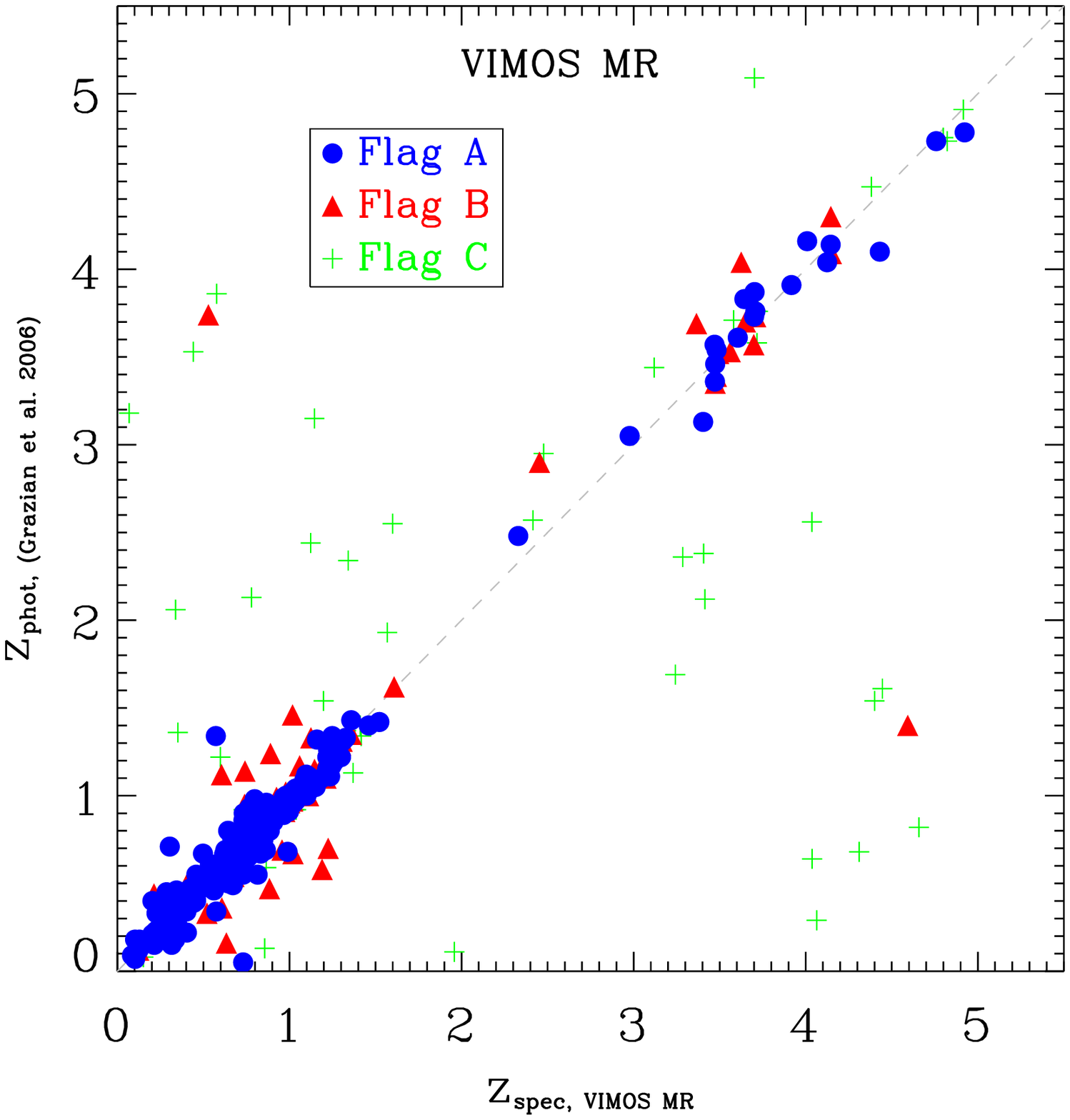}
\caption{Photometric redshift ($z_{phot}$ from Grazian et al. 2006) versus 
spectroscopic redshift ($z_{spec}$) of the VIMOS LR-Blue ({\em upper panel}) and VIMOS MR 
campaign ({\em lower panel}) for objects with quality flag A (blue circles), 
B (red triangles), and C (green crosses). The cross-correlation is limited
to coordinates matching within $0.3''$, i.e. a total of 645 and 523 objects 
for the LR-Blue and MR, respectively.}
\label{ph-spe}
\end{figure}

P09 used also another test to quantify the reliability of the redshifts based on
the simultaneous use of two independent photometric catalogs 
\citep[i.e. ][Daddi et al., private communication]{gra06}. However, results were 
in line with those obtained from the comparison with high-quality spectroscopic 
redshift from other surveys. Here, we plot in Figure \ref{ph-spe} the comparison 
between the GOODS/VIMOS spectroscopic redshifts and the GOODS-MUSIC photometric 
redshifts, which includes all the objects matching within an angular tolerance of 
$0.3''$. In the case of LR-Blue objects, we notice that a few objects, including 16 
objects with secure spectroscopic redshifts (i.e. flag A), have large discrepancies 
compared to their photometric estimates (see also Figure~6 in P09).
We checked the spectra of the 13 objects at $2\la z\la3.5$ with quality-A 
spectroscopic redshifts and large discrepancies ($z_{spec}-z_{phot}>1$) compared to the 
photometric redshifts. The average WFI $R$ magnitude of these objects is 24.4 and 
all the spectra have a relatively good S/N, enabling the clear identification of 
Ly-$\alpha$ in emission or absorption and several UV absorption lines. The spectra 
of these galaxies have also been used to produce the stacked spectra described 
in Sect.~\ref{stlya}.

\subsection{Success rate for VIMOS LR-Blue targets}

A summary of the final results of the VIMOS LR-Blue spectroscopic campaign is
presented in Table~\ref{tabLR}.
We measured redshifts for 66\% (62\% including also the secondary serendipitous 
objects) of the observed LR-Blue spectra. If only high-quality redshift 
determinations (i.e. A or B) are considered, the success rate of the LR-Blue survey
is 43\% for the original target sample and 38\%, if also secondary targets are 
considered. Serendipitous sources, which account for 21\% of the sample, are
usually faint neighbor of the primary targets and lie often at the edge of the 2D
spectrum. Moreover, they are not color-selected as the primary LR-Blue targets, 
therefore, they may often lie at redshifts that are not accessible to the wavelength
range covered by the chosen grism. 
The success rate for these objects is indeed relatively low ($\sim22\%$). 

\begin{figure}
\includegraphics[width=8 cm, angle=0]{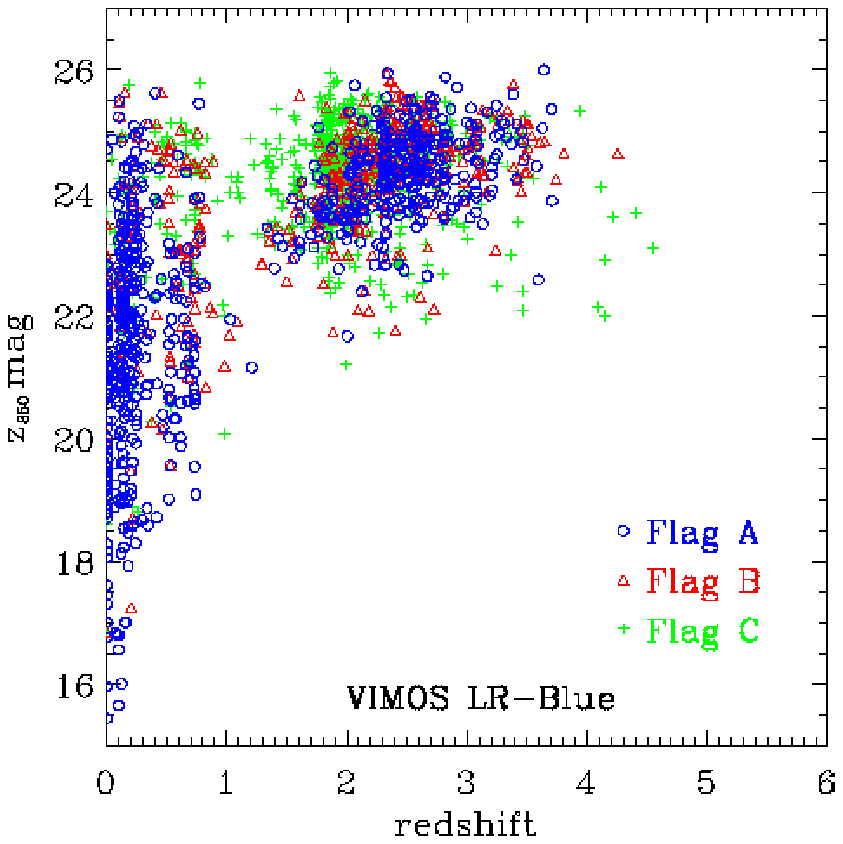}
\includegraphics[width=8 cm, angle=0]{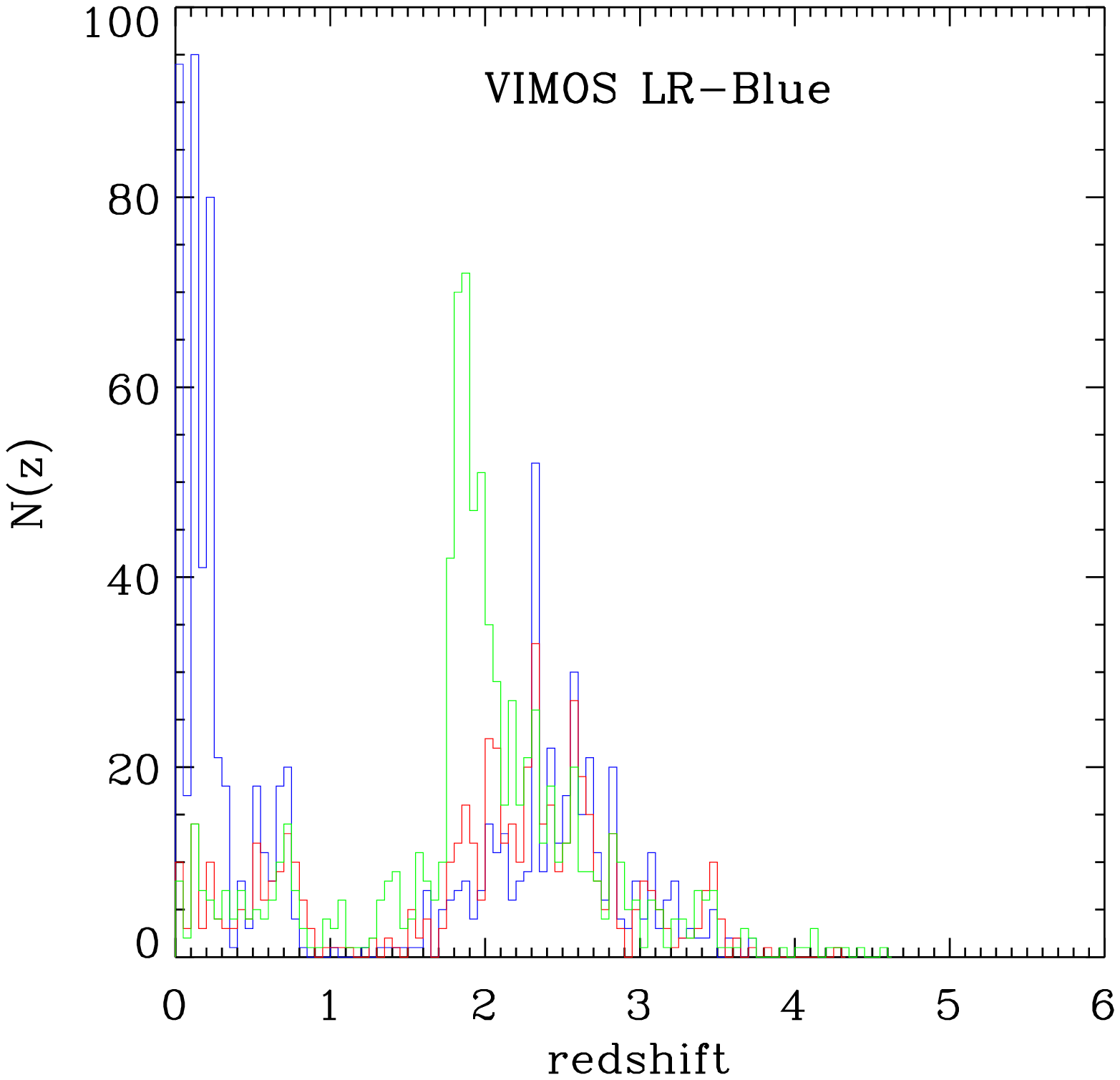}
\includegraphics[width=8 cm, angle=0]{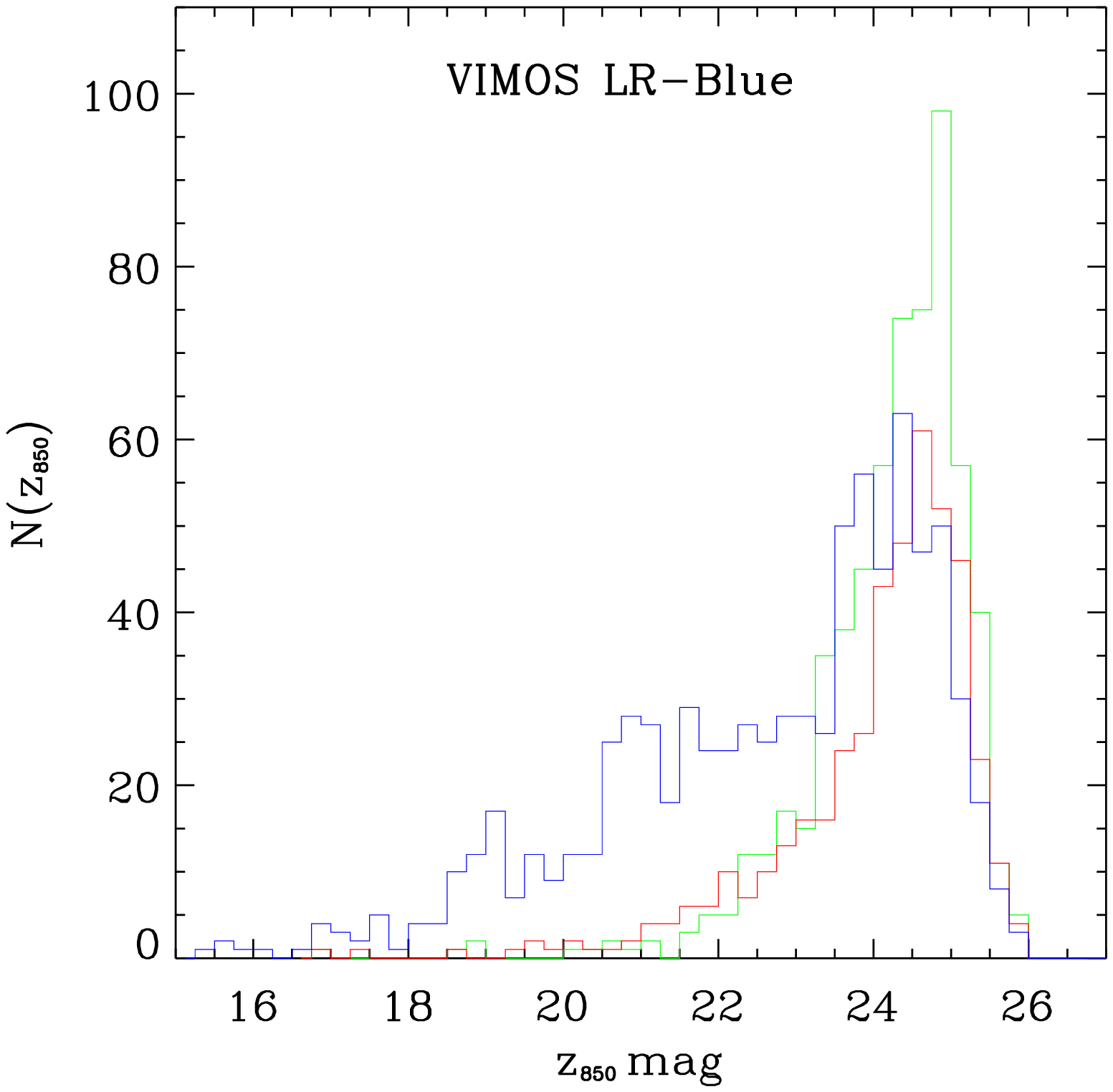}
\caption{VIMOS LR-Blue survey: GEMS $z_{850}$ magnitudes versus redshift 
(\textit{upper panel}), histogram of the redshift distribution (\textit{middle panel}), 
and histogram of the $z_{850}$ magnitude distribution (\textit{lower panel}).}
\label{zmag_LR}
\end{figure}

Figure \ref{zmag_LR} shows the $z_{850}$ magnitudes as a funtion of redshift 
together with the histograms of redshift and $z_{850}$ magnitude distribution. 
These plots show that objects with lower quality redshifts (C) tend to be very faint 
($z_{850}>24$) and have estimated redshifts in the range $1.8<z<2.2$. As already 
discussed in P09, objects with lower quality redshifts tend to have 
very faint magnitudes and have in general lower S/N spectra.
Moreover, the higher failure rate observed for $BzK$ galaxies and galaxies at 
$1.8<z<2.2$ may be explained by the fact that the Ly-$\alpha$ is often outside
the spectral range covered by the LR-Blue grism (i.e. $\lambda<3600\,\AA$) and 
by the fact that the VIMOS efficiency drops very quickly below $4000\,\AA$ 
(see also P09).

\begin{table}
\caption{Statistics of the GOODS/VIMOS LR-Blue campaign.}
\begin{center}
\begin{tabular}{l c c c c c}
\hline
\hline
\multicolumn{6}{c}{\textbf{LR-Blue}} \\
\textbf{P09$^*$} (6 masks)&    \textbf{A} &   \textbf{B} &   \textbf{C} &    \textbf{X} & \textbf{Total} \\
\hline
Num. of entries     &  617 & 302 & 550 &  810 &  2279 \\
Primary targets     &  544 & 265 & 415 &  526 &  1750 \\
Secondary targets   &   73 &  37 & 135 &  284 &   529 \\
\hline
\textbf{This work} (4 masks)&     &    &    &     &  \\
\hline
Num. of entries     &  236 & 240 & 297 &  582 &  1355 \\
Primary targets     &  213 & 208 & 244 &  466 &  1131 \\
Secondary targets   &   23 &  32 &  53 &  116 &   224 \\
\hline
\textbf{Total} (10 masks)&     &    &    &     &  \\
\hline
Num. of entries     &  853 & 542 & 847 & 1392 &  3634 \\
Primary targets     &  757 & 473 & 659 &  992 &  2881 \\
Secondary targets   &   96 &  69 & 188 &  400 &   753 \\
\hline
\end{tabular}
\end{center}
$^*$~For the first release (P09) the numbers of redshifts reported in this Table 
refer to those released in the final catalog. These numbers may differ from those 
reported in P09 since some redshifts have undergone revisions and corrections 
(see Sect.~\ref{tar} and Appendix~\ref{app1}).
\label{tabLR}
\end{table}

\subsection{Success rate for VIMOS MR targets}

We measured redshifts for 73\% (69\% including also the secondary serendipitous 
objects) of the observed MR spectra. In the VIMOS MR spectral campaign the 
overall success rate (i.e. redshift with quality A or B) is 58\% and reaches 63\%
if only primary targets are considered. In Table~\ref{tabMR} we summarize the 
final results of the VIMOS MR spectroscopic campaign.

\begin{figure}
\includegraphics[width=8 cm, angle=0]{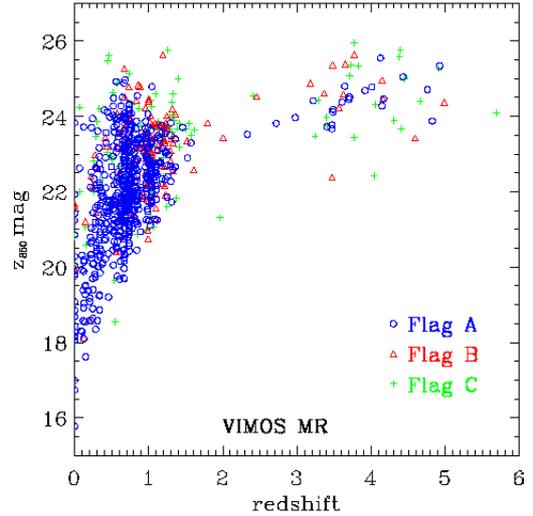}
\includegraphics[width=8 cm, angle=0]{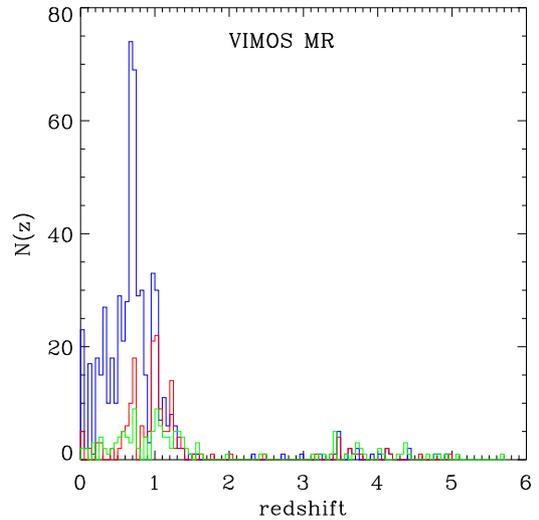}
\includegraphics[width=8 cm, angle=0]{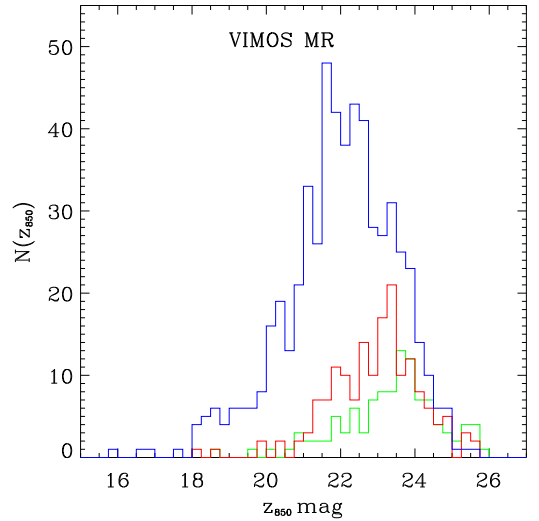}
\caption{VIMOS MR survey: GEMS $z_{850}$ magnitudes versus redshift 
(\textit{upper panel}), histogram of the redshift distribution (\textit{middle panel}), 
and histogram of the $z_{850}$ magnitude distribution (\textit{lower panel}).}
\label{zmag_MR}
\end{figure}

Figure \ref{zmag_MR} shows the $z_{850}$ magnitudes as a funtion of redshift 
for different redshift quality together with a histogram of the redshift distribution
and a histogram of the distribution of $z_{850}$ magnitudes for different redshift
quality. Here we notice that the C flags are more frequent at $z>0.8$, most probably 
due to the fact that above this redshift the main spectral features enter a wavelength 
range where both the OH sky emission lines and the CCD fringing are strong 
(i.e. at $\lambda>7500\,\AA$), making line identification more difficult.
We may also notice a slight trend with magnitude. As expected, lower quality flags 
peak at fainter magnitudes, due to the lower S/N of their spectra.

\begin{table}
\caption{Statistics of the GOODS/VIMOS MR campaign.}
\begin{center}
\begin{tabular}{l c c c c c}
\hline
\hline
\multicolumn{6}{c}{\textbf{MR}} \\
\textbf{P09$^*$} (6 masks)&    \textbf{A} &   \textbf{B} &   \textbf{C} &    \textbf{X} & \textbf{Total} \\
\hline
Num. of entries     &  458 & 112 &  85 &  295 &   950 \\
Primary targets     &  433 &  98 &  70 &  215 &   816 \\
Secondary targets   &   25 &  14 &  15 &   80 &   134 \\
\hline
\textbf{This work} (4 masks)&     &    &    &     &  \\
\hline
Num. of entries     &  170 &  88 &  63 &  147 &   468 \\
Primary targets     &  148 &  75 &  43 &  115 &   381 \\
Secondary targets   &   22 &  13 &  20 &  32  &    87 \\
\hline
\textbf{Total} (10 masks)&     &    &    &     &  \\
\hline
Num. of entries     &  628 & 200 & 148 &  442 &  1418 \\
Primary targets     &  581 & 173 & 113 &  330 &  1197 \\
Secondary targets   &   47 &  27 &  35 &  112 &   221 \\
\hline
\end{tabular}
\end{center}
$^*$~For the first release (P09) the numbers of redshifts reported in this Table 
refer to those released in the final catalog. These numbers may differ from those 
reported in P09 since some redshifts have undergone revisions and corrections 
(see Sect.~\ref{tar} and Appendix~\ref{app1}).
\label{tabMR}
\end{table}

\subsection{Data products and redshift catalogs}

The data products of the GOODS/VIMOS spectroscopic campaign released to the 
community\footnote{\texttt{http://archive.eso.org/cms/eso-data/data-
packages}} include, for each spectrum, the 1-dimensional flux-calibrated 
(in units of $10^{-18}$~erg~cm$^{-2}$~s$^{-1}$~$\AA^{-1}$) spectrum in FITS format and the 
corresponding plot in postscript format.

We created two separated redshift catalogs: one for the VIMOS LR-Blue and one 
for the VIMOS MR campaign. In the two catalogs (see Table \ref{cat}) 
we provide for each object: (column 1) the coordinate-based GOODS identification number,
where the coordinates used are those of the matching WFI object,
(column 2) the VIMOS identification number, (columns 3-4) the coordinates of the 
matching WFI object, (columns 5-6) the original VIMOS coordinates, (columns 7-10) 
$B$ and $R$ band WFI magnitudes with 
the corresponding errors, (columns 11-12) $z_{850}$ GEMS magnitudes and corresponding errors, (column 13)
redshift, (column 14) quality flag, (column 15) comments and identified spectral features, (column 16) 
a label for primary or secondary (i.e. serendipitous) objects, and (column 17)
a label to distiguish between the first (P09) and second VIMOS release (this work).

\begin{table*}
\caption{Spectroscopic redshift catalog for the GOODS/VIMOS LR-Blue campaign$^\dag$. 
Columns list the following information: (1) coordinate-based GOODS identification number, 
(2) VIMOS identification number, (3-4) coordinates in WFI reference 
astrometry, (5-6) original VIMOS coordinates, (7-10) $B$ and $R$ band WFI magnitudes with 
the corresponding errors, (11-12) $z_{850}$ GEMS magnitudes and corresponding errors, (13)
redshift, (14) quality flag, (15) comments and identified spectral features, (16) 
label for primary or secondary (i.e. serendipitous) objects, and (17)
label to distiguish between the first and second VIMOS release.}
\begin{center}
\begin{tabular}{l l c c c c c}
\hline
\hline
ID GOODS & ID VIMOS & RA$_{\mathrm{WFI}}$ & DEC$_{\mathrm{WFI}}$  & RA$_{\mathrm{VIMOS}}$  & DEC$_{\mathrm{VIMOS}}$  & $B$ mag \\
(1)                  & (2)                                 & (3)         & (4)          & (5)         & (6)          & (7)     \\
\hline
J033133.01-274243.9  & GOODS\_LRb\_002\_1\_q2\_4\_1	   & 52.8875408  & -27.7121947  & 52.8875408  & -27.7121947  & 23.034  \\
J033133.08-274301.0  & GOODS\_LRb\_003\_new\_1\_q2\_3\_1   & 52.8878231  & -27.7169564  & 52.8878330  & -27.7169530  & 25.590  \\
J033133.44-274350.7  & GOODS\_LRb\_001\_1\_q2\_3\_1	   & 52.8893390  & -27.7307466  & 52.8893390  & -27.7307470  & 19.698  \\
J033133.54-274303.6  & GOODS\_LRb\_001\_1\_q2\_6\_1	   & 52.8897394  & -27.7176759  & 52.8897400  & -27.7176700  & 21.711  \\
J033133.91-274349.5  & GOODS\_LRb\_001\_1\_q2\_3\_2	   & 52.8913004  & -27.7304179  & 52.8912850  & -27.7304150  & 22.733  \\
J033135.07-274256.4  & GOODS\_LRb\_002\_1\_q2\_8\_1	   & 52.8961455  & -27.7156782  & 52.8961455  & -27.7156782  & 21.322  \\
J033135.85-274312.2  & GOODS\_LRb\_003\_new\_1\_q2\_7\_1   & 52.8993653  & -27.7200705  & 52.8993610  & -27.7200740  & 21.715  \\
J033136.03-274328.9  & GOODS\_LRb\_001\_1\_q2\_11\_1       & 52.9001304  & -27.7246852  & 52.9001310  & -27.7246880  & 20.268  \\
J033136.08-274408.7  & GOODS\_LRb\_002\_1\_q2\_5\_1	   & 52.9003479  & -27.7357574  & 52.9003479  & -27.7357574  & 24.448  \\
J033136.09-274240.3  & GOODS\_LRb\_003\_new\_1\_q2\_10\_1  & 52.9003640  & -27.7111941  & 52.9003600  & -27.7112160  & 24.406  \\
J033136.18-274217.5  & GOODS\_LRb\_001\_1\_q2\_16\_1       & 52.9007486  & -27.7048485  & 52.9008030  & -27.7048170  & 20.468  \\
J033136.44-274421.9  & GOODS\_LRb\_001\_1\_q2\_8\_1	   & 52.9018278  & -27.7394201  & 52.9018100  & -27.7394120  & 25.134  \\
J033136.53-274415.4  & GOODS\_LRb\_003\_new\_1\_q2\_4\_1   & 52.9022064  & -27.7376119  & 52.9022140  & -27.7376100  & 22.879  \\
J033137.26-274220.6  & GOODS\_LRb\_003\_new\_1\_q2\_13\_1  & 52.9052472  & -27.7057199  & 52.9052810  & -27.7057650  & 24.442  \\
J033137.27-274553.8  & GOODS\_LRb\_002\_1\_q2\_2\_2	   & 52.9053051  & -27.7649414  & 52.9053051  & -27.7649414  & 24.747  \\
\hline
\end{tabular}
\end{center}
$^\dag$~Only a portion of the table is shown here for guidance regarding the form and 
content of the catalog. The entire table is available in electronic form at 
\texttt{http://archive.eso.org/cms/eso-data/data-packages}. The full table contains 17
columns of information on 3634 spectra.
\label{cat}
\end{table*}

For each grism we released two catalogs: one cleaned for duplicate observations of 
the same object, in which case we kept the best redshift estimate (with the only 
exception of objects having two quality-C redshifts, in which case we kept both 
estimates), and another containing the complete set of observations including duplicates.
The final VIMOS LR-Blue catalog contains 3634 entries for 3271 individual targets and 
the final VIMOS MR catalog contains 1418 entries for 1294 individual targets.

\section{Discussion}

\subsection{Redshift distribution and large scale structure}\label{zdist}

\begin{figure}
\centering
\includegraphics[width=7.5 cm, angle=0]{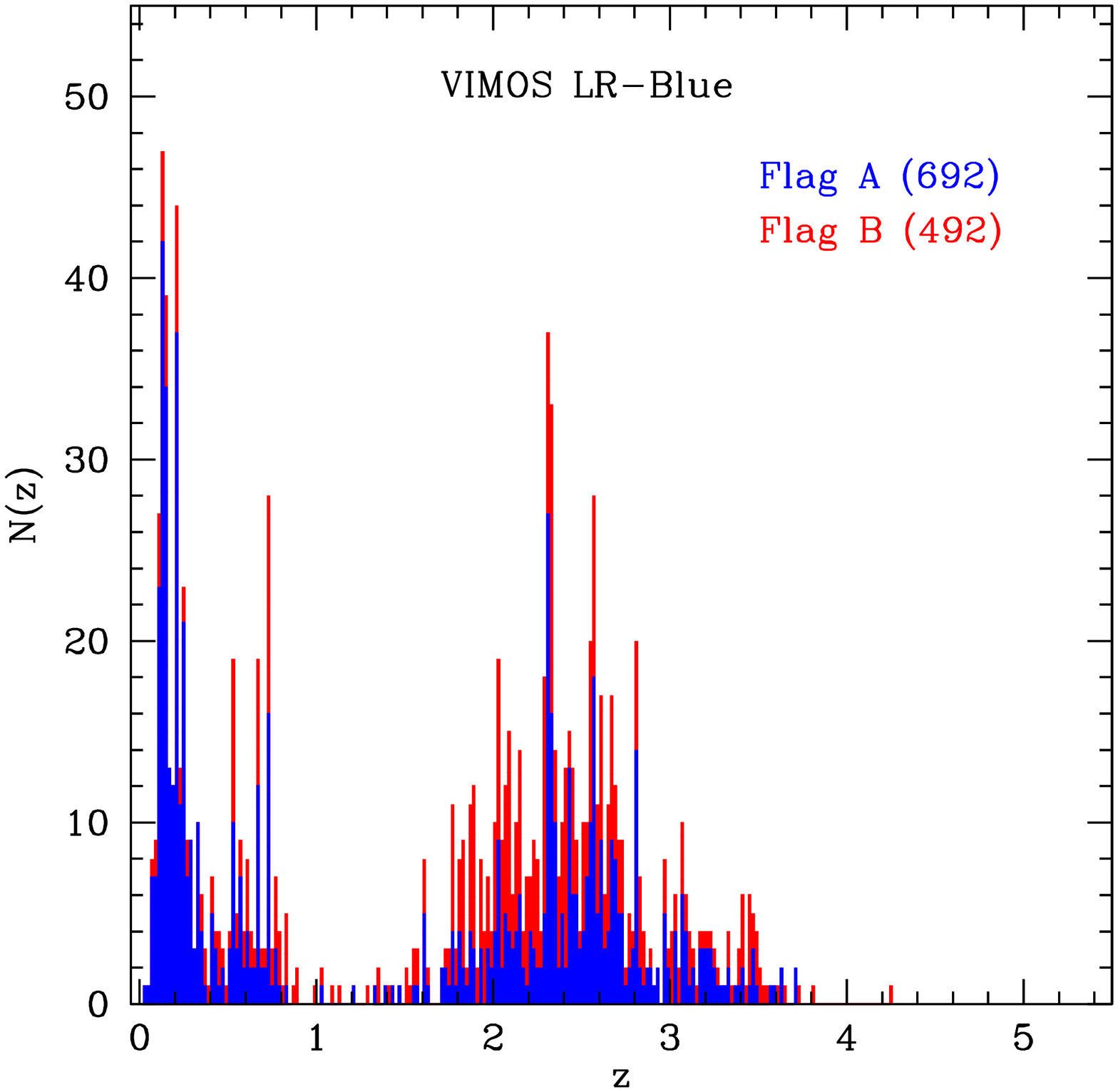}
\includegraphics[width=7.5 cm, angle=0]{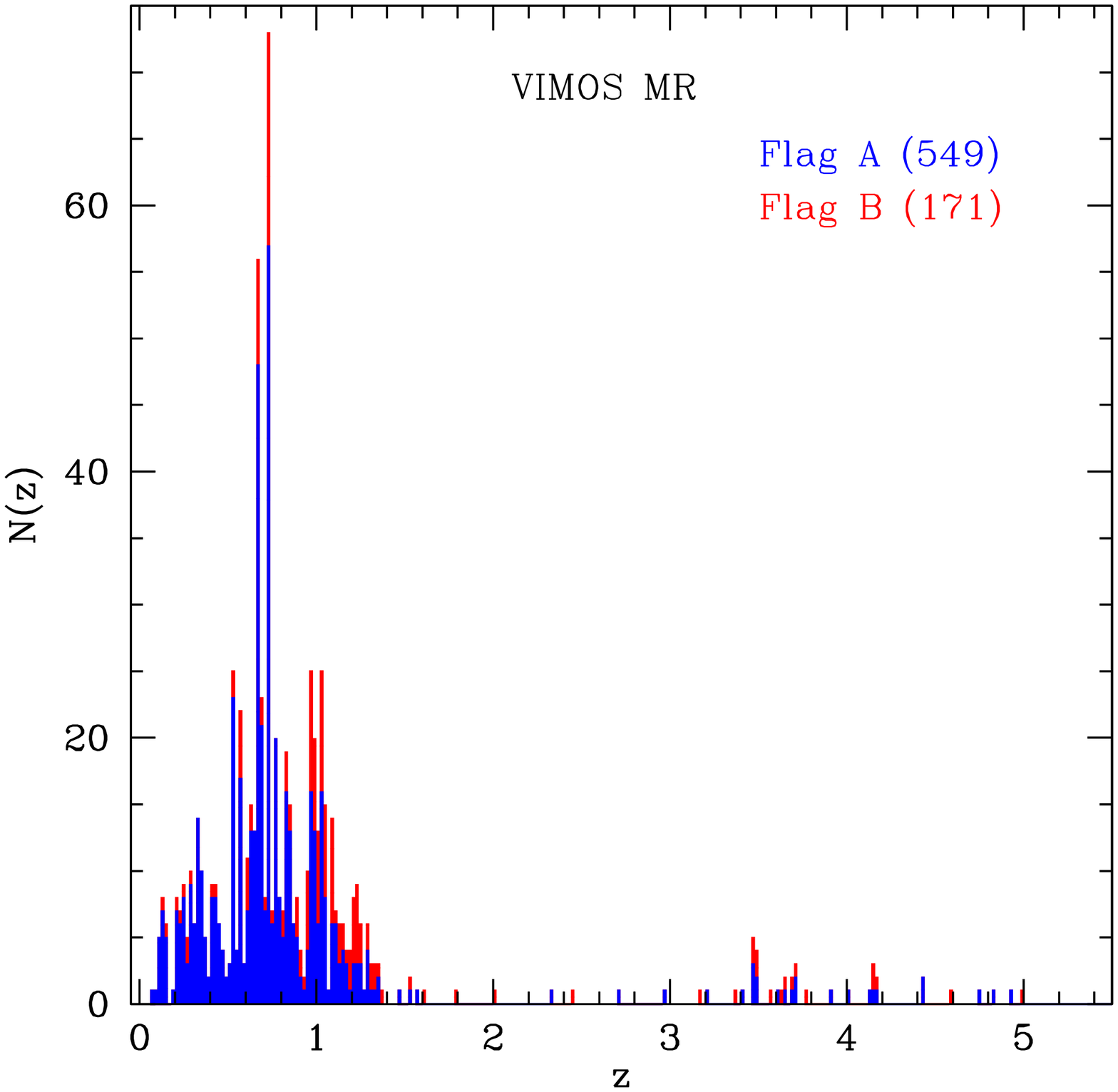}
\includegraphics[width=7.5 cm, angle=0]{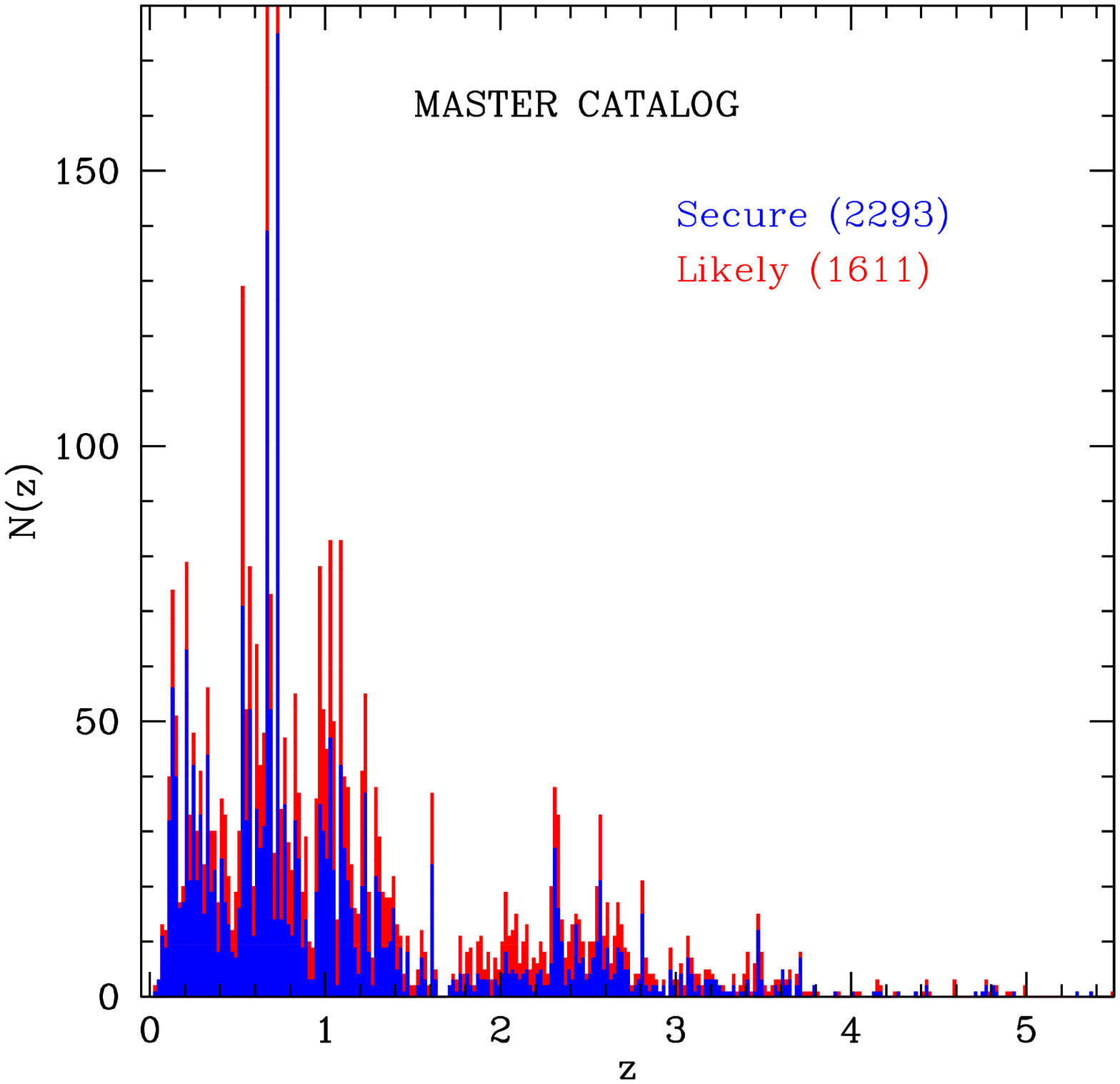}
\caption{Redshift distribution of the VIMOS LR-Blue survey ({\em upper panel}), 
VIMOS MR survey ({\em mid panel}), and master compilation of redshifts from
independent surveys in the CDFS ({\em lower panel}). Redshifts of the master 
compilation were subdivided into ``secure'' and ``likely'' as explained in the text 
(see Sect.~\ref{zdist}).
In the 3 histograms the size of the redshift bin is 0.02. Stars are not plotted.}
\label{zdb}
\end{figure}

The fine-grain redshift distribution of galaxies in the VIMOS LR-Blue and MR catalogs is
shown in Figure~\ref{zdb}, where only objects with redshift quality A and B
are plotted. We also plot the redshift distribution of the reference 
master catalog, described in Sect.~\ref{master}, which collects 4227 objects 
(including stars) having high-quality redshift determinations (confidence level $\ga60\%$) 
from all the publicly available spectroscopic surveys in the CDFS. In order to 
standardize as much as possible the quality flag used in different surveys, we subdivided
the master catalog into two subsamples: one including only ``secure'' redshifts 
(reliable at $\ga99\%$ c.l.) comparable to VIMOS quality-A redshifts and another one 
including ``likely'' redshift determinations (reliable at $\sim70-90\%$ c.l.) comparable to 
VIMOS quality-B redshifts.
A similar sample was used by P09 to assess the significance 
of the observed large scale structures. In P09 the significance of the observed peaks
of the redshift distribution was investigated through a procedure similar to that 
used by \citet{coh99} and \citet{gil03}: sources are distributed as 
a function of $V=c\ln(1+z)$, rather than $z$, where $\mathrm{d}V$ corresponds to the 
local velocity variation relative to the Hubble expansion. Simulations were used to 
assess the significance of the peaks in the redshift distribution.
The procedure used in P09 allowed to confirm the presence of 14 peaks.

Here, we use the results of the previous analysis to investigate the spatial 
distribution of galaxies belonging to each of the 14 confirmed peaks, plus 
three additional tentative structures at $z\simeq2.8$, $z\simeq3.5$, and  $z\simeq3.7$
in our new, extended, reference master sample (see Fig.~\ref{cones}; see also 
Table~\ref{peaks} and Fig.~\ref{654}--\ref{280} in the Appendix). 

\begin{figure*}
\centering
\includegraphics[width=16 cm, angle=0]{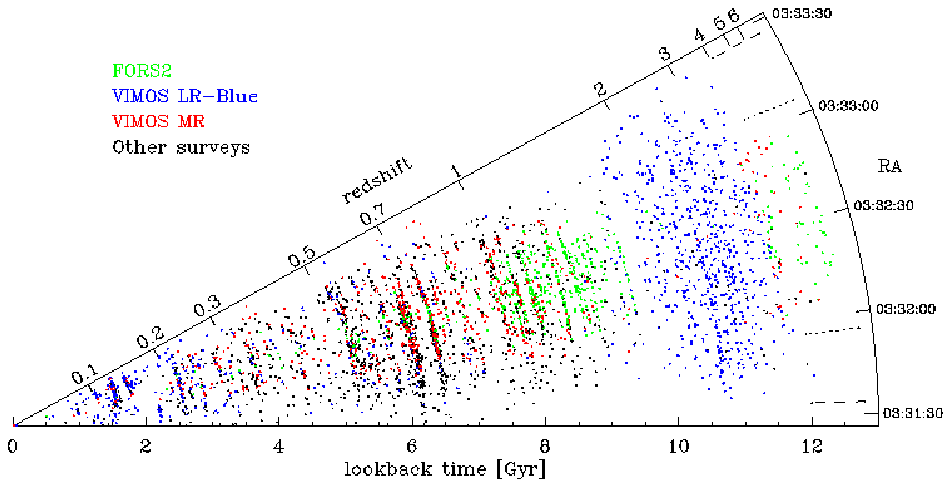}
\includegraphics[width=16 cm, angle=0]{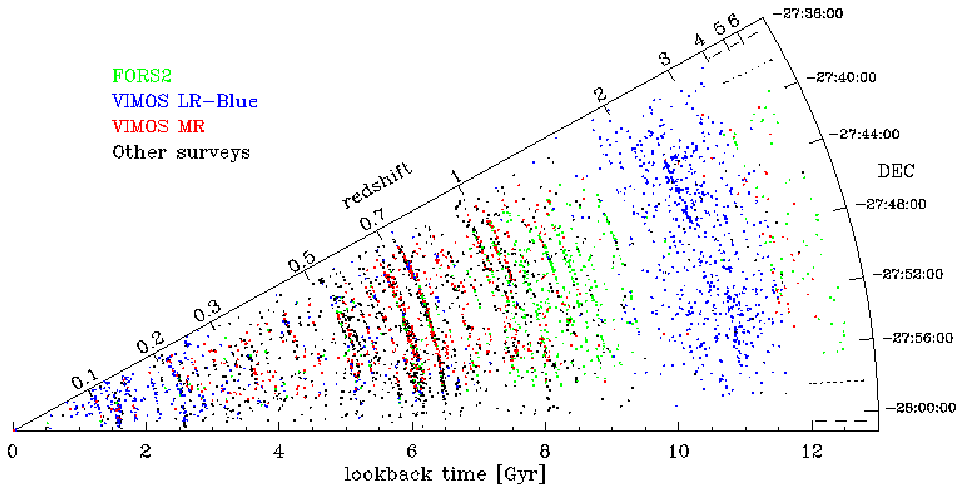}
\caption{Cone plots showing two projections of the spatial distribution of galaxies 
from the master catalog of redshifts of all the publicly available surveys in the 
CDFS described in the text (see Sect.~\ref{zdist} and Fig.~\ref{zdb}). 
Different spectroscopic surveys are color-coded. The \textit{dashed} and \textit{dotted lines} 
indicate the approximate size of the 2Ms CDFS and the GOODS area, respectively. 
The angle of the cones has been arbitrarily stretched to help visualization.}
\label{cones}
\end{figure*}

Three structures, having the size of small clusters/groups of galaxies, 
(i.e. at $z\simeq0.53$, $z\simeq0.67$, and $z\simeq0.73$.) have associated 
extended X-ray emission. All of them are well-known structures, also observed by 
other spectroscopic surveys (e.g. GOODS-FORS2, K20).

The galaxies belonging to the cluster-like structure at $z\simeq0.67$ are rather 
uniformly distributed on the whole CDFS area. 
Unfortunately, the extended X-ray source XID~645, most likely corresponding to the 
cluster core, lies in a region not entirely covered by the available 
spectroscopic surveys.

The second cluster-like structure, at $z\simeq0.73$, is mostly composed of early-type galaxies 
concentrated around the central cD galaxy. The two extended X-ray sources 
(XID 566, 594) associated with this structure are located approximately at the 
position of the central cD galaxy and $\sim5'$ to the north.

Another interesting cluster-like structure is the one at $z\sim1.22$.
The spatial distribution of the galaxies belonging to this structure 
is quite concentrated and elongated approximately along the east-west direction.
Most of the galaxies in this sturcture are concentrated on an area of 
approximately $5'$ radius, corresponding to $\sim2.5$~Mpc at $z=1.22$.

It is worth noticing that several structures, including those at $z\simeq2.3$ and 
$z\simeq2.6$, extend over the entire surveyed area. This indicates that the size
of these distant large structures must be of the order of $\sim14-15$~Mpc. 

At $z>3$ no over-densities were confirmed by the previous analysis in P09. However,
\citet{kan09} detected an overdensity of galaxies at $z\simeq3.7$ using
photometric redshifts.
In Figure~\ref{736} we  plot the spatial distribution of galaxies belonging 
to the two furthest density peaks in the GOODS/CDFS, at $z\sim3.48$ and $z\sim3.70$,
also observed by \citet{van09}.
Interestingly, the 29 galaxies at $z\sim3.48$ and the 11 galaxies at $z\sim3.70$ 
appear to be concentrated around the same area. 

\subsection{Photometric selection of LBGs at $z\simeq3$}\label{pholbg}

Several photometric techniques have been used to select galaxies at $1<z<3$ in 
the various GOODS-S spectroscopic surveys. In P09 the reliability of the $BzK$ 
and ``sub''-U-dropouts selection criteria have been checked against 
contamination due to foreground interlopers. 

\begin{figure}
\centering
\includegraphics[width=9.0 cm, angle=0]{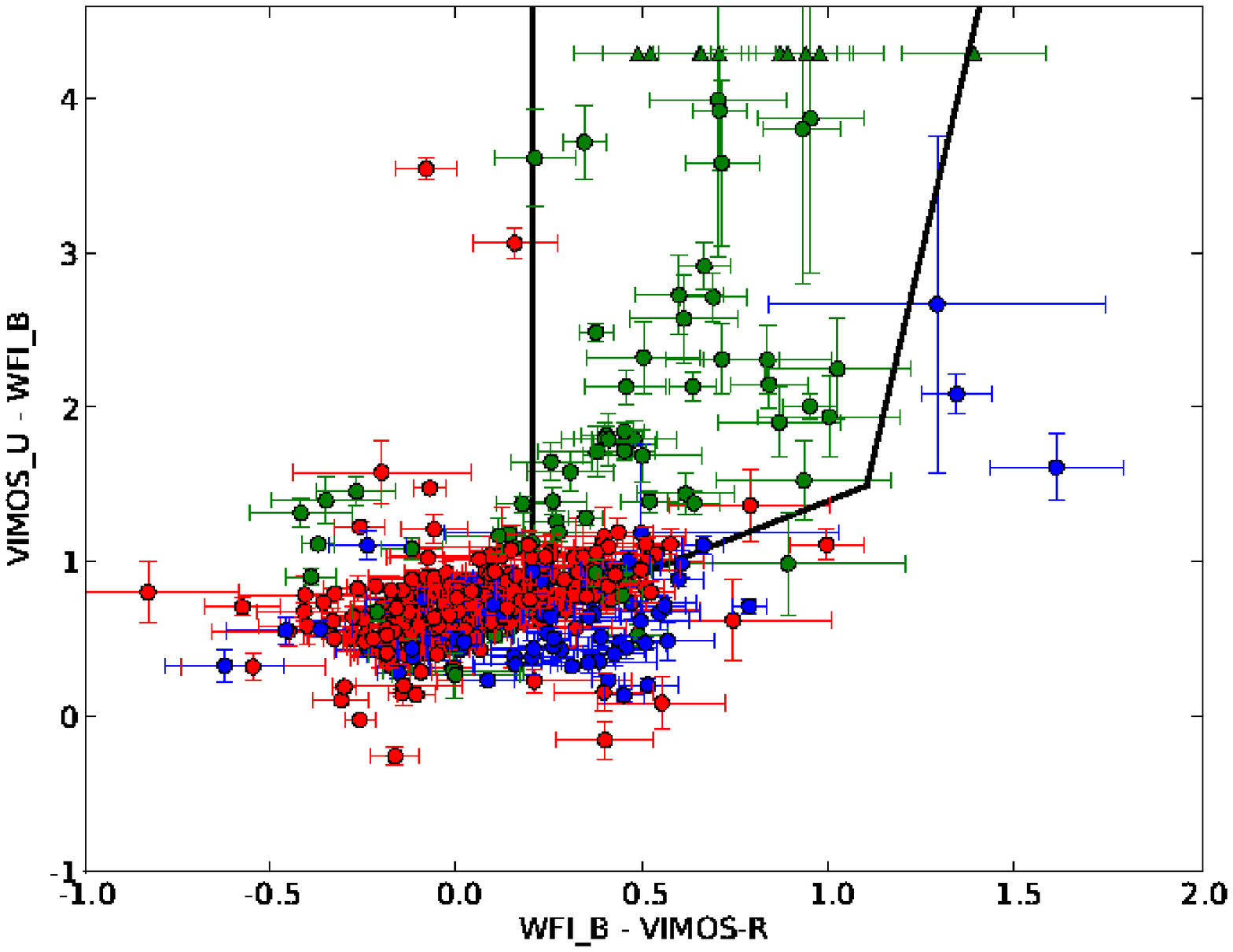}
\includegraphics[width=9.0 cm, angle=0]{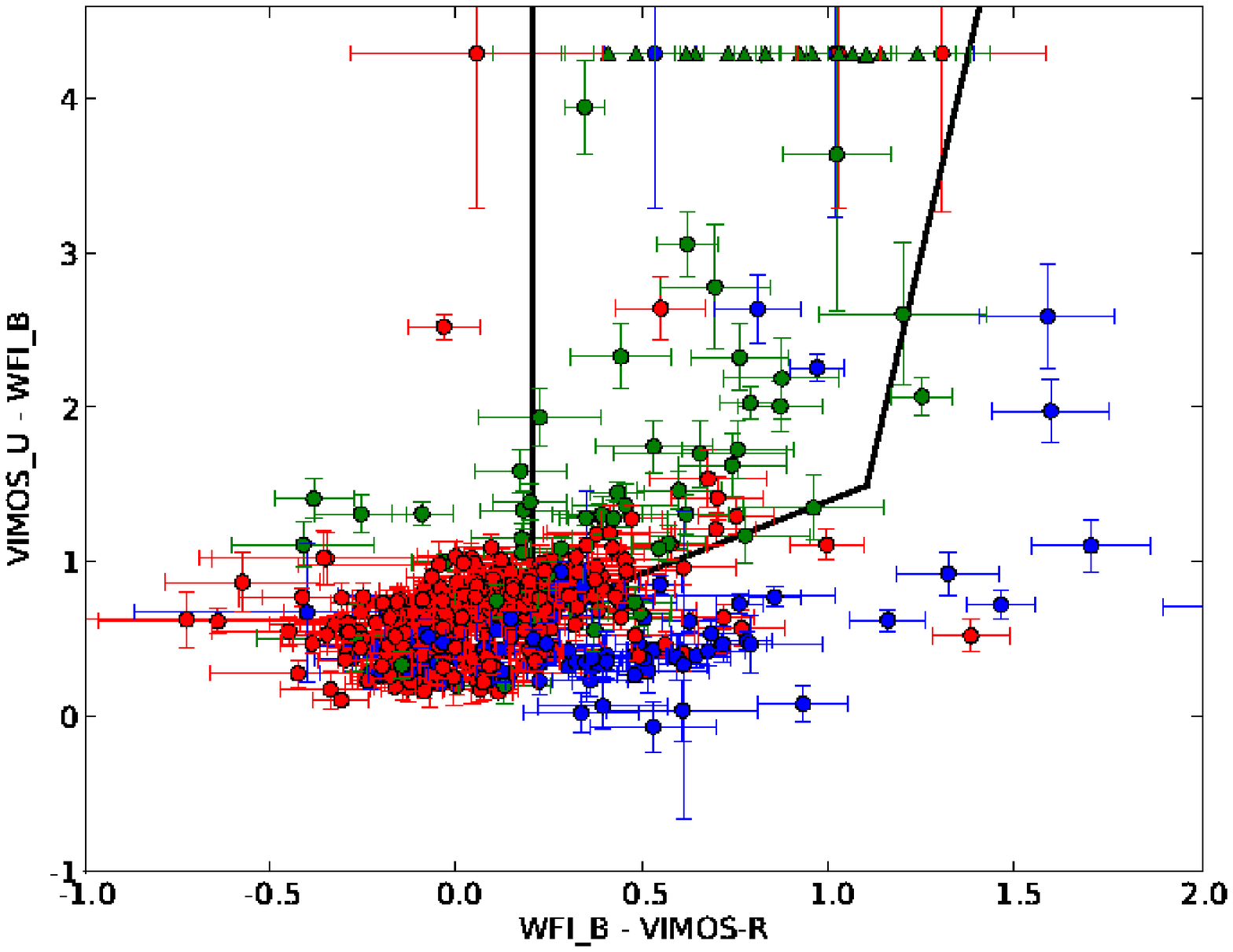}
\caption{$(U-B)-(B-R)$ color--color diagrams for VIMOS LR-Blue objects matching 
with $R$-band sources, $23.5\leq R_{MAG\_AUTO} \leq27.0$, and 
$\sigma(R_{MAG\_AUTO})\leq0.1$, and having redshift 
quality flag A {(\em upper panel)} and B {(\em lower panel)}. 
The redshift is color coded: objects at $z<1.8$ in blue, $1.8\leq z\leq2.8$ in red, 
and $z>2.8$ in green. Triangles denote lower limits at 1$\sigma$ c.l. 
(30 AB magnitude in $U$ band).}
\label{UbR}
\end{figure}

Recently, deep VIMOS $U$-band photometry have been used to test another 
efficient criterion for the selection of Lyman Break Galaxies \citep{non09}.
Here we use the $(U-B)$ and $(B-R)$ colors computed by \citet{non09} to
test the reliability of the $(U-B)-(B-R)$ selection criterion on
our VIMOS LR-Blue catalog of spectroscopic redshifts.
Figure \ref{UbR} shows the $(U-B)-(B-R)$ color--color diagram for VIMOS LR-Blue 
objects matching within an angular tolerance of $1''$ with $R$-band sources 
within the ACS image, having 
$23.5\leq R_{MAG\_AUTO} \leq27.0$, and error $\sigma(R_{MAG\_AUTO})\leq0.1$.
Error bars in the plots are as in \citet{non09}.
The selection box is analogous to that defined in \citet{non09} with the only
difference that here the WFI $B$-band filter is used instead of the ACS $B$ filter.
The plot of quality-A redshift estimates confirms that the selection box
is quite efficient in selecting galaxies at $z>2.8$. However, a small 
contamination by objects at smaller $z$ is still present at $(U-B)<1$.
VIMOS LR-Blue quality-B redshifts are less reliable and show more mixing which 
may be due to mismatches in spectroscopic redshift measurements.

\subsection{Ly-$\alpha$ EW vs UV Luminosity}\label{stlya}

After the completion of the GOODS/VIMOS spectroscopic campaign we identified 
a large number of LBGs: 288 in the LR-Blue campaign have a secure (i.e. quality flag=A) 
redshift at $2\la z\la4$ and 22 in the MR campaign have a secure at redshift 
$z\gse3$. Stacked spectra of all the LBGs available in the first release of the 
GOODS/VIMOS spectroscopic campaign were presented in P09. 

\begin{figure}
\centering
\includegraphics[width=9.0 cm, angle=0]{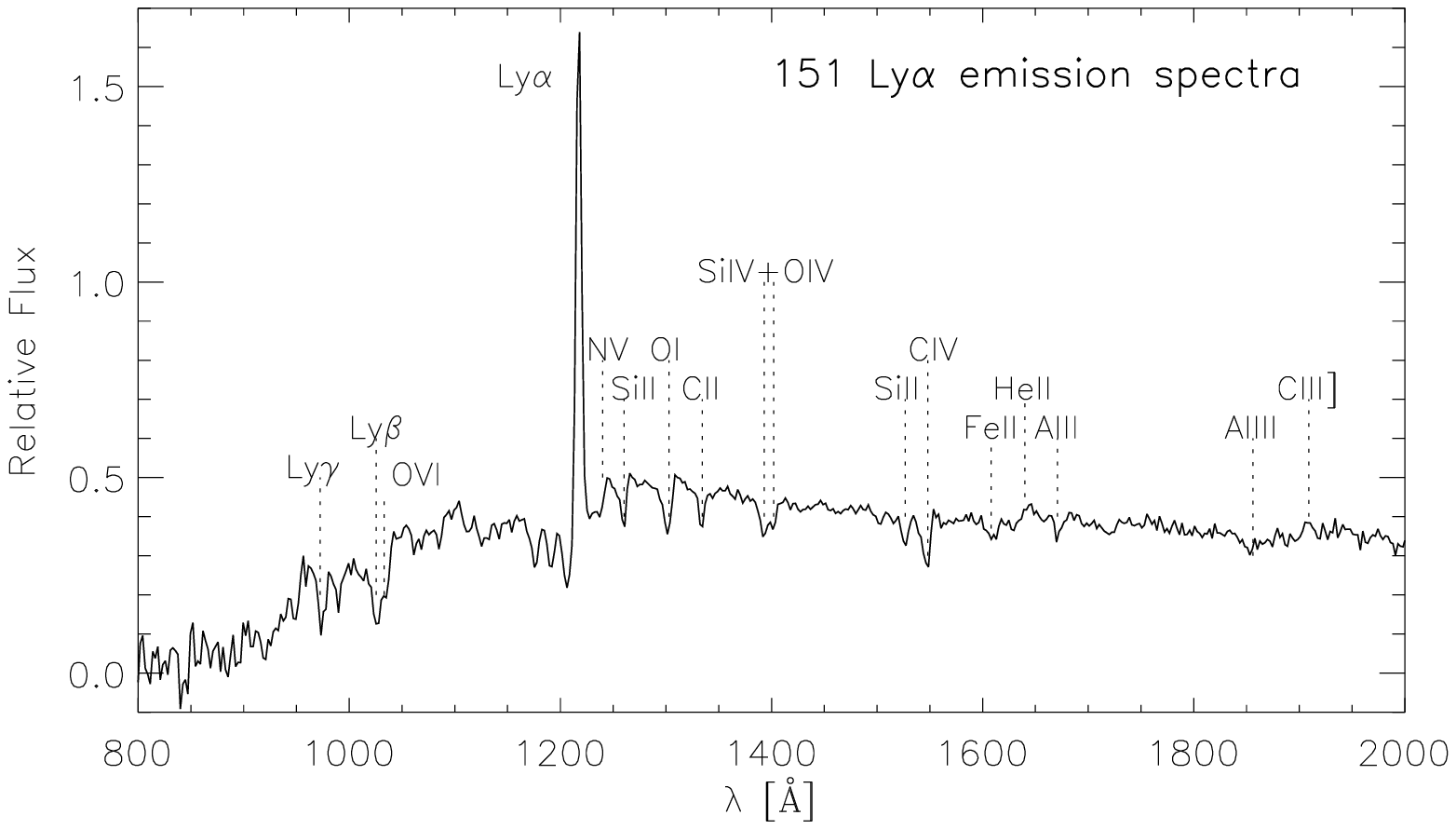}
\includegraphics[width=9.0 cm, angle=0]{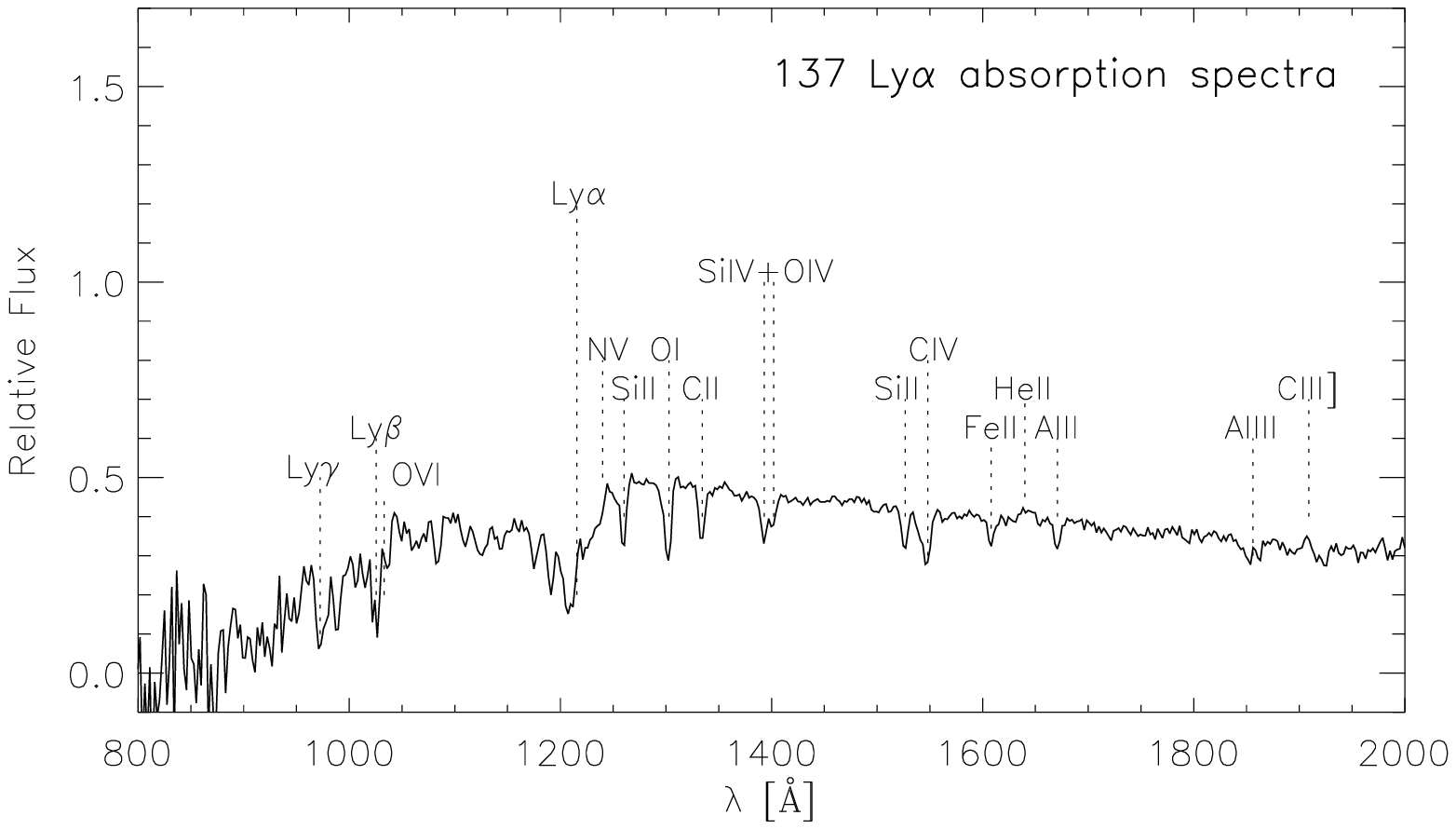}
\caption{Co-added LR-Blue spectra of 151 LBGs with Ly-$\alpha$ in emission 
{(\em upper panel)} and 137 with Ly-$\alpha$ in absorption {(\em lower panel)}.
Only high-quality (i.e. quality flag=A) spectra of LBGs at $2<z<4$ have been 
selected for stacking.}
\label{stLya}
\end{figure}

Here, we extend and 
refine the previous analysis by collecting 288 LBGs with high-quality spectra 
in the LR-Blue campaign. In Fig.~\ref{stLya} we plot the composite spectra of 
two sub-samples: 151 LBGs exhibiting Ly-$\alpha$ in emission (upper panel) and 137 
with Ly-$\alpha$ in absorption (lower panel). Low-ionization interstellar 
absorption lines appear more pronounced in the composite spectrum of LBGs with 
Ly-$\alpha$ in absorption against a stronger UV continuum. The continuum has also 
a redder spectral slope for this population compared to the Ly-$\alpha$ emitters, 
consistent with other studies of $z\simeq3$ LBGs samples \citep[e.g.][]{sha03,pen07}.
This becomes more evident when the spectra of Ly-$\alpha$ ``emitters'' are stacked
as a function of the Ly-$\alpha$ equivalent width (EW), as described below.

\begin{figure}
\centering
\includegraphics[width=9.0 cm, angle=0]{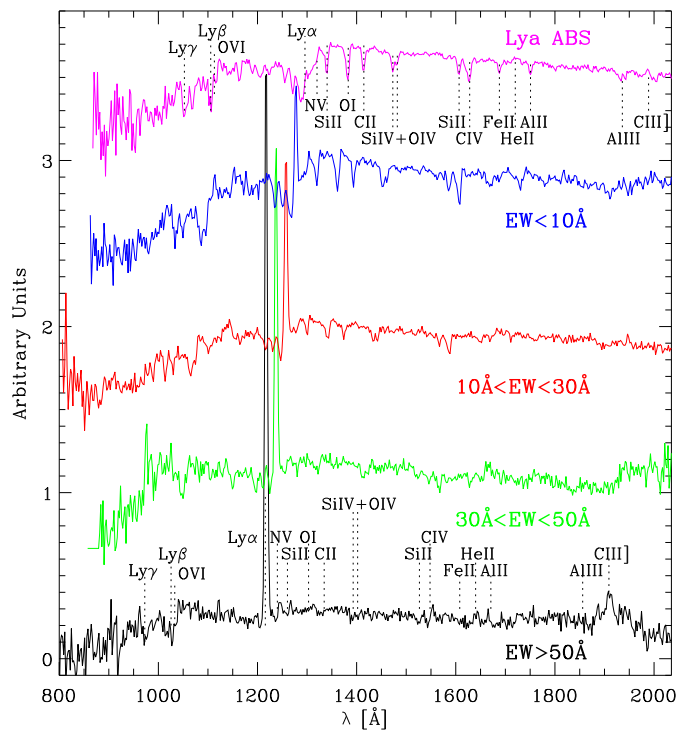}
\caption{Co-added LR-Blue spectra of LBGs with Ly-$\alpha$ in emission.
Spectra are stacked as a function of rest frame EW of the Ly-$\alpha$.
We defined 4 bins of EW as follows: EW$<10\,\AA$ (55 galaxies), 
$10\,\AA<$EW$<30\,\AA$ (62 galaxies), $30\,\AA<$EW$<50\,\AA$ (17 galaxies), and 
EW$>50\,\AA$ (15 galaxies). We also plot the stacked spectrum of the 137 LBGs with
Ly-$\alpha$ in absorption for a comparison. The spectra have been offset by an
arbitrary factor along both axes for easier viewing.}
\label{stEW}
\end{figure}

For the sub-sample of 151 Ly-$\alpha$ emitters we calculated the EW 
of the Ly-$\alpha$ and the absolute magnitude at $1450\,\AA$ ($M_{145}$).
The EW was measured from the spectra by estimating the average continuum level 
from a spectral band immediatly red-ward of the Ly-$\alpha$. Errors on the EWs were 
calculated by propagation of uncertainties, using the $1\sigma$-error on
the continuum.
The absolute $M_{145}$ magnitude was
derived from the $z_{850}$-band, assuming a template of a star forming galaxy from 
\citet{bru03}, with constant star formation and spectral slope $\beta\simeq-2$.
Fig.~\ref{stEW} shows a comparison between the composite spectra of Ly-$\alpha$ 
emitters stacked in 4 bins of EW: 55 galaxies with EW$<10\,\AA$, 
62 with $10\,\AA<$EW$<30\,\AA$, 17 with $30\,\AA<$EW$<50\,\AA$, and 
15 with EW$>50\,\AA$. A trend of stronger absorption lines and redder spectral slopes 
with decreasing EW of the Ly-$\alpha$ is clearly evident. For a comparison, we also 
plot the stacked spectrum of LBGs with Ly-$\alpha$ in absorption, which extends 
the observed trends toward lower EWs.

\begin{figure}
\centering
\includegraphics[width=9.0 cm, angle=0]{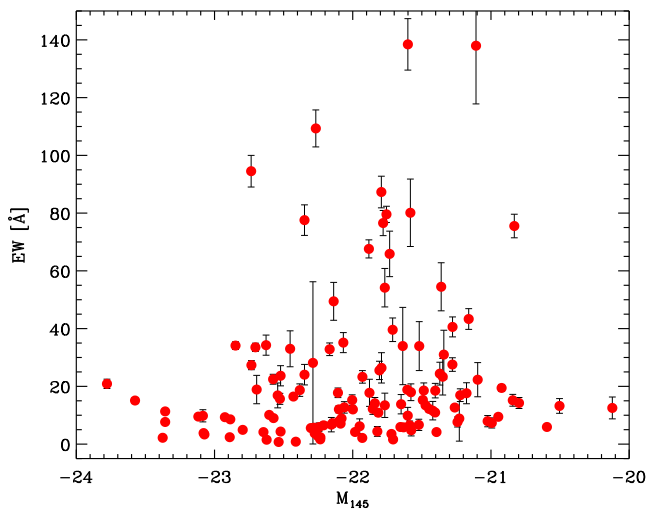}
\caption{Rest frame Ly-$\alpha$ EW as a function of UV luminosity (M$_{145}$, 
absolute magnitude at $1450\,\AA$). Error bars refer to $1\sigma$ confidence level.}
\label{ew-m}
\end{figure}

In the redshift range probed by our sample ($2\la z\la4$), we find no clear 
evidence for a correlation between the EW of the Ly-$\alpha$ and redshift.
In Fig.~\ref{ew-m} we plot the distribution of EWs versus the absolute magnitude 
at $1450\,\AA$. The plot shows the absence of large EW of Ly-$\alpha$ 
in bright galaxies, which confirms results based on different samples of Ly-$\alpha$ 
emitters and LBGs at $3\la z\la6$ 
\citep[e.g.][]{sha03,aji03,and06,and07,tap07,ver08,van09,pen09}.

The EW of the Ly-$\alpha$ line depends on the expansion velocity of the 
inter-stellar medium, the column density of neutral gas, the dust extinction, and 
the geometry (or the ``clumpiness'') of the medium.
A possible explanation for the absence of strong Ly-$\alpha$  emission lines in 
the more luminous galaxies might be the presence of a more dusty and metal rich medium,
residual of a recent, or still ongoing, burst of star formation and supernovae 
explosions in these galaxies.

\subsection{New redshifts of X-ray sources}\label{xray}

X-ray sources of the CDFS \citep{gia02,luo08}
and the ECDFS \citep{leh05} have been targeted for follow-up 
optical spectroscopy \citep[][Silverman et al., in preparation]{szo04}.
In this Section we present new (i.e. not yet published) redshifts for X-ray sources
in the CDFS and ECDFS.
We cross-correlated our VIMOS catalogs with both the 2Ms CDFS catalog of X-ray sources 
\citep{luo08} and the catalog of the ECDFS \citep{leh05}.
For the 2Ms CDFS catalog, we found 62 X-ray sources matching within an angular 
tolerance of $1''$ objects having redshift determination in the 
VIMOS LR-Blue catalog and 42 in the VIMOS MR catalog.
For the ECDFS catalog, we found 57 X-ray sources matching within a positional
tolerance of $1''$ objects having redshift determination in the VIMOS LR-Blue catalog and 
17 in the VIMOS MR catalog. 

In total 114 X-ray sources have been observed in the GOODS/VIMOS spectroscopic 
campaign. In Table~\ref{tabcdfs} we list 12 new (not yet published) redshifts, either 
unknown before or improving previous estimates, obtained from the last 8 VIMOS masks
released in this paper. Spectra and finding charts of these sources can be found 
in the Appendix~\ref{app2}.

\begin{table*}
\caption{List of 12 new spectroscopic redshifts for X-ray sources of the 2Ms 
CDFS catalog \citep{luo08} and of the ECDFS catalog \citep{leh05}. Columns list 
the following information:  (1) coordinate-based GOODS identification number, 
(2) VIMOS identification number, (3-4) coordinates in WFI reference 
astrometry, (5) X-ray identification number from \citep{luo08}, (6) X-ray 
identification number from \citep{leh05}, (7) redshift, (8) quality flag, and (9)
comments and identified spectral features.}
\begin{center}
\begin{footnotesize}
\begin{tabular}{l l c c r r c c l}
\hline
\hline
GOODS\_ID           & VIMOS\_ID		& RA  & DEC 	 & \multicolumn{2}{c}{XID} & z   &  Q   & Comments			         \\
   (1)      	    & (2)               & (3) & (4)      & (5) & (6) & (7) & (8)  & (9)         			         \\ 
\hline
J033143.17-274131.0 & GOODS\_LRb\_002\_1\_q2\_33\_1	   & 52.929871 &  -27.691935      & -   & 135  &  2.9981 &  C  & Ly$\alpha$(em),CIV(em)			\\
J033146.45-274123.6 & GOODS\_LRb\_002\_1\_q2\_44\_1	   & 52.943559 &  -27.689879      & -   & 160  &  0.6646 &  B  & OII					\\
J033155.33-274313.6 & GOODS\_LRb\_002\_1\_q2\_61\_1	   & 52.980564 &  -27.720448      & -   & 225  &  $0.8610^a$ &  B  & MgII,FeII				\\
J033214.98-274225.0 & GOODS\_LRb\_002\_1\_q1\_6\_1	   & 53.062400 &  -27.706952      & 145 & 340  &  $2.6087^*$ &  B  & Ly$\alpha$(abs),OI,CII,AlIII		\\
J033216.13-275644.0 & GOODS\_LRb\_002\_1\_q3\_65\_1	   & 53.067204 &  -27.945558      & 157 & 344  &  $2.6667^b$ &  A  & Ly$\alpha$(em),SiIV,CIV (BLAGN)  	\\
J033218.25-275224.9 & GOODS\_MR\_dec06\_2\_q2\_19\_1	   & 53.076059 &  -27.873578      & 175 & -    &  0.7396 &  A  & OII,H$\beta$,OIII		    \\
J033220.05-274447.3 & GOODS\_LRb\_002\_1\_q1\_8\_1	   & 53.083537 &  -27.746460      & 188 & 357  &  $1.8930^*$ &  B  & CIV,CIII] (BLAGN)			\\
J033225.69-273941.2 & GOODS\_LRb\_dec06\_3\_q1\_2\_1	   & 53.107044 &  -27.661458      & -   & 370  &  0.0010 &  A  & Star					    \\
J033244.61-274835.9 & GOODS\_LRb\_002\_1\_q4\_66\_1	   & 53.185862 &  -27.809984      & 372 & -    &  $2.5860^*$ &  A  & Ly$\alpha$(em),OI,CIV,FeII		\\
J033250.25-275251.9 & GOODS\_MR\_dec06\_3\_q3\_2\_2	   & 53.209362 &  -27.881094      & 402 & 480  &  3.4742 &  B  & Ly$\alpha$(em)				    \\
J033304.81-274731.7 & GOODS\_LRb\_dec06\_1\_q4\_69\_1	   & 53.270059 &  -27.792149      & 445 & -    &  2.0265 &  A  & Ly$\alpha$(em),CIV,HeII,CIII] (BLAGN)	\\
J033306.80-274626.6 & GOODS\_LRb\_dec06\_1\_q4\_79\_2	   & 53.278332 &  -27.774065      & -   & 571  &  1.7048 &  A  & SiIV,SiII,CIV,FeII,AlII  		    \\
\hline																					
\end{tabular}	
\end{footnotesize}
\end{center}																				
$^*$ Redshift more reliable compared to previously published estimates. \\
$^a$ Redshift more reliable then the second available measurement: 
GOODS\_LRb\_dec06\_1\_q2\_15\_2 $z=2.3634$ flag C. \\
$^b$ Redshift resulting from average of two measurements: 
GOODS\_LRb\_002\_1\_q3\_65\_1 $z=2.6652$ flag A and 
GOODS\_LRb\_dec06\_1\_q3\_16\_1 $z=2.6681$ flag A.
\label{tabcdfs}
\end{table*}

\section{Conclusions}

After the completion of the two complementary ESO/GOODS spectroscopic campaigns 
carried out with FORS2 \citep{van05,van06,van08} and VIMOS (P09 and this work), 
a very large sample of galaxies in the CDFS has been spectroscopically targeted. 
In this paper we presented the final data release, including also data from the 
previous release (P09), of the GOODS/VIMOS spectroscopic campaign, which was organized
in two separated campaign: one using the LR-Blue grism and one using the MR grism.

The main outcome for the LR-Blue campaign can be summarized as follows:
\begin{itemize}
\item a total of 3634 spectra has been extracted, providing 2242 redshift 
measurements with a typical $\sigma_{z}\simeq0.0012$ ($\sim255\,\rm{km\,s}^{-1}$). 
We obtained a redshift determination for 2040 out of the 3271 individual 
targets observed. We assigned a quality flag to the redshift measurements, which
provides an estimate of their reliability. The reliability of VIMOS LR-Blue 
redshifts with quality flag A, B, and C is estimated to be approximately
$100\%$, $60\%$, and $20\%$ confidence level, respectively.
The number of redshifts determinations above the $60\%$ c.l. (i.e. flag A+B) 
amounts to 1395.
\end{itemize}

For the MR campaign the main results can be summarized as follows:
\begin{itemize}
\item a total of 1418 spectra has been extracted, providing 976 redshift 
measurements with a typical $\sigma_{z}\simeq0.00040$ ($\sim120\,\rm{km\,s}^{-1}$). 
We obtained a redshift determination for 882 out of the 1294 individual 
targets observed. VIMOS MR quality flag-A, -B, and -C redshifts are estimated to 
be reliable at approximately $100\%$, $95\%$, and $60\%$ confidence level, respectively.
\end{itemize}

We complemented our VIMOS spectroscopic catalog with all existing spectroscopic 
redshifts publicly available in the CDFS and obtained a redshift master catalog 
with 7332 entries. We used good-quality redshifts (c.l. $>60\%$)
to investigate the spatial distribution of galaxies in 16 peaks of the redshift
distribution, tracing large scale structures out to $z\simeq3.7$. 

Stacked spectra of LBGs were produced in a few bins of Ly-$\alpha$ EW.
We found evidence for a lack of bright objects with high EW of the Ly-$\alpha$, 
which confirms results based on different samples of Ly-$\alpha$ emitters and 
LBGs at $3\la z\la6$ \citep[e.g.][]{sha03,aji03,and06,and07,tap07,ver08,van09,pen09}. 

Additionally, we obtained new redshifts for 12 X-ray sources of the CDFS and ECDFS.
These sources also appear in the new catalog of X-ray sources detected in the ECDFS, 
which includes the identification of optical and near-IR counterparts 
(Silverman et al., in preparation).

The reduced spectra and the redshift catalogs are released to the community and 
can be retrieved in electronic form at 
\textit{http://archive.eso.org/cms/eso-data/data-packages}.
These data, in combination with the other spectroscopic campaigns in the GOODS-S field, 
represent an essential contribution to achieve the scientific goals of 
GOODS, providing the lookback time across which the evolution of galaxy 
masses, morphologies, clustering, and star formation can be traced.

\begin{acknowledgements}
We acknowledge the ESO staff in Paranal and Garching for the help
in the development of this program (ESO program 171.A-3045 {\it The Great
Observatories Origins Deep Survey: ESO Public Observations of the
SIRTF Legacy/HST Treasury/Chandra Deep Field South.}). 
We wish to thank Marcella Brusa for many helpful suggestions and Remco Slijkhuis and
Joerg Retzlaff for their work on the GOODS/VIMOS relesase.
We thank the anonymous referee for the valuable comments and suggestions.
\end{acknowledgements}

\bibliographystyle{aa}

\begin{thebibliography}{40}
\expandafter\ifx\csname natexlab\endcsname\relax\def\natexlab#1{#1}\fi

\bibitem[{{Ajiki} {et~al.}(2003){Ajiki}, {Taniguchi}, {Fujita}, {Shioya},
  {Nagao}, {Murayama}, {Yamada}, {Umeda}, \& {Komiyama}}]{aji03}
{Ajiki}, M., {Taniguchi}, Y., {Fujita}, S.~S., {et~al.} 2003, \aj, 126, 2091

\bibitem[{{Ando} {et~al.}(2006){Ando}, {Ohta}, {Iwata}, {Akiyama}, {Aoki}, \&
  {Tamura}}]{and06}
{Ando}, M., {Ohta}, K., {Iwata}, I., {et~al.} 2006, \apjl, 645, L9

\bibitem[{{Ando} {et~al.}(2007){Ando}, {Ohta}, {Iwata}, {Akiyama}, {Aoki}, \&
  {Tamura}}]{and07}
{Ando}, M., {Ohta}, K., {Iwata}, I., {et~al.} 2007, \pasj, 59, 717

\bibitem[{{Bruzual} \& {Charlot}(2003)}]{bru03}
{Bruzual}, G. \& {Charlot}, S. 2003, \mnras, 344, 1000

\bibitem[{{Bunker} {et~al.}(2003){Bunker}, {Stanway}, {Ellis}, {McMahon}, \&
  {McCarthy}}]{bun03}
{Bunker}, A.~J., {Stanway}, E.~R., {Ellis}, R.~S., {McMahon}, R.~G., \&
  {McCarthy}, P.~J. 2003, \mnras, 342, L47

\bibitem[{{Cimatti} {et~al.}(2002){Cimatti}, {Mignoli}, {Daddi}, {Pozzetti},
  {Fontana}, {Saracco}, {Poli}, {Renzini}, {Zamorani}, {Broadhurst},
  {Cristiani}, {D'Odorico}, {Giallongo}, {Gilmozzi}, \& {Menci}}]{cim02}
{Cimatti}, A., {Mignoli}, M., {Daddi}, E., {et~al.} 2002, \aap, 392, 395

\bibitem[{{Cohen} {et~al.}(1999){Cohen}, {Blandford}, {Hogg}, {Pahre}, \&
  {Shopbell}}]{coh99}
{Cohen}, J.~G., {Blandford}, R., {Hogg}, D.~W., {Pahre}, M.~A., \& {Shopbell},
  P.~L. 1999, \apj, 512, 30

\bibitem[{{Cristiani} {et~al.}(2000){Cristiani}, {Appenzeller}, {Arnouts},
  {Nonino}, {Arag{\'o}n-Salamanca}, {Benoist}, {da Costa}, {Dennefeld},
  {Rengelink}, {Renzini}, {Szeifert}, \& {White}}]{cri00}
{Cristiani}, S., {Appenzeller}, I., {Arnouts}, S., {et~al.} 2000, \aap, 359,
  489

\bibitem[{{Croom} {et~al.}(2001){Croom}, {Warren}, \& {Glazebrook}}]{cro01}
{Croom}, S.~M., {Warren}, S.~J., \& {Glazebrook}, K. 2001, \mnras, 328, 150

\bibitem[{{Daddi} {et~al.}(2004){Daddi}, {Cimatti}, {Renzini}, {Fontana},
  {Mignoli}, {Pozzetti}, {Tozzi}, \& {Zamorani}}]{dad04}
{Daddi}, E., {Cimatti}, A., {Renzini}, A., {et~al.} 2004, \apj, 617, 746

\bibitem[{{Dickinson} {et~al.}(2003){Dickinson}, {Giavalisco}, \& {The Goods
  Team}}]{dic03}
{Dickinson}, M., {Giavalisco}, M., \& {The Goods Team}. 2003, in The Mass of
  Galaxies at Low and High Redshift, ed. R.~{Bender} \& A.~{Renzini}, 324--+

\bibitem[{{Dickinson} {et~al.}(2004){Dickinson}, {Stern}, {Giavalisco},
  {Ferguson}, {Tsvetanov}, {Chornock}, {Cristiani}, {Dawson}, {Dey},
  {Filippenko}, {Moustakas}, {Nonino}, {Papovich}, {Ravindranath}, {Riess},
  {Rosati}, {Spinrad}, \& {Vanzella}}]{dic04}
{Dickinson}, M., {Stern}, D., {Giavalisco}, M., {et~al.} 2004, \apjl, 600, L99

\bibitem[{{Giacconi} {et~al.}(2002){Giacconi}, {Zirm}, {Wang}, {Rosati},
  {Nonino}, {Tozzi}, {Gilli}, {Mainieri}, {Hasinger}, {Kewley}, {Bergeron},
  {Borgani}, {Gilmozzi}, {Grogin}, {Koekemoer}, {Schreier}, {Zheng}, \&
  {Norman}}]{gia02}
{Giacconi}, R., {Zirm}, A., {Wang}, J., {et~al.} 2002, \apjs, 139, 369

\bibitem[{{Giavalisco} {et~al.}(2004){Giavalisco}, {Ferguson}, {Koekemoer},
  {Dickinson}, {Alexander}, {Bauer}, {Bergeron}, {Biagetti}, {Brandt},
  {Casertano}, {Cesarsky}, {Chatzichristou}, {Conselice}, {Cristiani}, {Da
  Costa}, {Dahlen}, {de Mello}, {Eisenhardt}, {Erben}, {Fall}, {Fassnacht},
  {Fosbury}, {Fruchter}, {Gardner}, {Grogin}, {Hook}, {Hornschemeier}, {Idzi},
  {Jogee}, {Kretchmer}, {Laidler}, {Lee}, {Livio}, {Lucas}, {Madau},
  {Mobasher}, {Moustakas}, {Nonino}, {Padovani}, {Papovich}, {Park},
  {Ravindranath}, {Renzini}, {Richardson}, {Riess}, {Rosati}, {Schirmer},
  {Schreier}, {Somerville}, {Spinrad}, {Stern}, {Stiavelli}, {Strolger},
  {Urry}, {Vandame}, {Williams}, \& {Wolf}}]{gia04}
{Giavalisco}, M., {Ferguson}, H.~C., {Koekemoer}, A.~M., {et~al.} 2004, \apjl,
  600, L93

\bibitem[{{Gilli} {et~al.}(2003){Gilli}, {Cimatti}, {Daddi}, {Hasinger},
  {Rosati}, {Szokoly}, {Tozzi}, {Bergeron}, {Borgani}, {Giacconi}, {Kewley},
  {Mainieri}, {Mignoli}, {Nonino}, {Norman}, {Wang}, {Zamorani}, {Zheng}, \&
  {Zirm}}]{gil03}
{Gilli}, R., {Cimatti}, A., {Daddi}, E., {et~al.} 2003, \apj, 592, 721

\bibitem[{{Grazian} {et~al.}(2006){Grazian}, {Fontana}, {de Santis}, {Nonino},
  {Salimbeni}, {Giallongo}, {Cristiani}, {Gallozzi}, \& {Vanzella}}]{gra06}
{Grazian}, A., {Fontana}, A., {de Santis}, C., {et~al.} 2006, \aap, 449, 951

\bibitem[{{Kang} \& {Im}(2009)}]{kan09}
{Kang}, E. \& {Im}, M. 2009, \apjl, 691, L33

\bibitem[{{Le F{\`e}vre} {et~al.}(2005){Le F{\`e}vre}, {Vettolani}, {Garilli},
  {Tresse}, {Bottini}, {Le Brun}, {Maccagni}, {Picat}, {Scaramella},
  {Scodeggio}, {Zanichelli}, {Adami}, {Arnaboldi}, {Arnouts}, {Bardelli},
  {Bolzonella}, {Cappi}, {Charlot}, {Ciliegi}, {Contini}, {Foucaud},
  {Franzetti}, {Gavignaud}, {Guzzo}, {Ilbert}, {Iovino}, {McCracken}, {Marano},
  {Marinoni}, {Mathez}, {Mazure}, {Meneux}, {Merighi}, {Paltani}, {Pell{\`o}},
  {Pollo}, {Pozzetti}, {Radovich}, {Zamorani}, {Zucca}, {Bondi}, {Bongiorno},
  {Busarello}, {Lamareille}, {Mellier}, {Merluzzi}, {Ripepi}, \&
  {Rizzo}}]{lef05}
{Le F{\`e}vre}, O., {Vettolani}, G., {Garilli}, B., {et~al.} 2005, \aap, 439,
  845

\bibitem[{{LeFevre} {et~al.}(2003){LeFevre}, {Saisse}, {Mancini}, {Brau-Nogue},
  {Caputi}, {Castinel}, {D'Odorico}, {Garilli}, {Kissler-Patig}, {Lucuix},
  {Mancini}, {Pauget}, {Sciarretta}, {Scodeggio}, {Tresse}, \&
  {Vettolani}}]{lef03}
{LeFevre}, O., {Saisse}, M., {Mancini}, D., {et~al.} 2003, in Society of
  Photo-Optical Instrumentation Engineers (SPIE) Conference Series, Vol. 4841,
  Society of Photo-Optical Instrumentation Engineers (SPIE) Conference Series,
  ed. M.~{Iye} \& A.~F.~M. {Moorwood}, 1670--1681

\bibitem[{{Lehmer} {et~al.}(2005){Lehmer}, {Brandt}, {Alexander}, {Bauer},
  {Schneider}, {Tozzi}, {Bergeron}, {Garmire}, {Giacconi}, {Gilli}, {Hasinger},
  {Hornschemeier}, {Koekemoer}, {Mainieri}, {Miyaji}, {Nonino}, {Rosati},
  {Silverman}, {Szokoly}, \& {Vignali}}]{leh05}
{Lehmer}, B.~D., {Brandt}, W.~N., {Alexander}, D.~M., {et~al.} 2005, \apjs,
  161, 21

\bibitem[{{Luo} {et~al.}(2008){Luo}, {Bauer}, {Brandt}, {Alexander}, {Lehmer},
  {Schneider}, {Brusa}, {Comastri}, {Fabian}, {Finoguenov}, {Gilli},
  {Hasinger}, {Hornschemeier}, {Koekemoer}, {Mainieri}, {Paolillo}, {Rosati},
  {Shemmer}, {Silverman}, {Smail}, {Steffen}, \& {Vignali}}]{luo08}
{Luo}, B., {Bauer}, F.~E., {Brandt}, W.~N., {et~al.} 2008, \apjs, 179, 19

\bibitem[{{Mignoli} {et~al.}(2005){Mignoli}, {Cimatti}, {Zamorani}, {Pozzetti},
  {Daddi}, {Renzini}, {Broadhurst}, {Cristiani}, {D'Odorico}, {Fontana},
  {Giallongo}, {Gilmozzi}, {Menci}, \& {Saracco}}]{mig05}
{Mignoli}, M., {Cimatti}, A., {Zamorani}, G., {et~al.} 2005, \aap, 437, 883

\bibitem[{{Nonino} {et~al.}(2009){Nonino}, {Dickinson}, {Rosati}, {Grazian},
  {Reddy}, {Cristiani}, {Giavalisco}, {Kuntschner}, {Vanzella}, {Daddi},
  {Fosbury}, \& {Cesarsky}}]{non09}
{Nonino}, M., {Dickinson}, M., {Rosati}, P., {et~al.} 2009, \apjs, 183, 244

\bibitem[{{Pentericci} {et~al.}(2009){Pentericci}, {Grazian}, {Fontana},
  {Castellano}, {Giallongo}, {Salimbeni}, \& {Santini}}]{pen09}
{Pentericci}, L., {Grazian}, A., {Fontana}, A., {et~al.} 2009, \aap, 494, 553

\bibitem[{{Pentericci} {et~al.}(2007){Pentericci}, {Grazian}, {Fontana},
  {Salimbeni}, {Santini}, {de Santis}, {Gallozzi}, \& {Giallongo}}]{pen07}
{Pentericci}, L., {Grazian}, A., {Fontana}, A., {et~al.} 2007, \aap, 471, 433

\bibitem[{{Popesso} {et~al.}(2009){Popesso}, {Dickinson}, {Nonino}, {Vanzella},
  {Daddi}, {Fosbury}, {Kuntschner}, {Mainieri}, {Cristiani}, {Cesarsky},
  {Giavalisco}, {Renzini}, \& {The Goods Team}}]{pop09}
{Popesso}, P., {Dickinson}, M., {Nonino}, M., {et~al.} 2009, \aap, 494, 443

\bibitem[{{Ravikumar} {et~al.}(2007){Ravikumar}, {Puech}, {Flores}, {Proust},
  {Hammer}, {Lehnert}, {Rawat}, {Amram}, {Balkowski}, {Burgarella}, {Cassata},
  {Cesarsky}, {Cimatti}, {Combes}, {Daddi}, {Dannerbauer}, {di Serego
  Alighieri}, {Elbaz}, {Guiderdoni}, {Kembhavi}, {Liang}, {Pozzetti},
  {Vergani}, {Vernet}, {Wozniak}, \& {Zheng}}]{rav07}
{Ravikumar}, C.~D., {Puech}, M., {Flores}, H., {et~al.} 2007, \aap, 465, 1099

\bibitem[{{Renzini} {et~al.}(2003){Renzini}, {Cesarsky}, {Cristiani}, {da
  Costa}, {Fosbury}, {Hook}, {Leibundgut}, {Rosati}, \& {Vandame}}]{ren03}
{Renzini}, A., {Cesarsky}, C., {Cristiani}, S., {et~al.} 2003, in The Mass of
  Galaxies at Low and High Redshift, ed. R.~{Bender} \& A.~{Renzini}, 332--+

\bibitem[{{Scodeggio} {et~al.}(2005){Scodeggio}, {Franzetti}, {Garilli},
  {Zanichelli}, {Paltani}, {Maccagni}, {Bottini}, {Le Brun}, {Contini},
  {Scaramella}, {Adami}, {Bardelli}, {Zucca}, {Tresse}, {Ilbert}, {Foucaud},
  {Iovino}, {Merighi}, {Zamorani}, {Gavignaud}, {Rizzo}, {McCracken}, {Le
  F{\`e}vre}, {Picat}, {Vettolani}, {Arnaboldi}, {Arnouts}, {Bolzonella},
  {Cappi}, {Charlot}, {Ciliegi}, {Guzzo}, {Marano}, {Marinoni}, {Mathez},
  {Mazure}, {Meneux}, {Pell{\`o}}, {Pollo}, {Pozzetti}, \& {Radovich}}]{sco05}
{Scodeggio}, M., {Franzetti}, P., {Garilli}, B., {et~al.} 2005, \pasp, 117,
  1284

\bibitem[{{Shapley} {et~al.}(2003){Shapley}, {Steidel}, {Pettini}, \&
  {Adelberger}}]{sha03}
{Shapley}, A.~E., {Steidel}, C.~C., {Pettini}, M., \& {Adelberger}, K.~L. 2003,
  \apj, 588, 65

\bibitem[{{Stanway} {et~al.}(2004){Stanway}, {Bunker}, {McMahon}, {Ellis},
  {Treu}, \& {McCarthy}}]{sta04}
{Stanway}, E.~R., {Bunker}, A.~J., {McMahon}, R.~G., {et~al.} 2004, \apj, 607,
  704

\bibitem[{{Strolger} {et~al.}(2004){Strolger}, {Riess}, {Dahlen}, {Livio},
  {Panagia}, {Challis}, {Tonry}, {Filippenko}, {Chornock}, {Ferguson},
  {Koekemoer}, {Mobasher}, {Dickinson}, {Giavalisco}, {Casertano}, {Hook},
  {Blondin}, {Leibundgut}, {Nonino}, {Rosati}, {Spinrad}, {Steidel}, {Stern},
  {Garnavich}, {Matheson}, {Grogin}, {Hornschemeier}, {Kretchmer}, {Laidler},
  {Lee}, {Lucas}, {de Mello}, {Moustakas}, {Ravindranath}, {Richardson}, \&
  {Taylor}}]{str04}
{Strolger}, L.-G., {Riess}, A.~G., {Dahlen}, T., {et~al.} 2004, \apj, 613, 200

\bibitem[{{Szokoly} {et~al.}(2004){Szokoly}, {Bergeron}, {Hasinger}, {Lehmann},
  {Kewley}, {Mainieri}, {Nonino}, {Rosati}, {Giacconi}, {Gilli}, {Gilmozzi},
  {Norman}, {Romaniello}, {Schreier}, {Tozzi}, {Wang}, {Zheng}, \&
  {Zirm}}]{szo04}
{Szokoly}, G.~P., {Bergeron}, J., {Hasinger}, G., {et~al.} 2004, \apjs, 155,
  271

\bibitem[{{Tapken} {et~al.}(2007){Tapken}, {Appenzeller}, {Noll}, {Richling},
  {Heidt}, {Meink{\"o}hn}, \& {Mehlert}}]{tap07}
{Tapken}, C., {Appenzeller}, I., {Noll}, S., {et~al.} 2007, \aap, 467, 63

\bibitem[{{van der Wel} {et~al.}(2004){van der Wel}, {Franx}, {van Dokkum}, \&
  {Rix}}]{van04}
{van der Wel}, A., {Franx}, M., {van Dokkum}, P.~G., \& {Rix}, H.-W. 2004,
  \apjl, 601, L5

\bibitem[{{Vanzella} {et~al.}(2008){Vanzella}, {Cristiani}, {Dickinson},
  {Giavalisco}, {Kuntschner}, {Haase}, {Nonino}, {Rosati}, {Cesarsky},
  {Ferguson}, {Fosbury}, {Grazian}, {Moustakas}, {Rettura}, {Popesso},
  {Renzini}, {Stern}, \& {The Goods Team}}]{van08}
{Vanzella}, E., {Cristiani}, S., {Dickinson}, M., {et~al.} 2008, \aap, 478, 83

\bibitem[{{Vanzella} {et~al.}(2005){Vanzella}, {Cristiani}, {Dickinson},
  {Kuntschner}, {Moustakas}, {Nonino}, {Rosati}, {Stern}, {Cesarsky}, {Ettori},
  {Ferguson}, {Fosbury}, {Giavalisco}, {Haase}, {Renzini}, {Rettura}, {Serra},
  \& {The Goods Team}}]{van05}
{Vanzella}, E., {Cristiani}, S., {Dickinson}, M., {et~al.} 2005, \aap, 434, 53

\bibitem[{{Vanzella} {et~al.}(2006){Vanzella}, {Cristiani}, {Dickinson},
  {Kuntschner}, {Nonino}, {Rettura}, {Rosati}, {Vernet}, {Cesarsky},
  {Ferguson}, {Fosbury}, {Giavalisco}, {Grazian}, {Haase}, {Moustakas},
  {Popesso}, {Renzini}, {Stern}, \& {The Goods Team}}]{van06}
{Vanzella}, E., {Cristiani}, S., {Dickinson}, M., {et~al.} 2006, \aap, 454, 423

\bibitem[{{Vanzella} {et~al.}(2009){Vanzella}, {Giavalisco}, {Dickinson},
  {Cristiani}, {Nonino}, {Kuntschner}, {Popesso}, {Rosati}, {Renzini}, {Stern},
  {Cesarsky}, {Ferguson}, \& {Fosbury}}]{van09}
{Vanzella}, E., {Giavalisco}, M., {Dickinson}, M., {et~al.} 2009, \apj, 695,
  1163

\bibitem[{{Verhamme} {et~al.}(2008){Verhamme}, {Schaerer}, {Atek}, \&
  {Tapken}}]{ver08}
{Verhamme}, A., {Schaerer}, D., {Atek}, H., \& {Tapken}, C. 2008, \aap, 491, 89

\end{thebibliography}

\appendix

\section{Corrections to previous release}\label{app1}

We revised some of the redshifts from the first release of the GOODS/VIMOS survey 
for objects that were observed more than once in the GOODS/VIMOS
survey or in other surveys. After comparing their spectra, 7 redshift 
determinations were modified as follows:
\begin{itemize}
 \item GOODS\_LRb\_001\_1\_q1\_51\_1 at $z=2.2077$ (flag A) in the previous release, 
was observed also with FORS2 (GDS~J033226.67-274013.4). The low-S/N UV absorption 
lines were misclassified in the VIMOS spectrum. Instead, thanks to the comparison 
with the FORS2 spectrum, CIII] in emission could be identified. The new redshift 
obtained, $z=1.6045$, with quality flag B, is in agreement with the FORS2 measurement 
($z=1.612$, flag A).
 \item GOODS\_LRb\_003\_new\_q1\_61\_1 at $z=0.067$ (flag A) in the previous release.
A possibly broad CIII] emission line was previously identified as [OII]. This object 
was also observed by Silverman et al. (in preparation), where the measured redshift 
is $z=1.089$. Our new estimate of the redshift based on CIII] and FeII lines is 
in agreement with the more recent measurement. We find $z=1.0890$ with quality flag B.
 \item GOODS\_LRb\_003\_new\_q3\_11\_1 at $z=2.6114$ (flag B) in the previous release. This object was 
also observed in another mask (GOODS\_LRb\_003\_new\_2\_q3\_30\_1, $z=2.5831$, flag B).
Both estimates are based on the detection of a Ly-$\alpha$ in emission. However, 
we noticed that the former measurement required a slight adjustment (i.e. the Ly-$\alpha$
was correctly identified, but slightly misaligned). The new resulting redshift, 
$z=2.5882$, with quality flag B, is in better agreement with the second VIMOS measurement.
 \item GOODS\_LRb\_003\_new\_1\_q3\_68\_1 at $z=0.7297$, flag C based on 
the identification of [OII]. The redshift is correct and it is confermed by a second 
observation in a MR mask (GOODS\_MR\_new\_1\_d\_q2\_32\_1 at $z=0.7308$, flag A). We 
simply upgraded the quality flag from C to B in this case. 
 \item GOODS\_LRb\_003\_new\_2\_q1\_42\_1 at $z=2.1084$ (flag B) in the previous release. The broad MgII 
emission line was misclassified as Ly-$\alpha$. This object was also observed in one
of the new reduced masks (GOODS\_LRb\_dec06\_3\_q1\_46\_2, $z=0.7358$, flag A). Two
spectral features are clearly identified: a broad emission line from MgII and [OII].
Therefore, the new redshift assigned is $z=0.7357$ with quality flag A.
 \item GOODS\_MR\_new\_1\_d\_q3\_2\_1 at $z=0.5107$ (flag B) in the previous release. Two spectral 
features were misclassified: [OIII] and H$\beta$. Thanks to the comparison with a second 
observation of this object (GOODS\_MR\_001\_q4\_3\_1 $z=1.0338$, flag A), we identified 
[OII] and MgII instead. The new redshift estimate is $z=1.0349$ with quality flag B.
\end{itemize}

The following redshifts were also revised:
\begin{itemize}
 \item GOODS\_LRb\_001\_1\_q2\_66\_1 at $z=0.240$ (flag B) in the previous release.
The broad MgII emission line was misclassified as [OII]. After revising the spectrum
we could identify some typical AGN features (e.g. a broad MgII emission line and broad 
Fe bump at $\sim$2960~$\AA$) plus [OII] and Ca H and K.
The new redshift obtained is $z=0.6632$ (flag A).
 \item GOODS\_LRb\_001\_1\_q3\_5\_3 at $z=0.076$ (flag B) in the previous release.
The spectrum shows only one visible emission line at $\sim4000$~$\AA$, which was 
previously interpreted as [OII]. For this object we find more likely the identification
as Ly-$\alpha$ in emission, which gives a new tentative redshift of $z=2.2980$ (flag C).
 \item GOODS\_LRb\_001\_q4\_44\_1 and GOODS\_LRb\_003\_new\_q4\_28\_1 at $z=1.6285$ 
(flag B) in the previous release. We reviewed the redshifts of the two spectra
for this object, which were based on the identification of two very broad emission lines
(i.e. CIV and CIII]). In addition, we identified several absorption lines 
(e.g. FeII, AlIII) that helped further calibrating the redshifts. The new measurements 
are $z=1.609$ (flag A) and $z=1.608$ (flag A) for GOODS\_LRb\_001\_q4\_44\_1 
and GOODS\_LRb\_003\_new\_q4\_28\_1, respectively.
 \item GOODS\_LRb\_003\_new\_1\_q2\_46\_2 is the same object as 
GOODS\_LRb\_001\_1\_q2\_66\_1 (see above). The new redshift measurement is $z=0.6649$ (flag A).
 \item GOODS\_LRb\_003\_new\_1\_q4\_63\_2 at $z=0.585$ (flag C) in the previous release.
The spectrum is red, but quite noisy and may resemble that of an elliptical galaxy. 
However, after the inspection of the WFI $R$-image and the ACS $z$-image of 
this object, we found that a more likely explanation might be given in terms of 
a star. Therefore, we assign $z=0.000$ and flag B to this object.
 \item GOODS\_MR\_new\_2\_b\_q4\_15\_1 at $z=5.2936$ (flag C) in the previous release.
We could identify Ca H and K, Fe I and possibly [OII]. The new redshift assigned to 
this object is $z=0.9250$ with quality flag B.
\end{itemize}

\section{Large scale structure}\label{app3}

\begin{figure*}
\centering
\includegraphics[width=3.0 cm, angle=0]{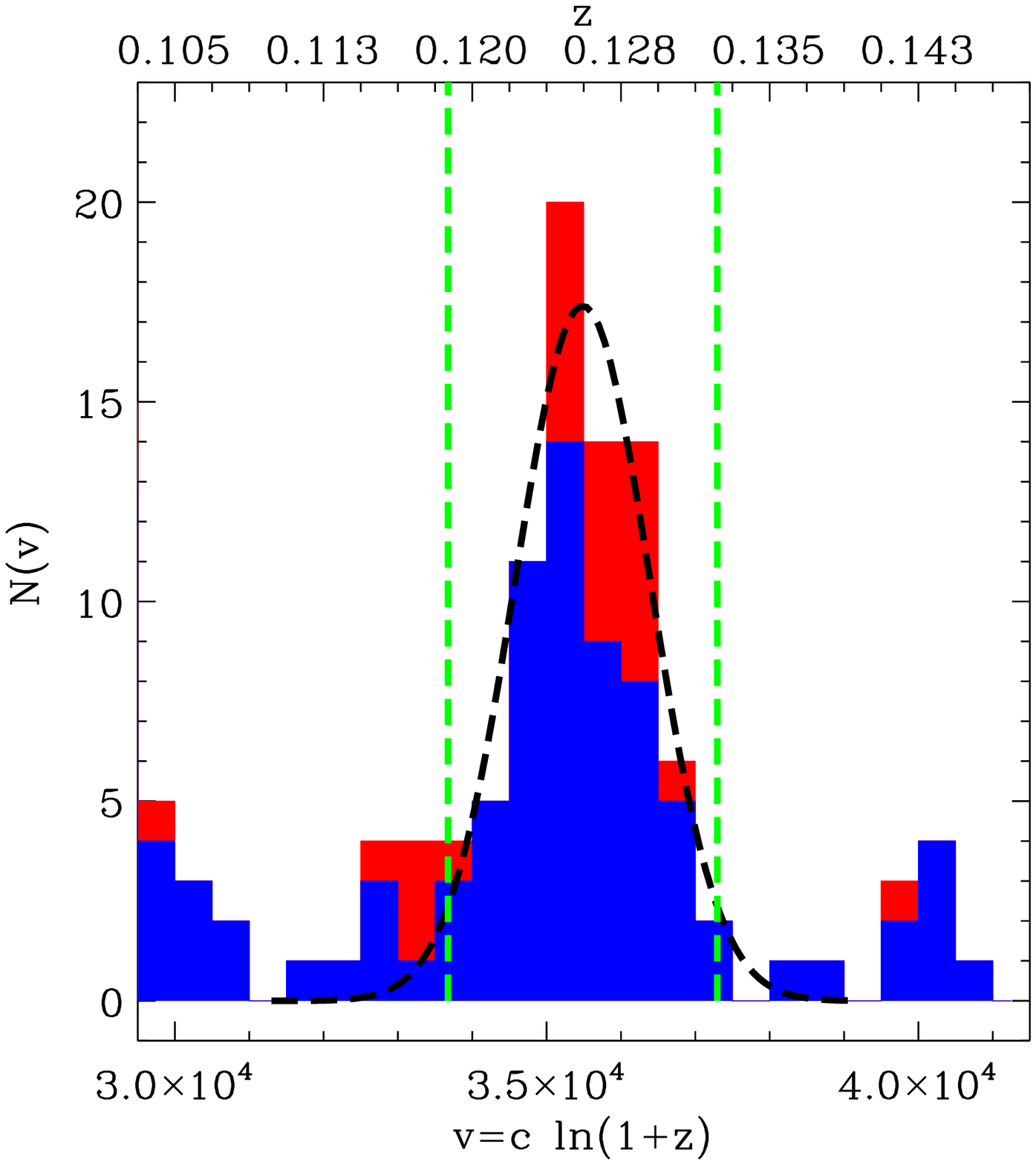}
\includegraphics[width=6.0 cm, angle=0]{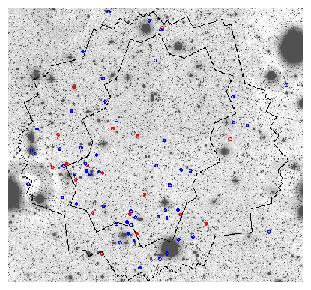}
\includegraphics[width=3.0 cm, angle=0]{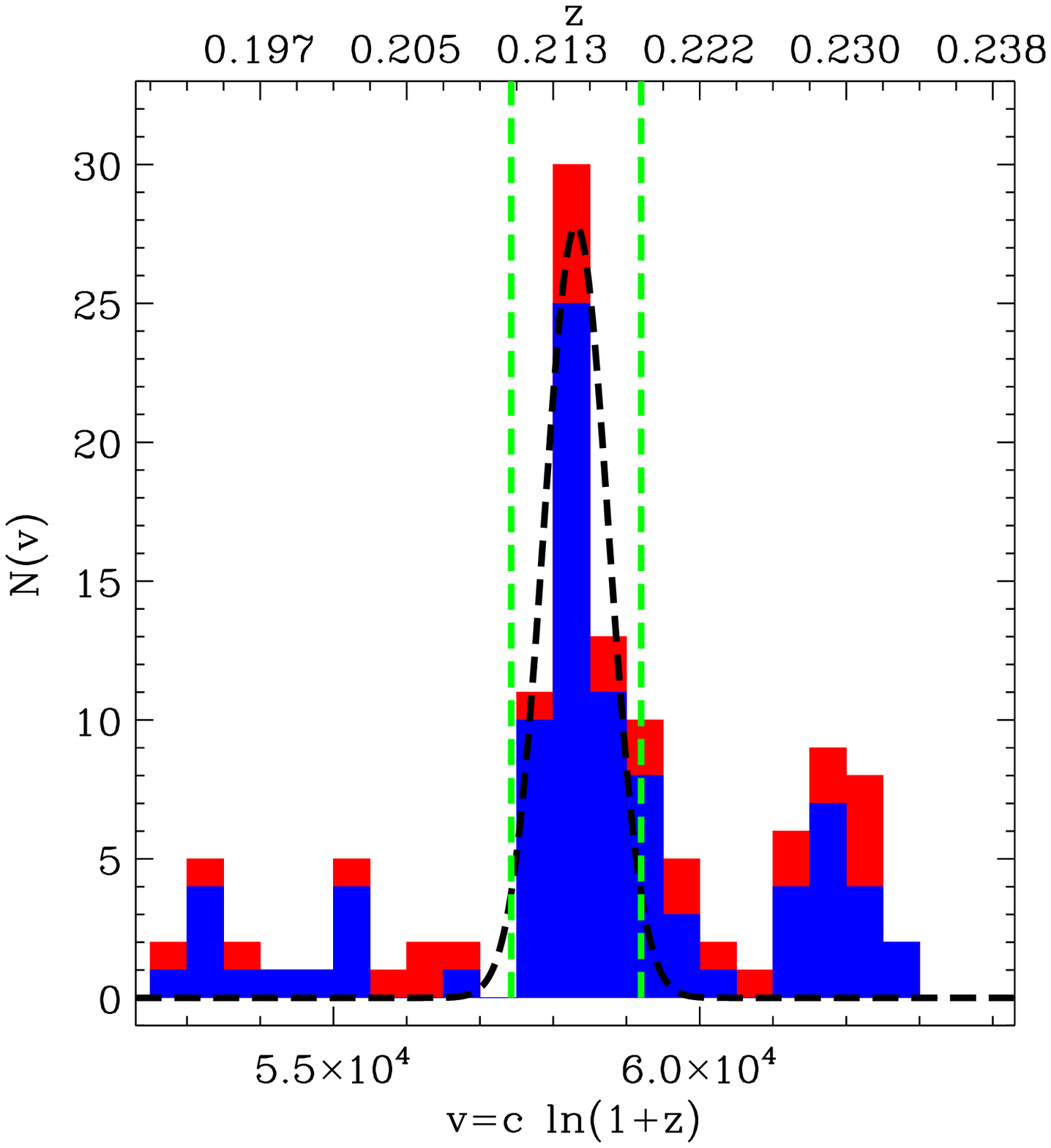}
\includegraphics[width=6.0 cm, angle=0]{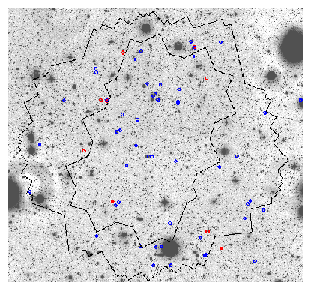}
\includegraphics[width=3.0 cm, angle=0]{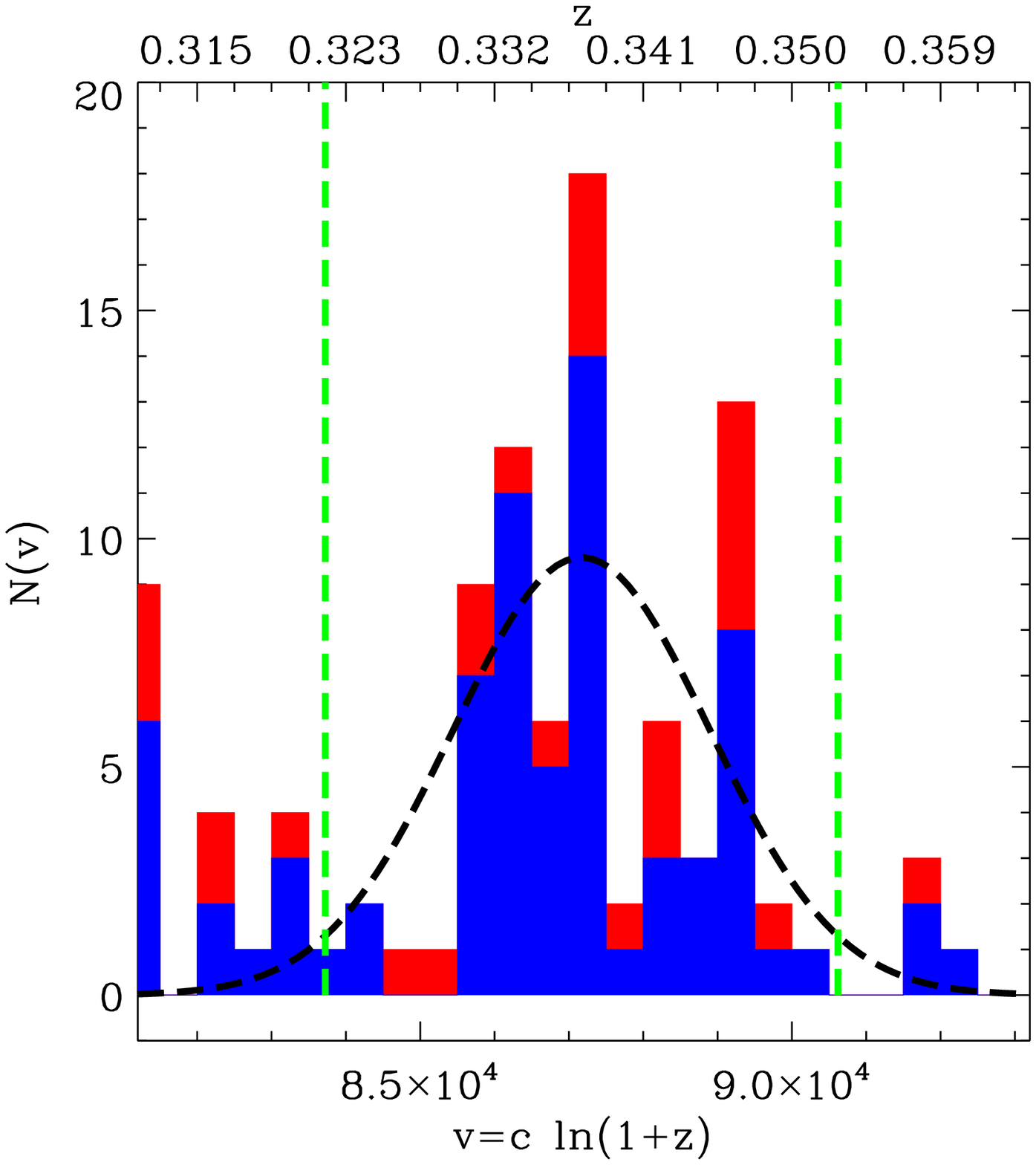}
\includegraphics[width=6.0 cm, angle=0]{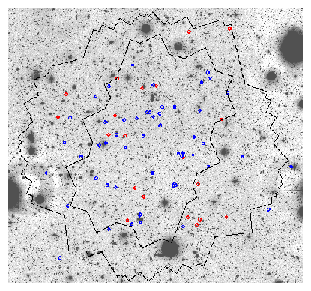}
\includegraphics[width=3.0 cm, angle=0]{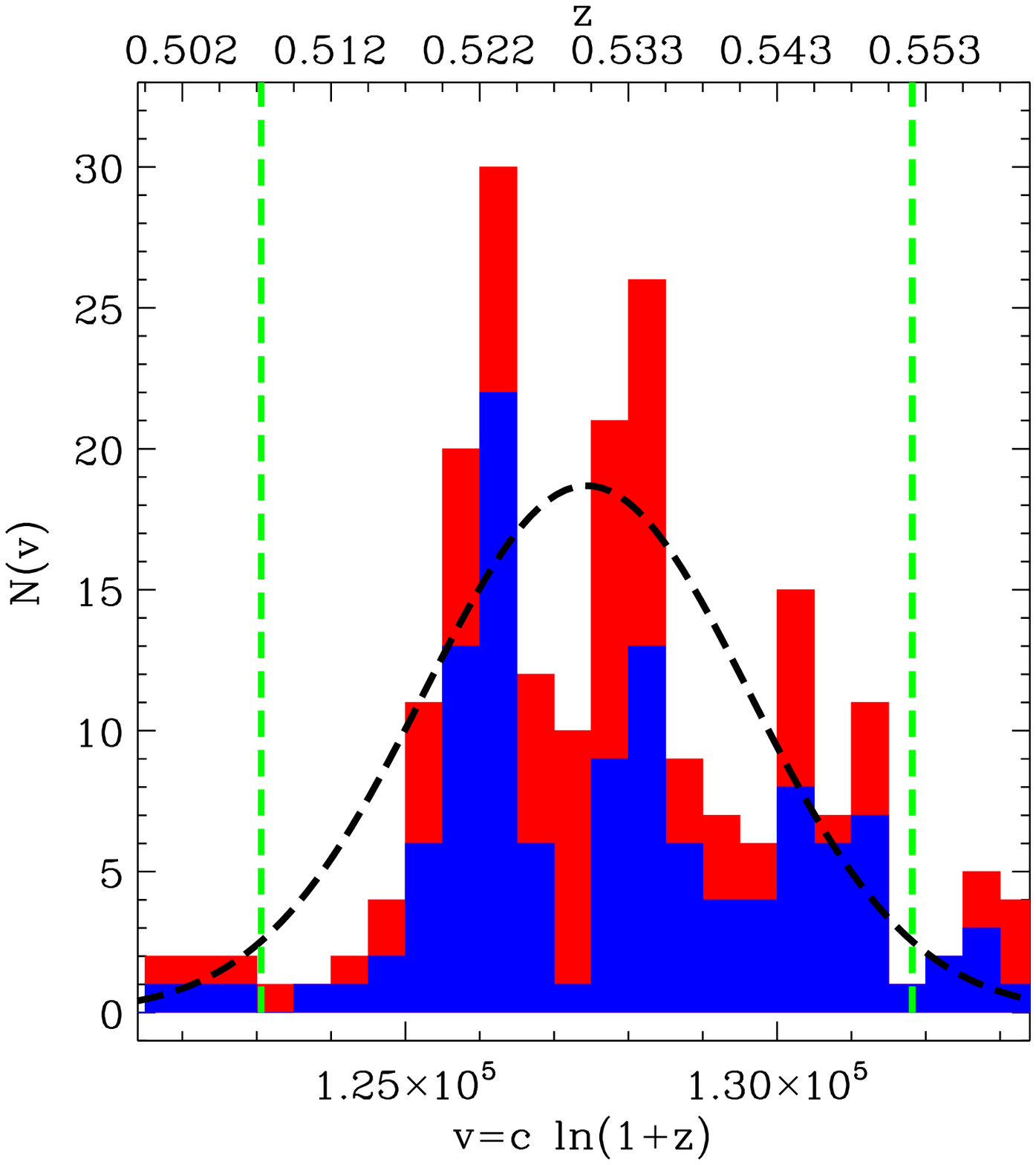}
\includegraphics[width=6.0 cm, angle=0]{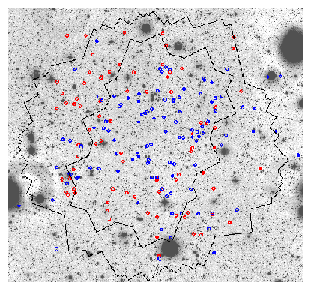}
\includegraphics[width=3.0 cm, angle=0]{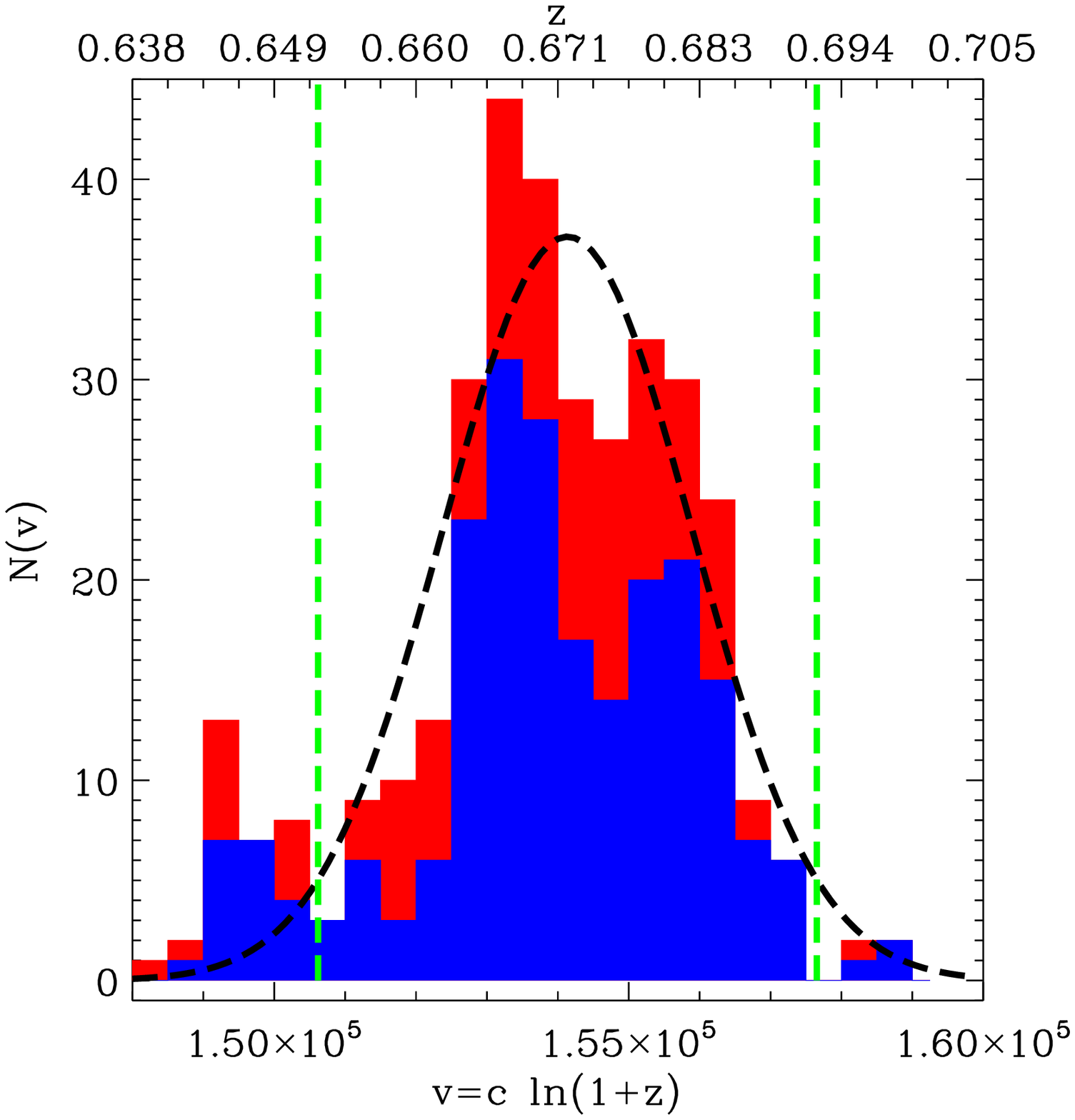}
\includegraphics[width=6.0 cm, angle=0]{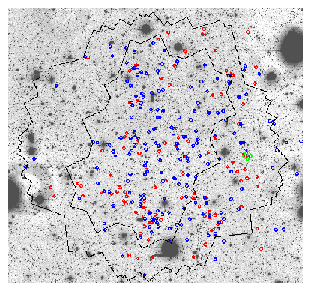}
\includegraphics[width=3.0 cm, angle=0]{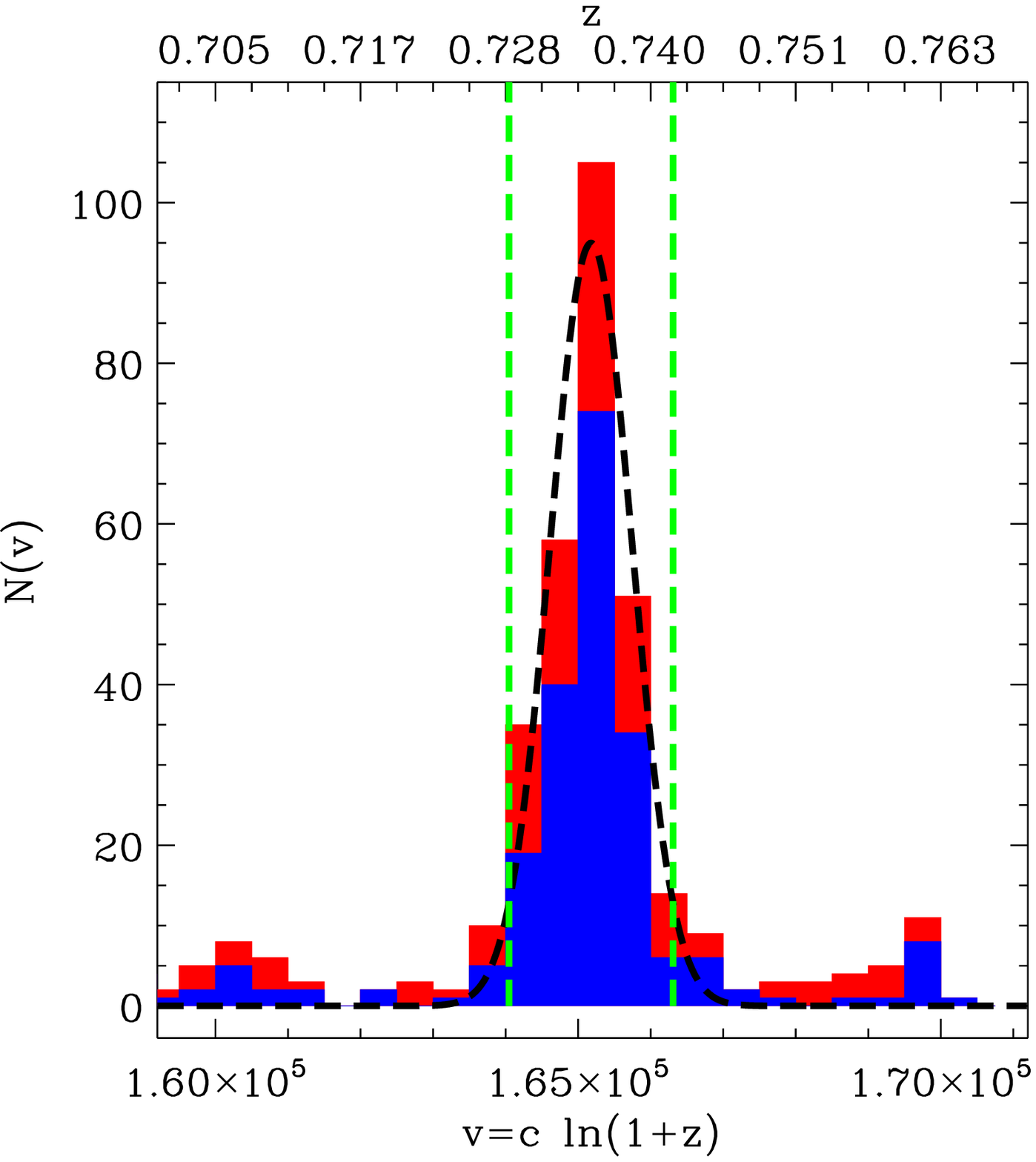}
\includegraphics[width=6.0 cm, angle=0]{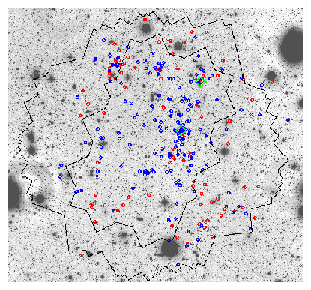}
\caption{Histograms of the velocity distribution, $V=c\ln(1+z)$, and spatial 
distribution of galaxies belonging to each of the density peak listed in 
Table~\ref{peaks}. For each structure, we display ``secure'' redshifts 
($>99$\% c.l.) in \textit{blue} and ``likely'' redshifts ($\sim70-90$\% c.l.) 
in \textit{red}. The histograms show the Gaussian fit to the peak of the velocity distribution 
(\textit{black dashed curve}) and the $2\sigma$-intervals around the peak 
selected for the spatial analysis (\textit{vertical green dashed lines}). All the histograms 
are binned to $\Delta V=500$~km/s. The green diamonds show the position of the extended 
X-ray sources (i.e. XID 645 at $z\sim0.67$; XID 566 and XID 594 at $z\sim0.73$; 
XID 249 at $z\sim0.97$). The black contours indicate the GOODS area 
and the field coverage of the 2Ms Chandra exposure.}
\label{654}
\end{figure*}

\begin{figure*}
\centering
\includegraphics[width=3.0 cm, angle=0]{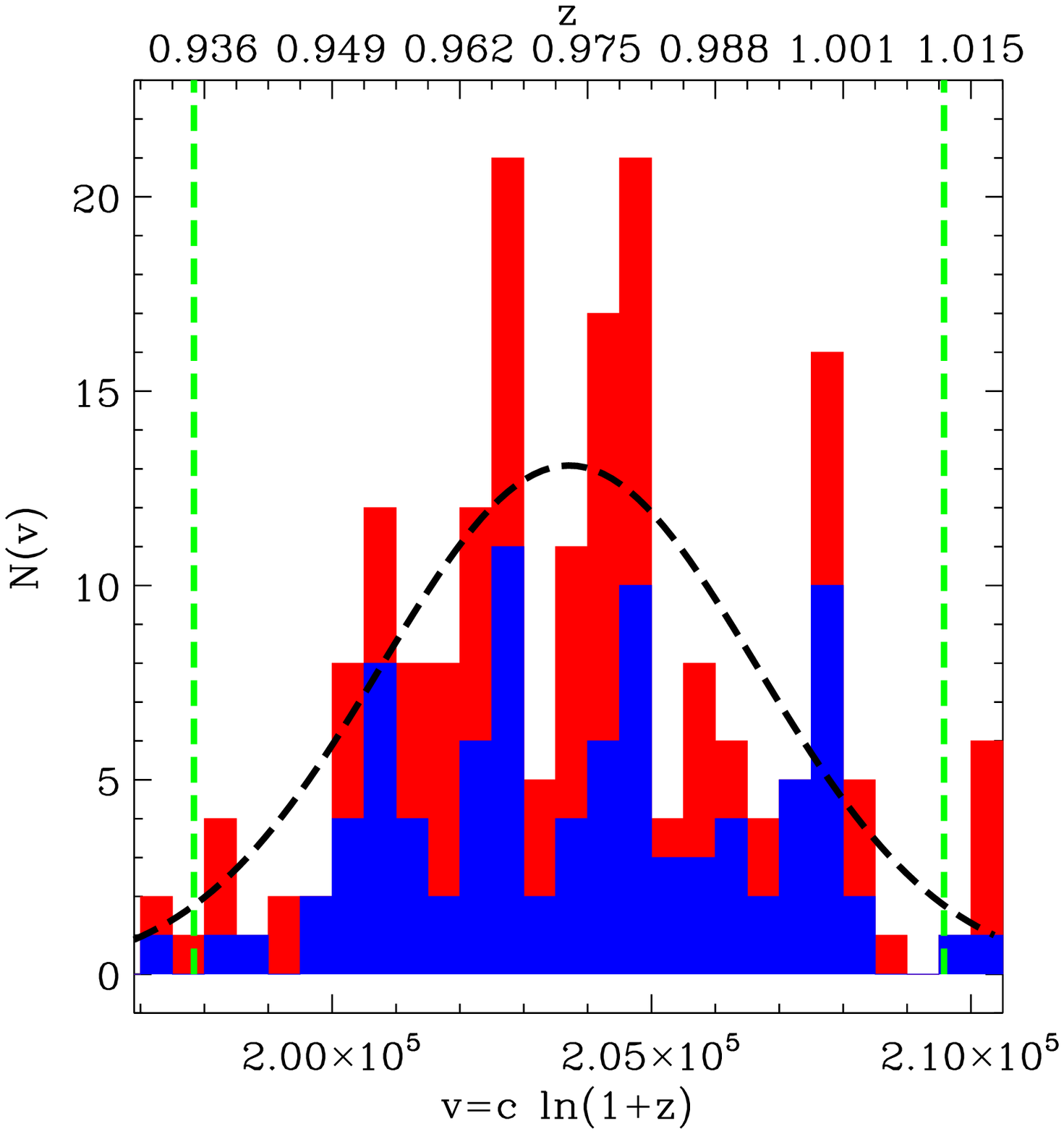}
\includegraphics[width=6.0 cm, angle=0]{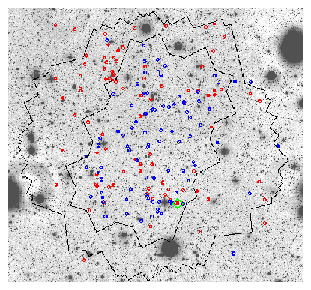}
\includegraphics[width=3.0 cm, angle=0]{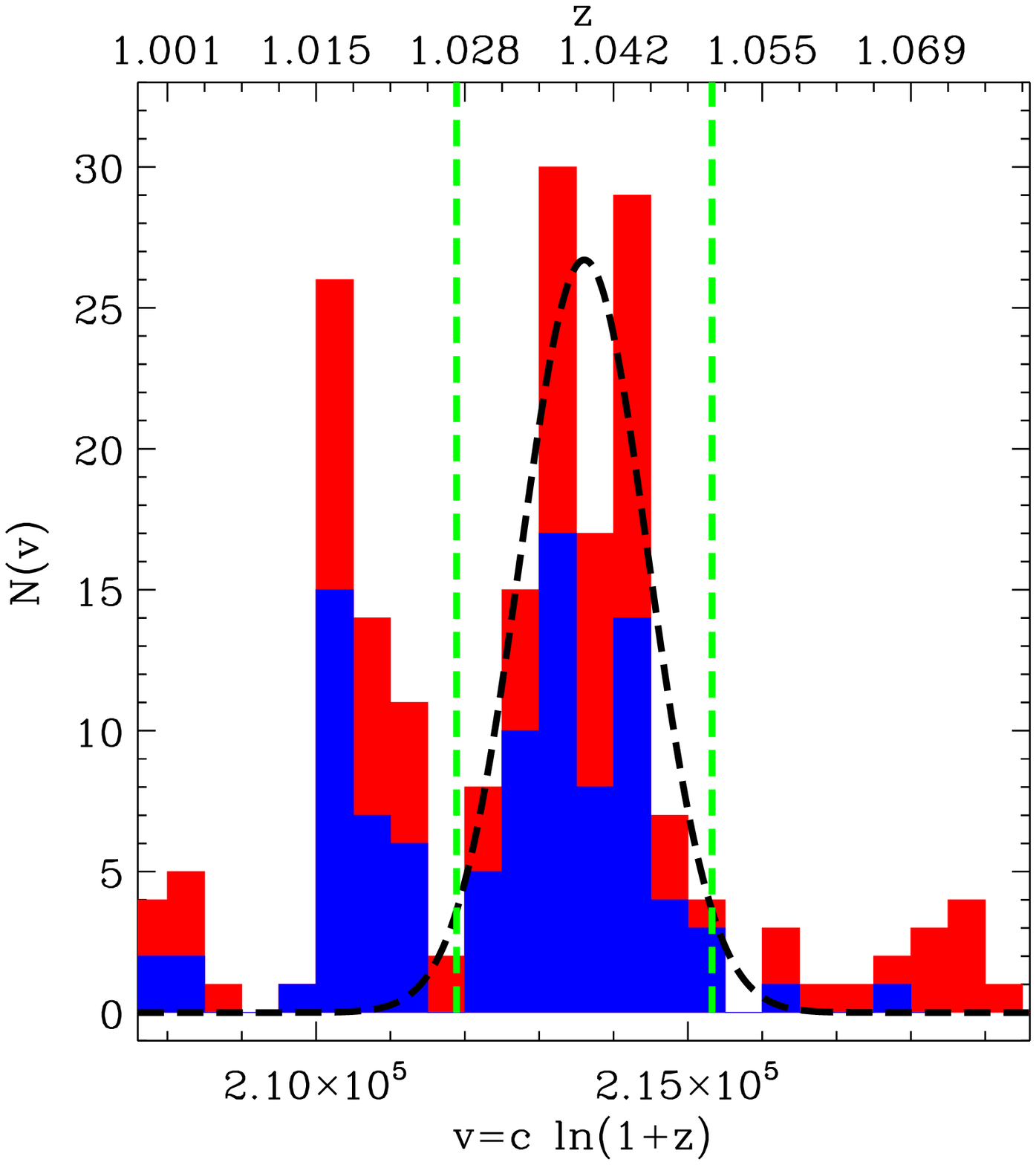}
\includegraphics[width=6.0 cm, angle=0]{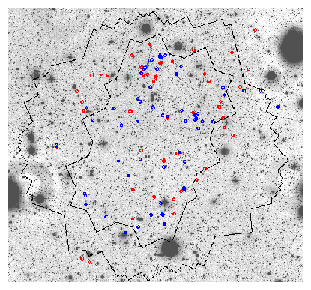}
\includegraphics[width=3.0 cm, angle=0]{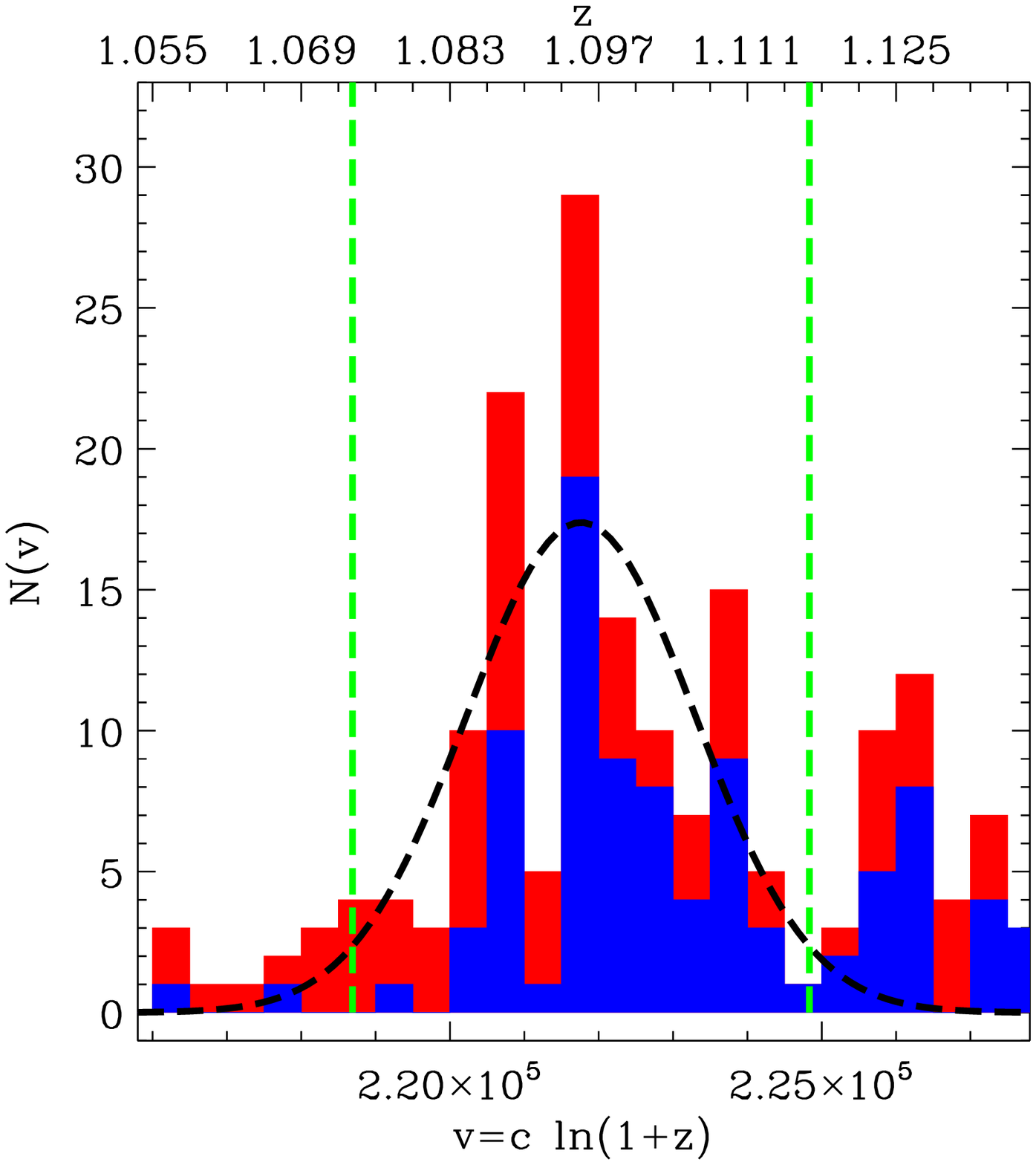}
\includegraphics[width=6.0 cm, angle=0]{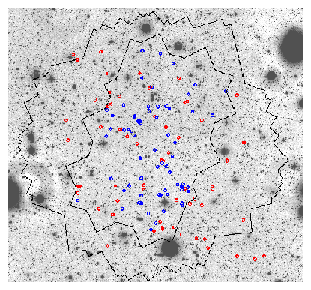}
\includegraphics[width=3.0 cm, angle=0]{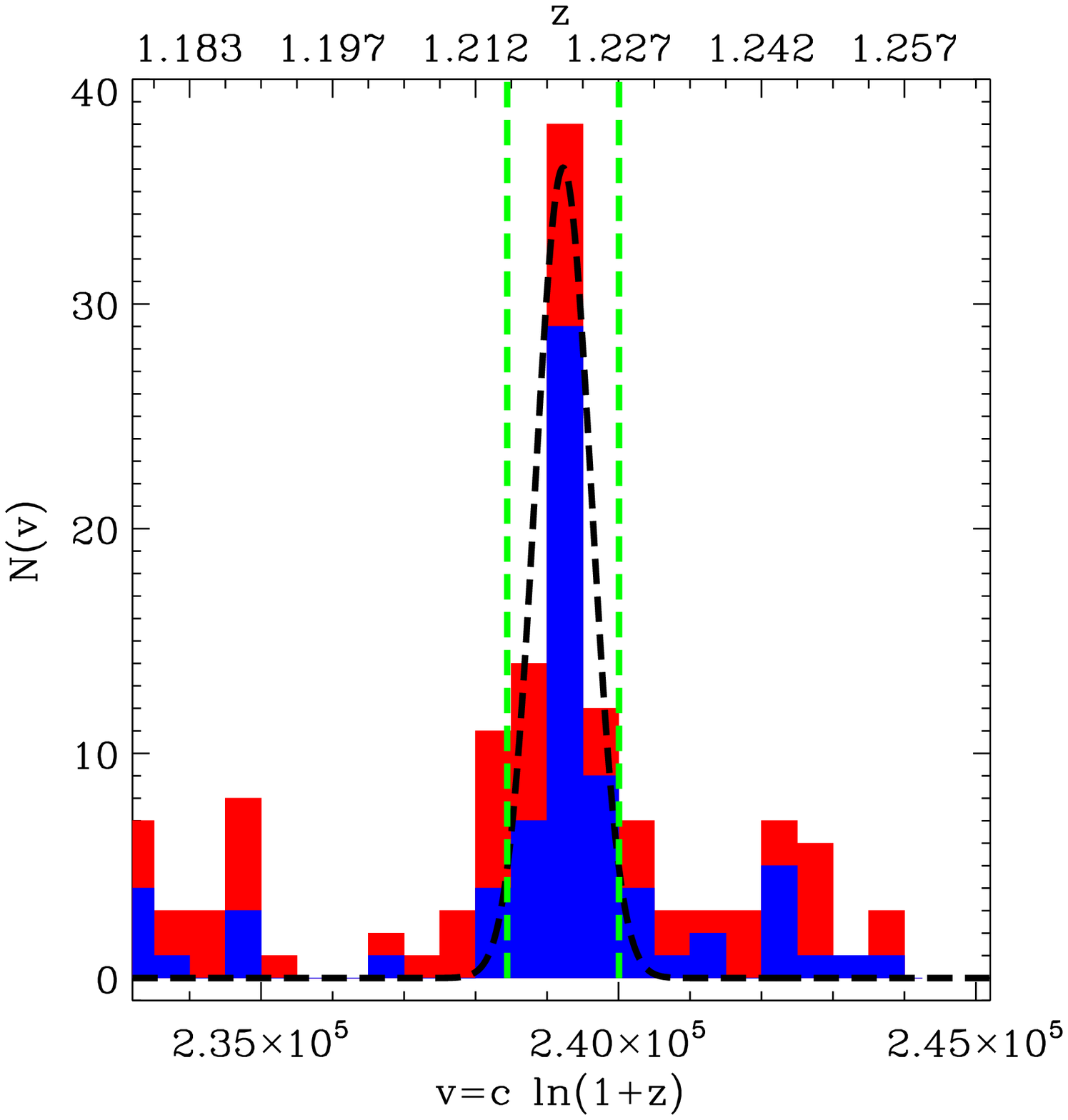}
\includegraphics[width=6.0 cm, angle=0]{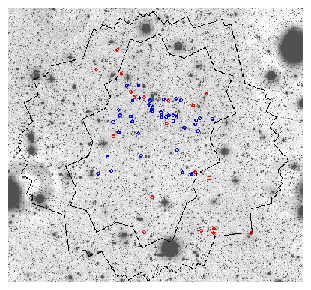}
\includegraphics[width=3.0 cm, angle=0]{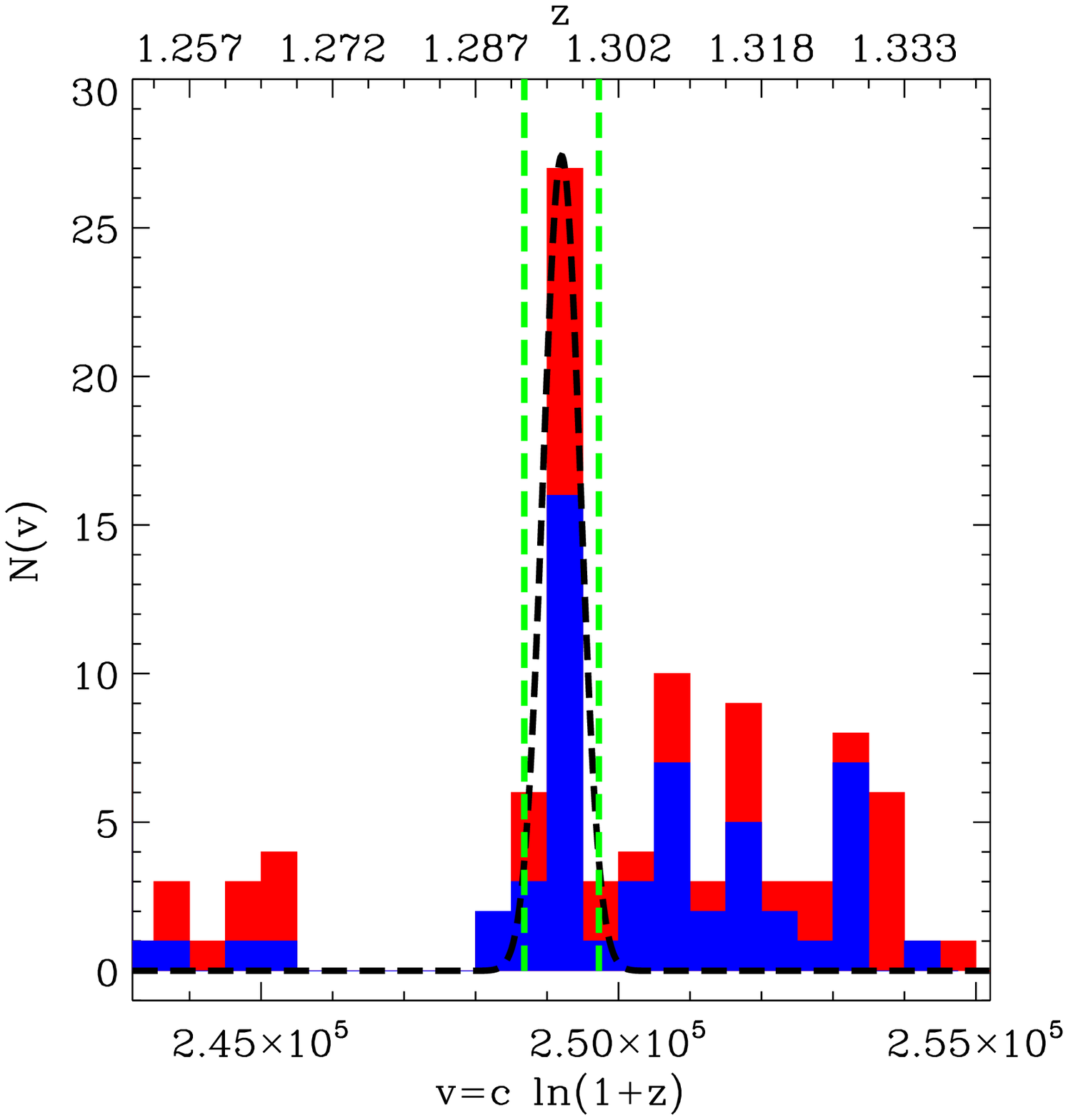}
\includegraphics[width=6.0 cm, angle=0]{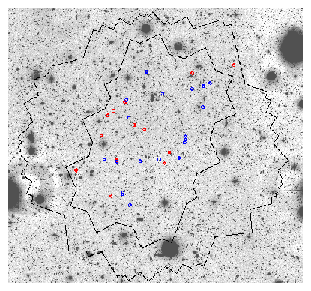}
\includegraphics[width=3.0 cm, angle=0]{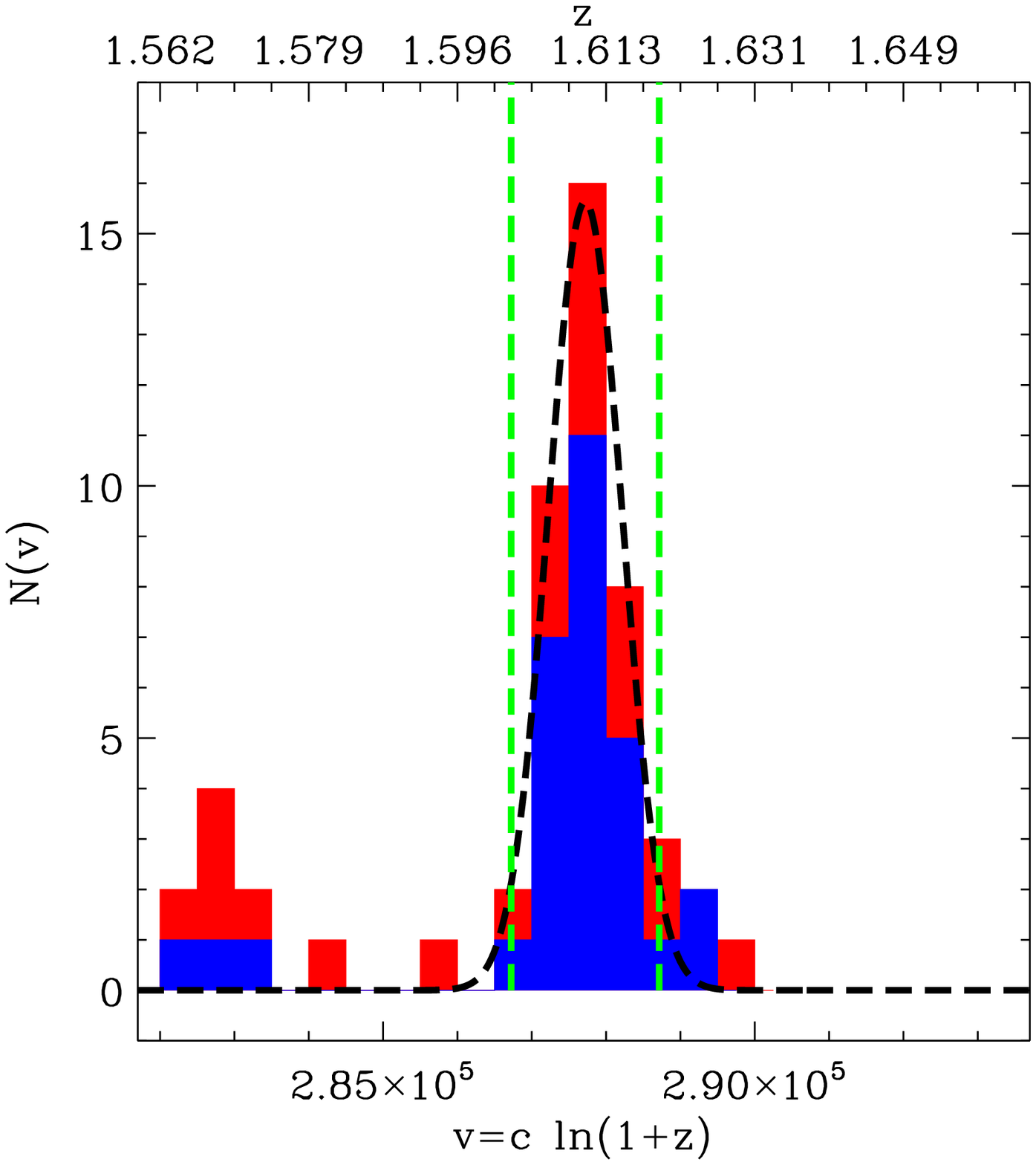}
\includegraphics[width=6.0 cm, angle=0]{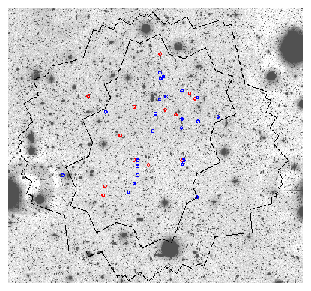}
\caption{Histograms of the velocity distribution, $V=c\ln(1+z)$, and spatial 
distribution of galaxies belonging to each of the density peak listed in 
Table~\ref{peaks}. For each structure, we display ``secure'' redshifts 
($>99$\% c.l.) in \textit{blue} and ``likely'' redshifts ($\sim70-90$\% c.l.) 
in \textit{red}. The histograms show the Gaussian fit to the peak of the velocity distribution 
(\textit{black dashed curve}) and the $2\sigma$-intervals around the peak 
selected for the spatial analysis (\textit{vertical green dashed lines}). All the histograms 
are binned to $\Delta V=500$~km/s. The black contours indicate the GOODS area 
and the field coverage of the 2Ms Chandra exposure.}
\label{736}
\end{figure*}

\begin{figure*}
\centering
\includegraphics[width=3.0 cm, angle=0]{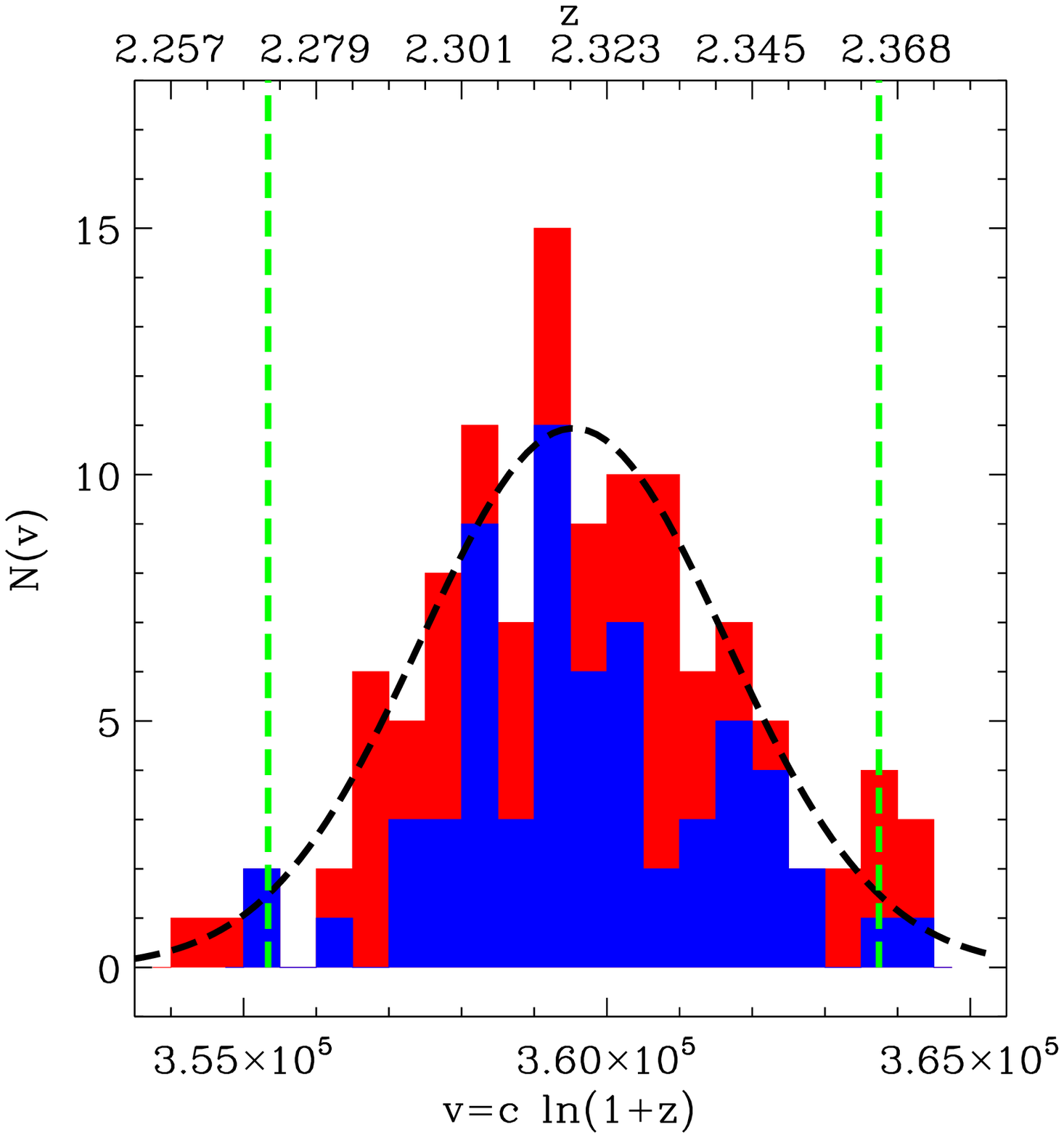}
\includegraphics[width=6.0 cm, angle=0]{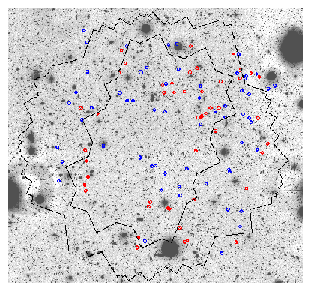}
\includegraphics[width=3.0 cm, angle=0]{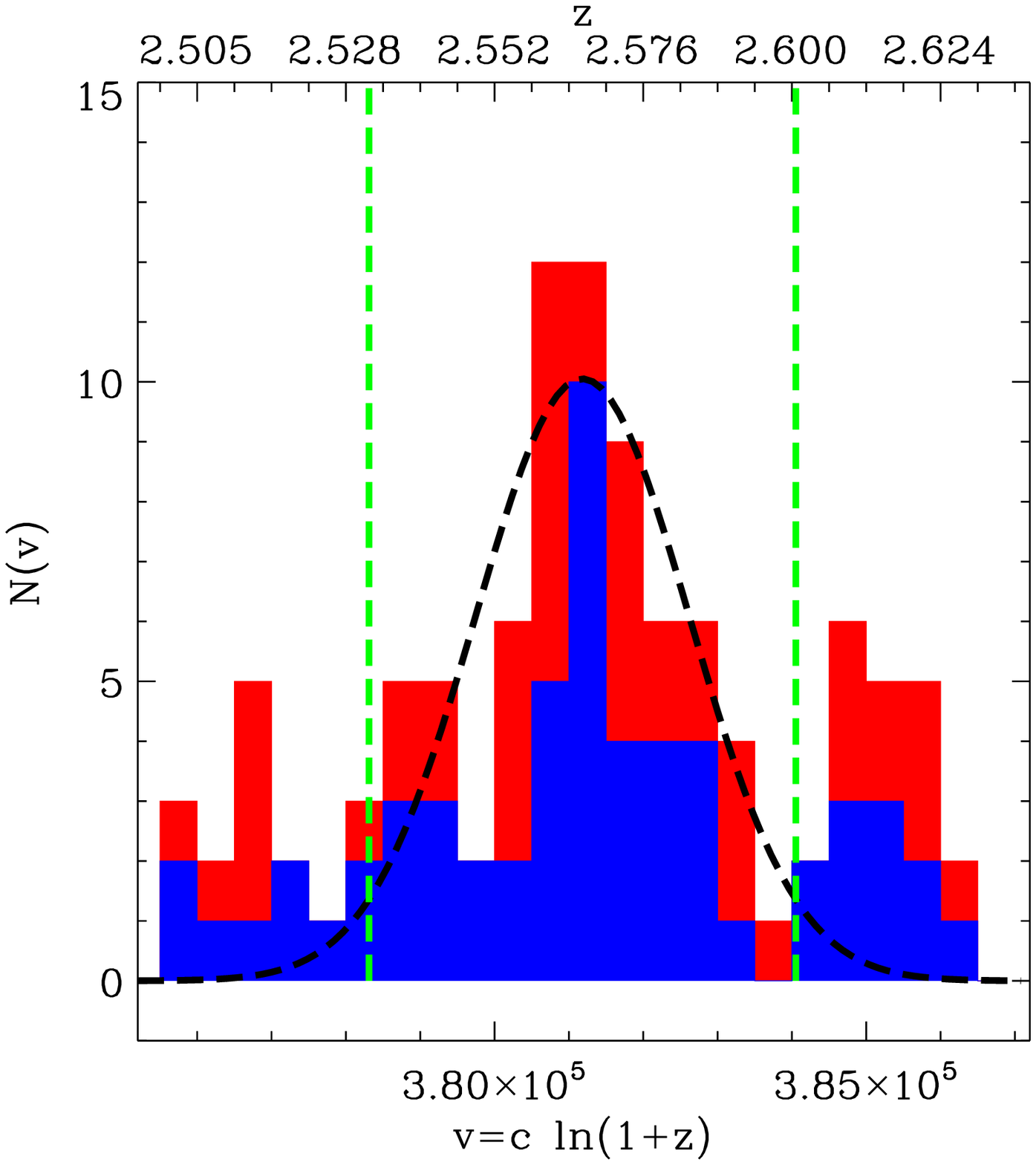}
\includegraphics[width=6.0 cm, angle=0]{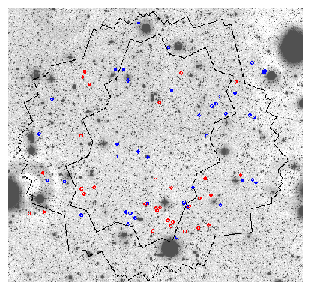}
\includegraphics[width=3.0 cm, angle=0]{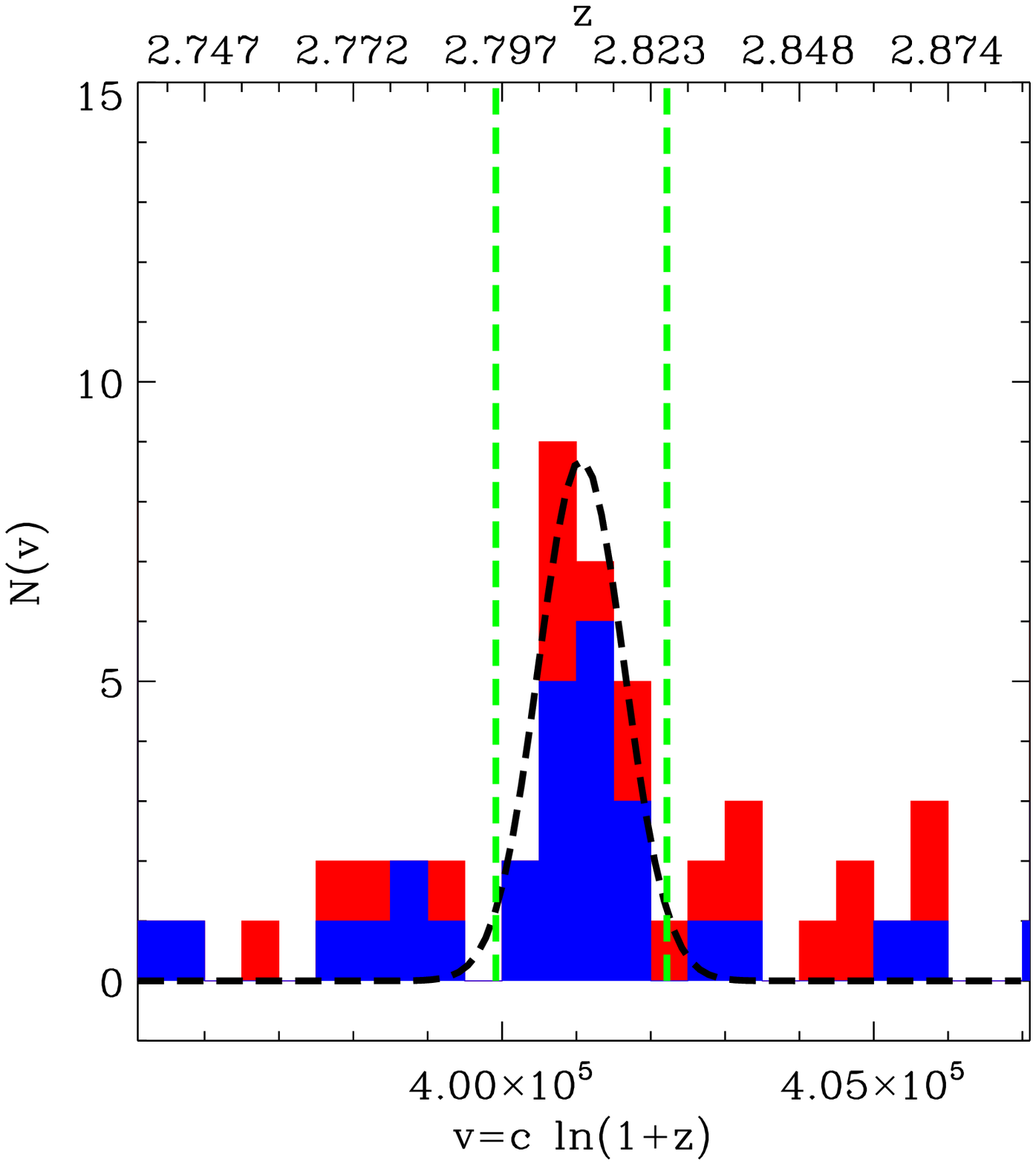}
\includegraphics[width=6.0 cm, angle=0]{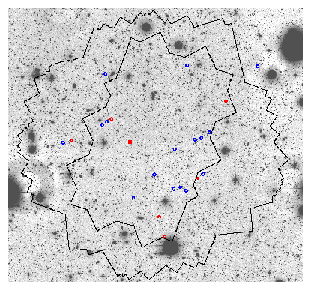}
\includegraphics[width=3.0 cm, angle=0]{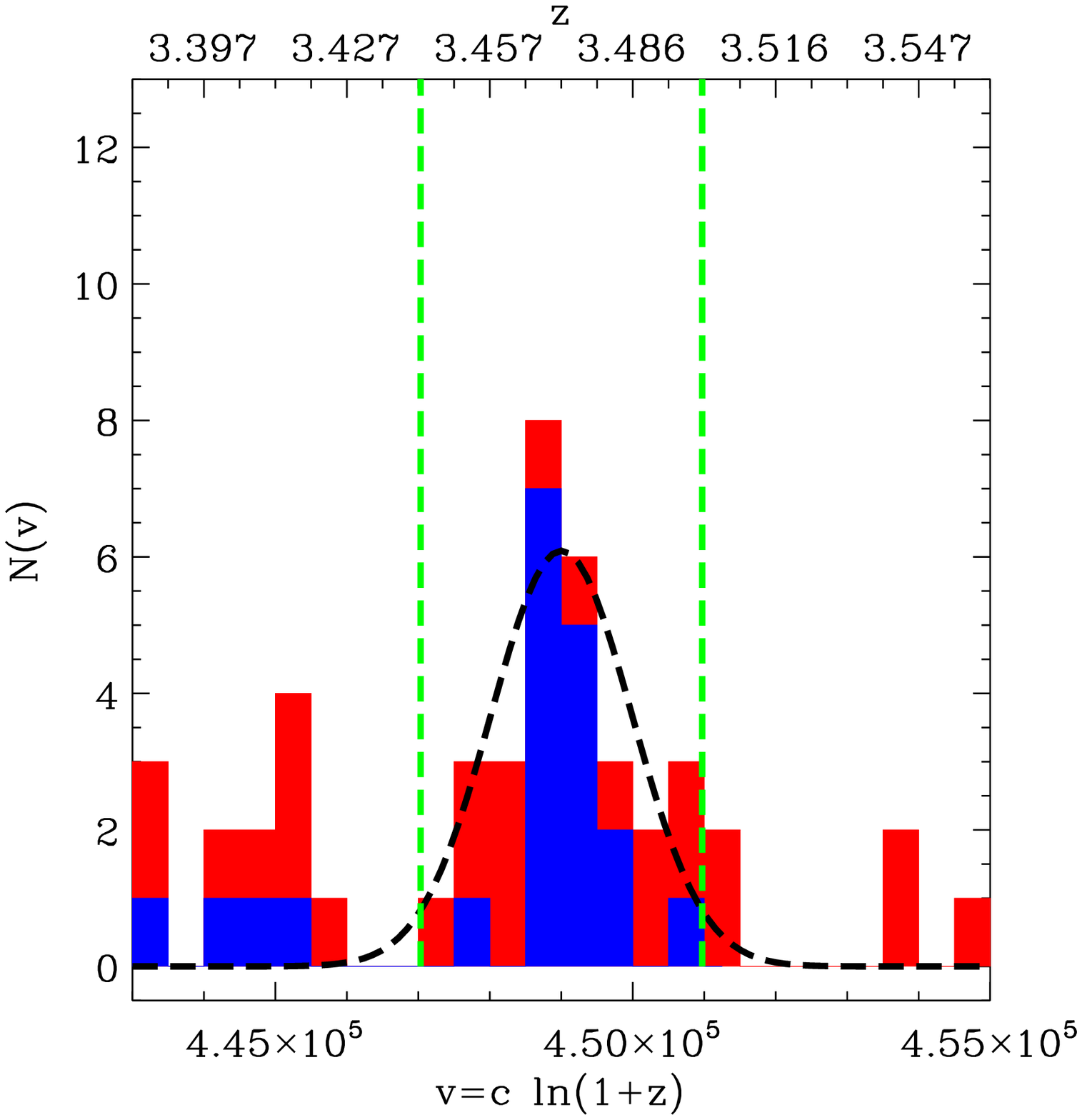}
\includegraphics[width=6.0 cm, angle=0]{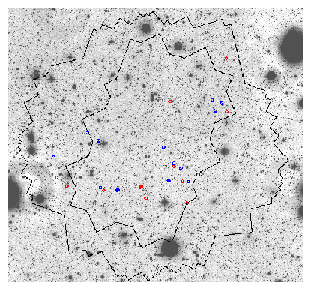}
\includegraphics[width=3.0 cm, angle=0]{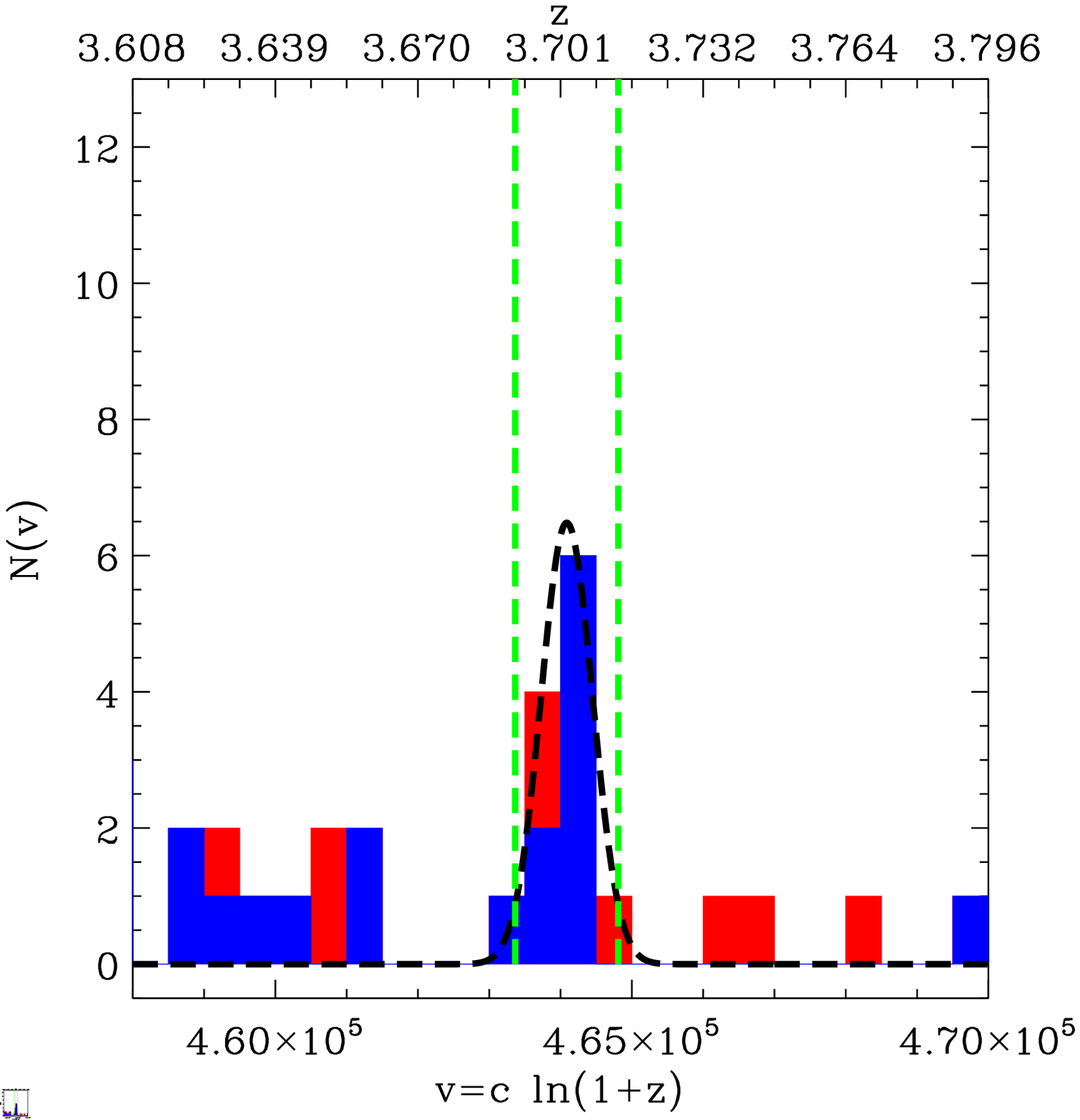}
\includegraphics[width=6.0 cm, angle=0]{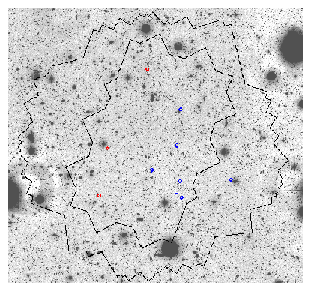}
\caption{Histograms of the velocity distribution, $V=c\ln(1+z)$, and spatial 
distribution of galaxies belonging to each of the density peak listed in 
Table~\ref{peaks}. For each structure, we display ``secure'' redshifts 
($>99$\% c.l.) in \textit{blue} and ``likely'' redshifts ($\sim70-90$\% c.l.) 
in \textit{red}. The histograms show the Gaussian fit to the peak of the velocity distribution 
(\textit{black dashed curve}) and the $2\sigma$-intervals around the peak 
selected for the spatial analysis (\textit{vertical green dashed lines}). All the histograms 
are binned to $\Delta V=500$~km/s. The black contours indicate the GOODS area 
and the field coverage of the 2Ms Chandra exposure.}
\label{280}
\end{figure*}

Table~\ref{peaks} and Fig.~\ref{654}--\ref{280} show the results of the analysis of 
the spatial distribution of the 14 confirmed density peaks, plus three additional 
tentative structures at $z\simeq2.8$, $z\simeq3.5$, and $z\simeq3.7$, observed 
in the master catalog of redshifts of the CDFS, described in Sect.~\ref{zdist}.

\begin{table}[t]
\caption{List of observed peaks in the redshift distribution of the reference 
master catalog. The columns list: (1) the average $<z>$ of each peak, (2)
the velocity dispersion $\sigma_V$, (3) the redshift range corresponding 
to $\pm2\sigma_V$, selected for the analysis of the spatial distribution 
(see Fig.~\ref{654}, \ref{736}, and \ref{280}), (4) the number of galaxies in the selected redshift 
interval, and (5) the identification numbers of extended X-ray sources 
\citep[from ][]{gia02} associated with the peak.}
\begin{center}
\begin{tabular}{c c c c c}
\hline
\hline
$<z>$ & $\sigma_V$~[km/s] &  $z_{min}-z_{max}$ & $N$  & XID \\
(1)      & (2)     &   (3)                 &  (4)    & (5)    \\
\hline
0.126 &   906	   &  $0.119-0.133$ &  73     & --  \\
0.215 &   443	   &  $0.211-0.218$ &  61     & --  \\
0.338 &  1724	   &  $0.322-0.353$ &  76     & --  \\
0.530 &  2190	   &  $0.508-0.552$ & 193     & --  \\
0.672 &  1758	   &  $0.653-0.692$ & 306     & 645 \\
0.735 &   565	   &  $0.728-0.742$ & 257     & 566, 594  \\
0.973 &  2936	   &  $0.935-1.012$ & 181     & 249 \\
1.039 &   859	   &  $1.028-1.051$ & 110     & --  \\
1.095 &  1536	   &  $1.074-1.117$ & 129     & --  \\
1.221 &   391	   &  $1.215-1.227$ &  65     & --  \\
1.296 &   261	   &  $1.292-1.300$ &  35     & --  \\
1.611 &   497	   &  $1.602-1.620$ &  36     & --  \\
2.318 &  2102	   &  $2.272-2.365$ & 108     & --  \\
2.566 &  1435	   &  $2.532-2.601$ &  72     & --  \\
2.811 &   575      &  $2.796-2.825$ &  23     & --  \\
3.471 &   984	   &  $3.442-3.501$ &  29     & --  \\
3.702 &   360	   &  $3.691-3.714$ &  11     & --  \\
\hline
\end{tabular}
\end{center}
\label{peaks}
\end{table}
\newpage

\newpage

\newpage

\section{Spectra of X-ray sources with new redshift determinations}\label{app2}

\begin{figure*}
\centering
\includegraphics[width=8.0 cm, angle=0]{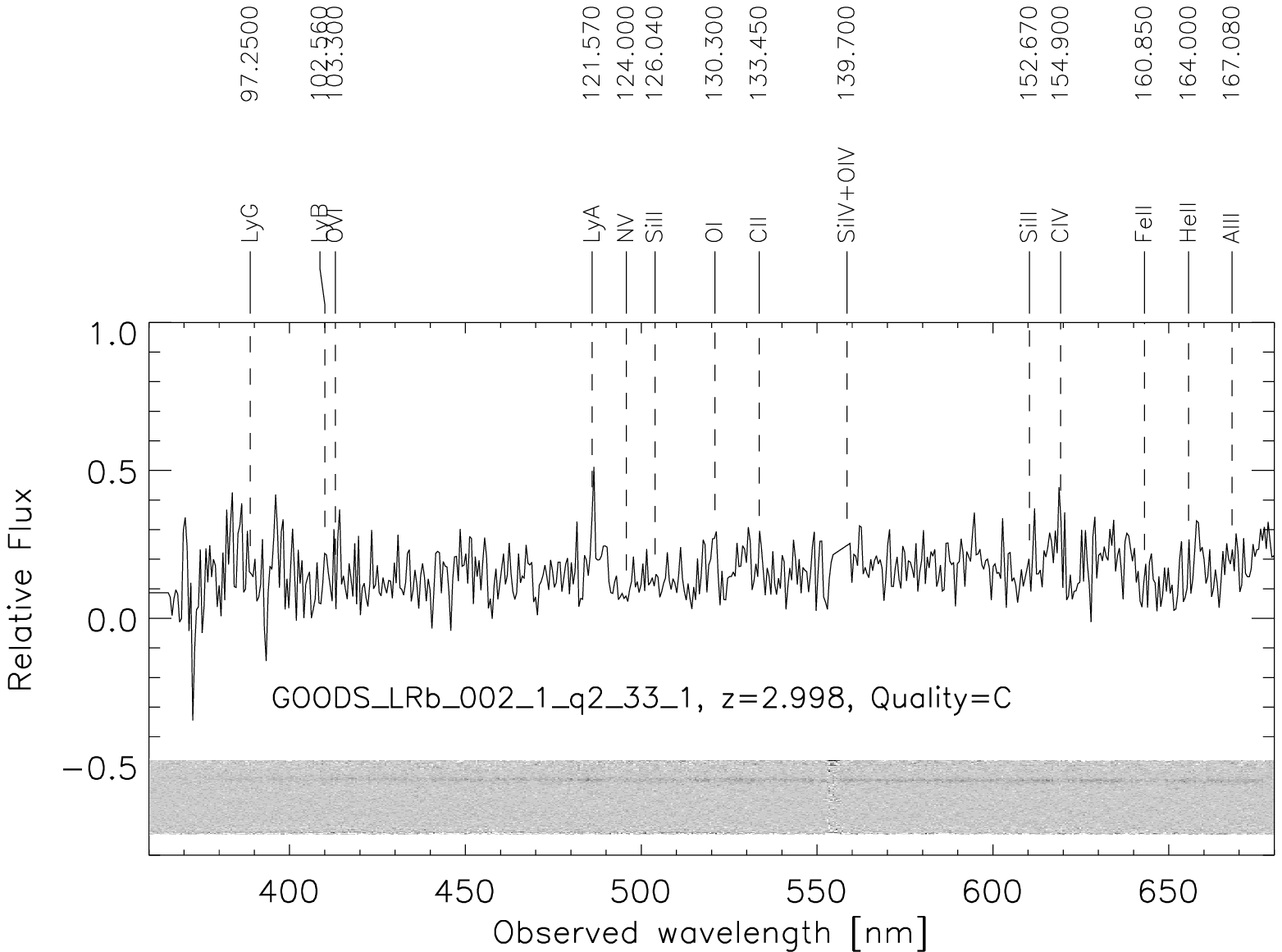}
\includegraphics[width=5.0 cm, angle=0]{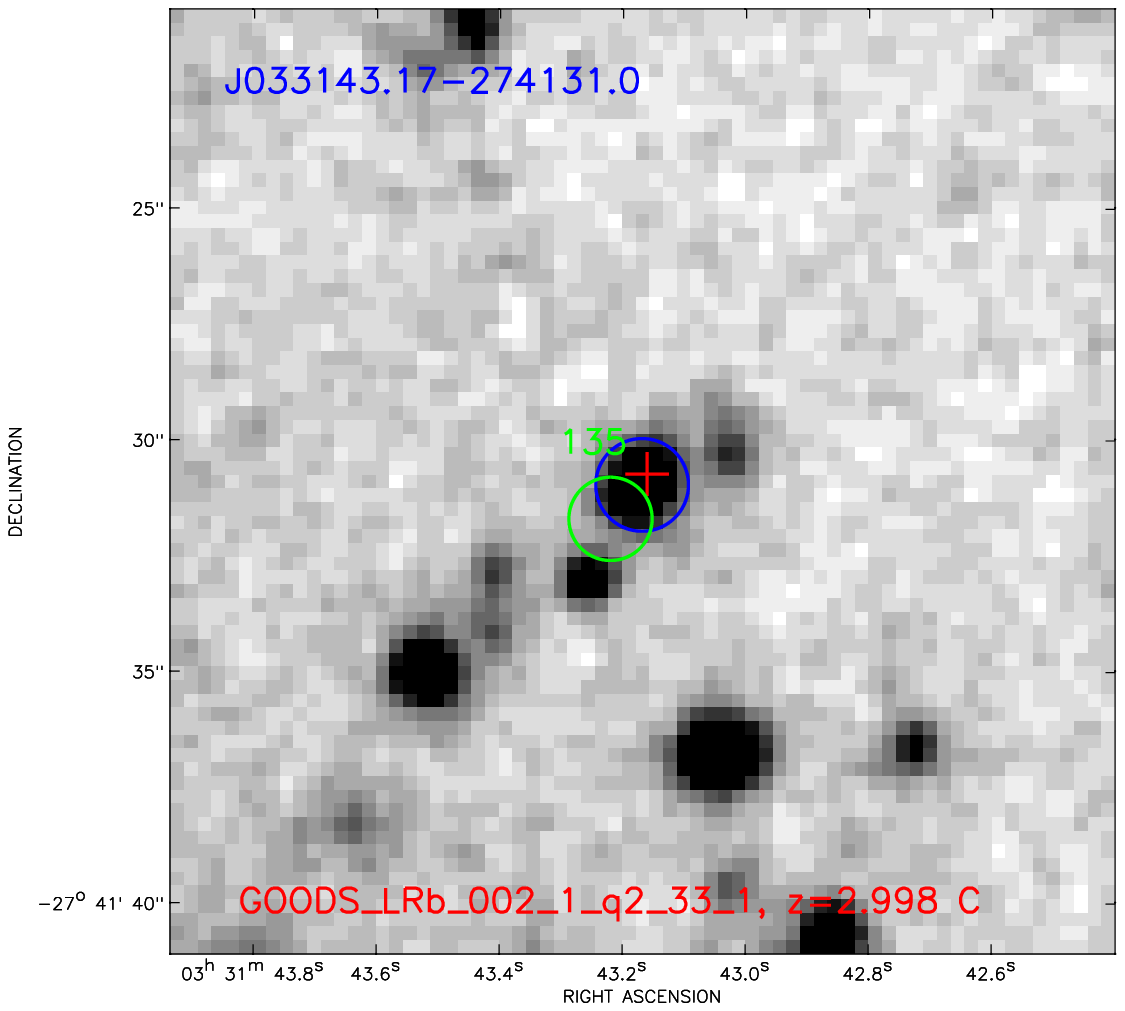}
\includegraphics[width=8.0 cm, angle=0]{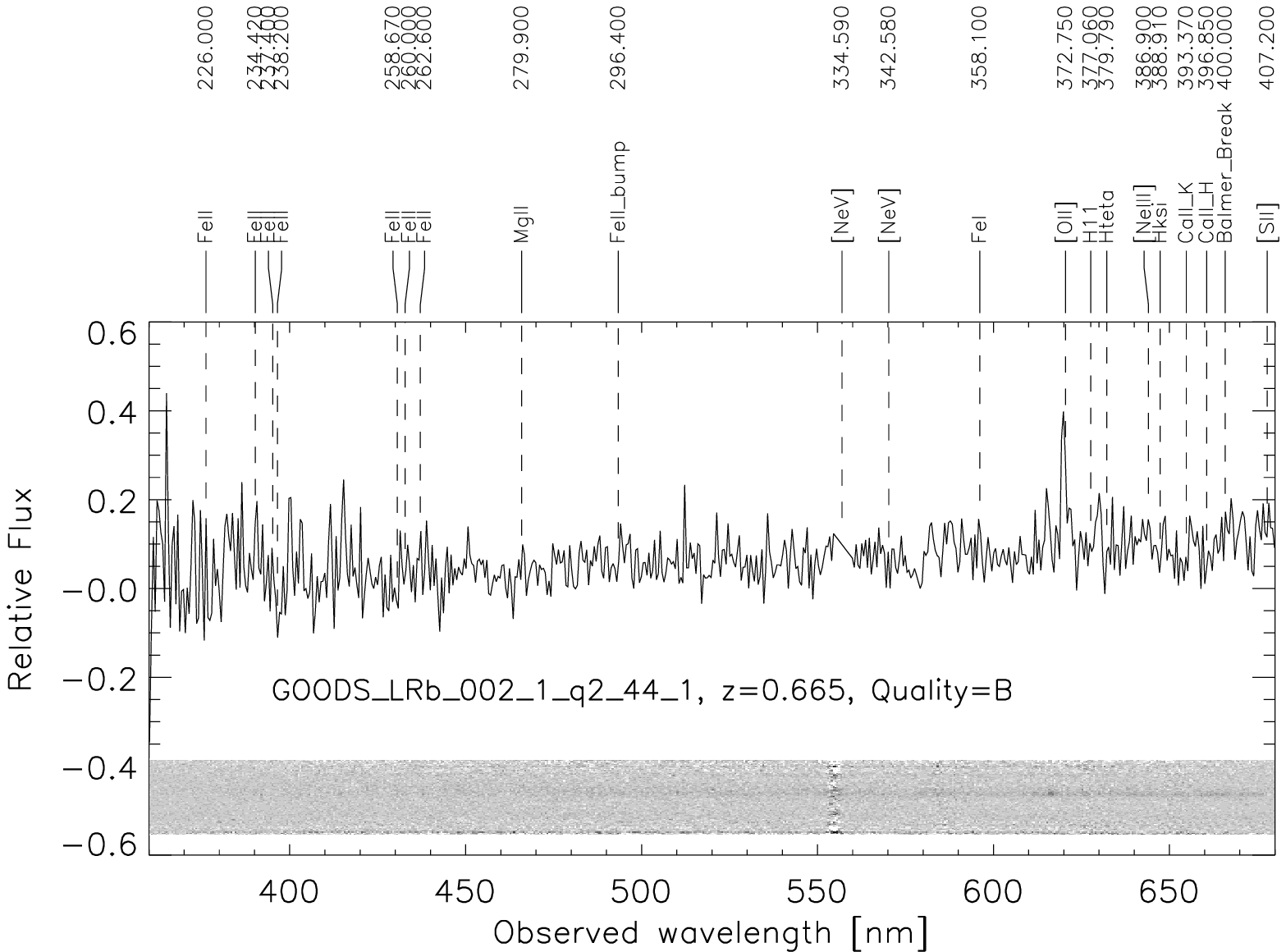}
\includegraphics[width=5.0 cm, angle=0]{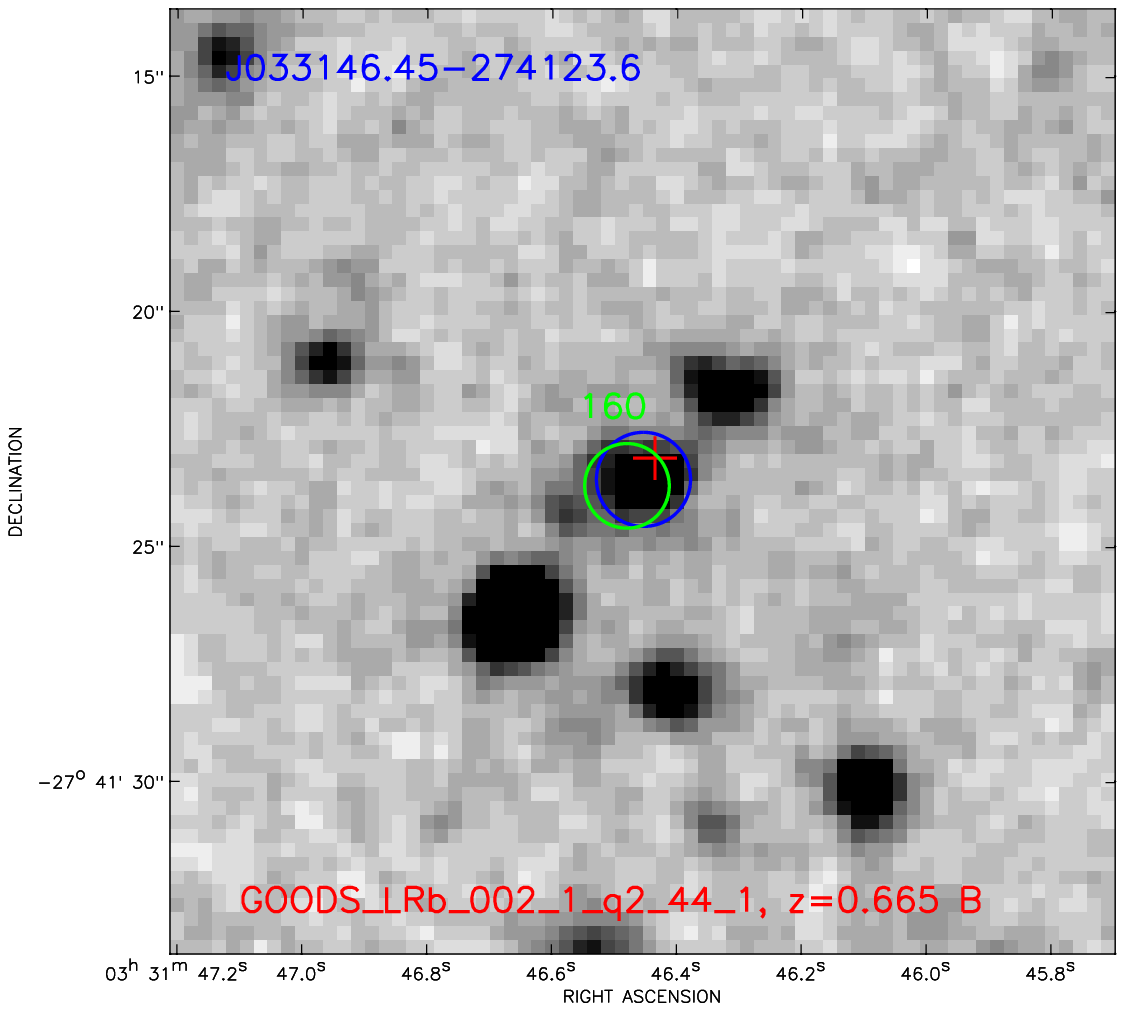}
\includegraphics[width=8.0 cm, angle=0]{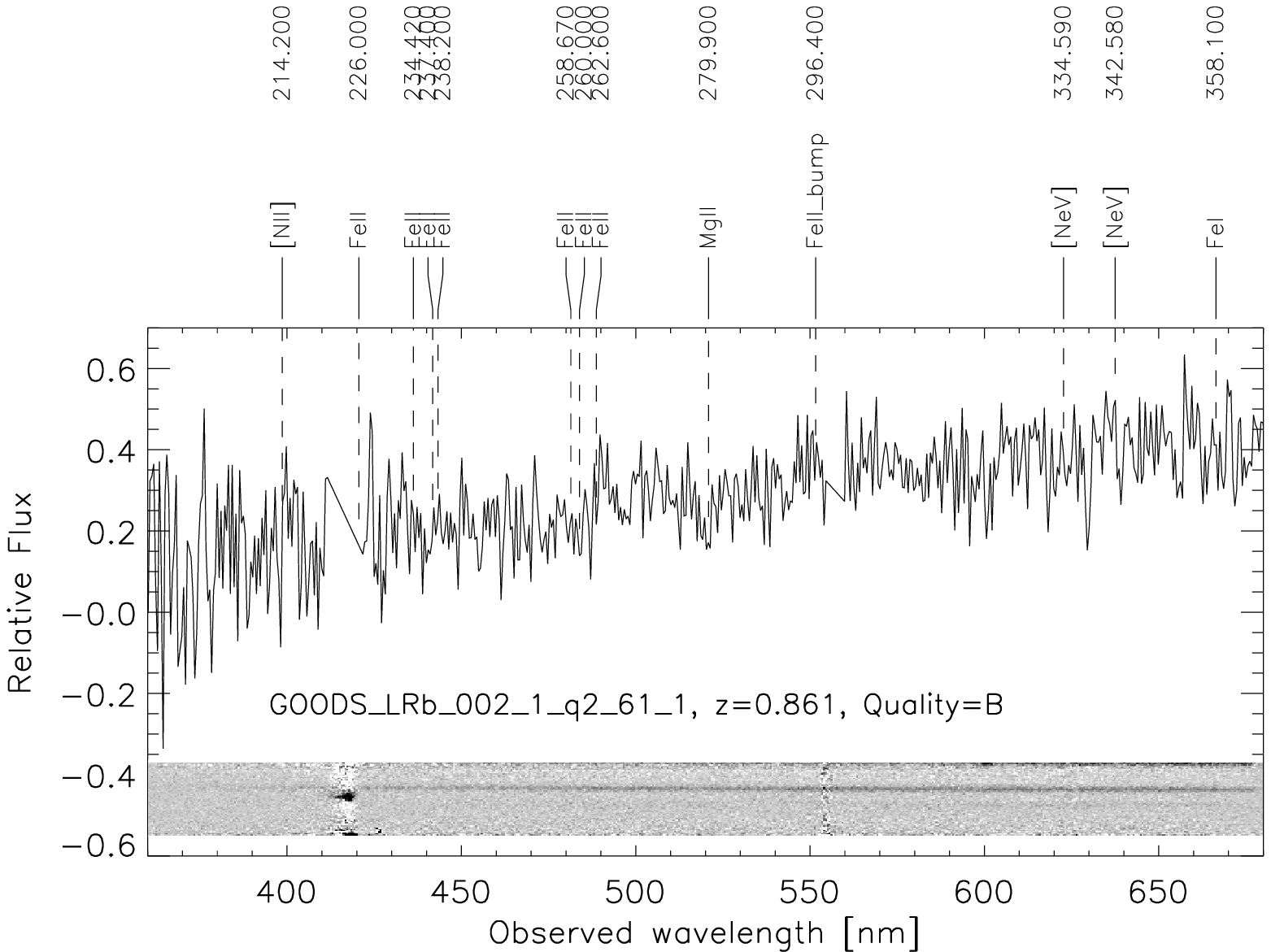}
\includegraphics[width=5.0 cm, angle=0]{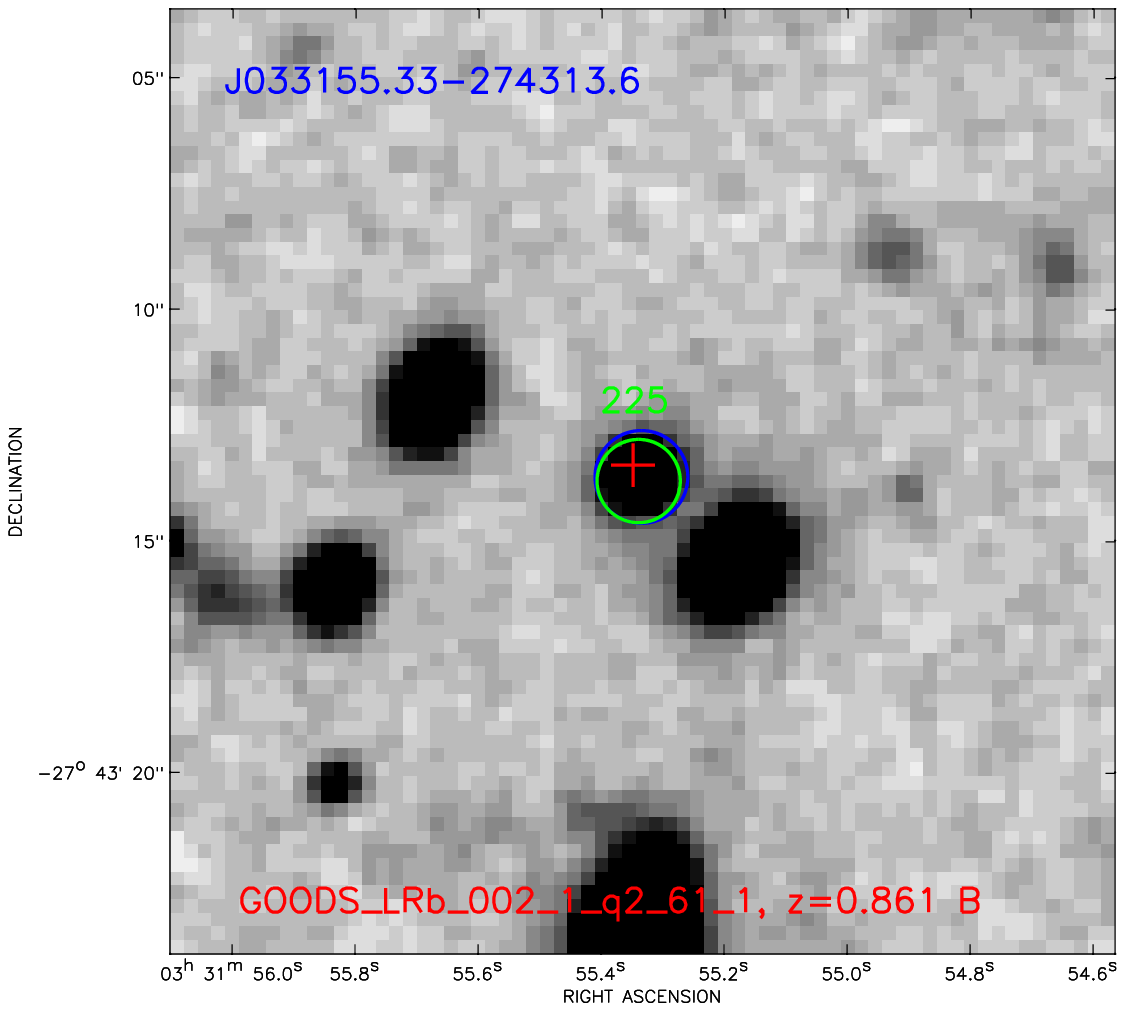}
\includegraphics[width=8.0 cm, angle=0]{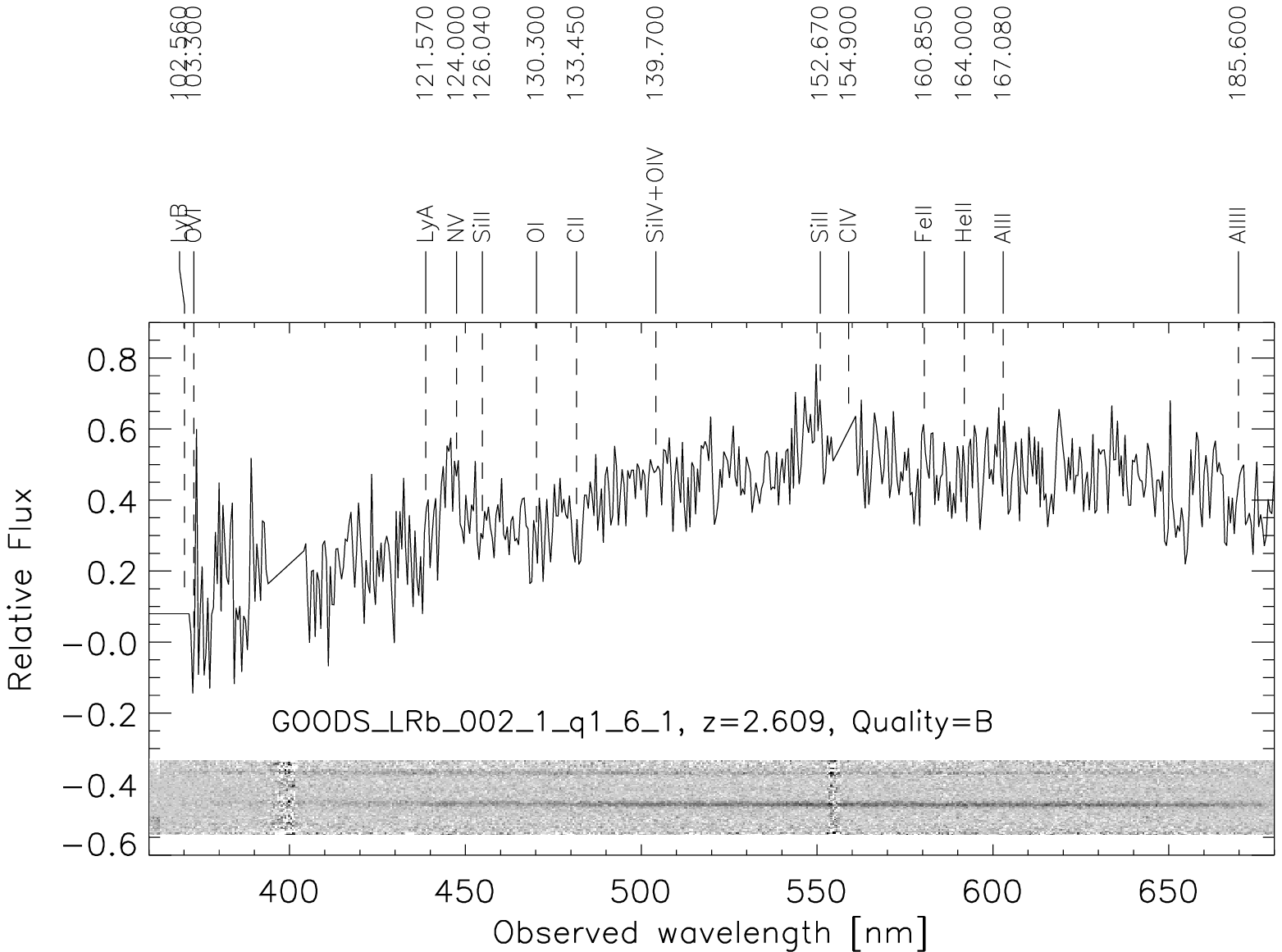}
\includegraphics[width=5.0 cm, angle=0]{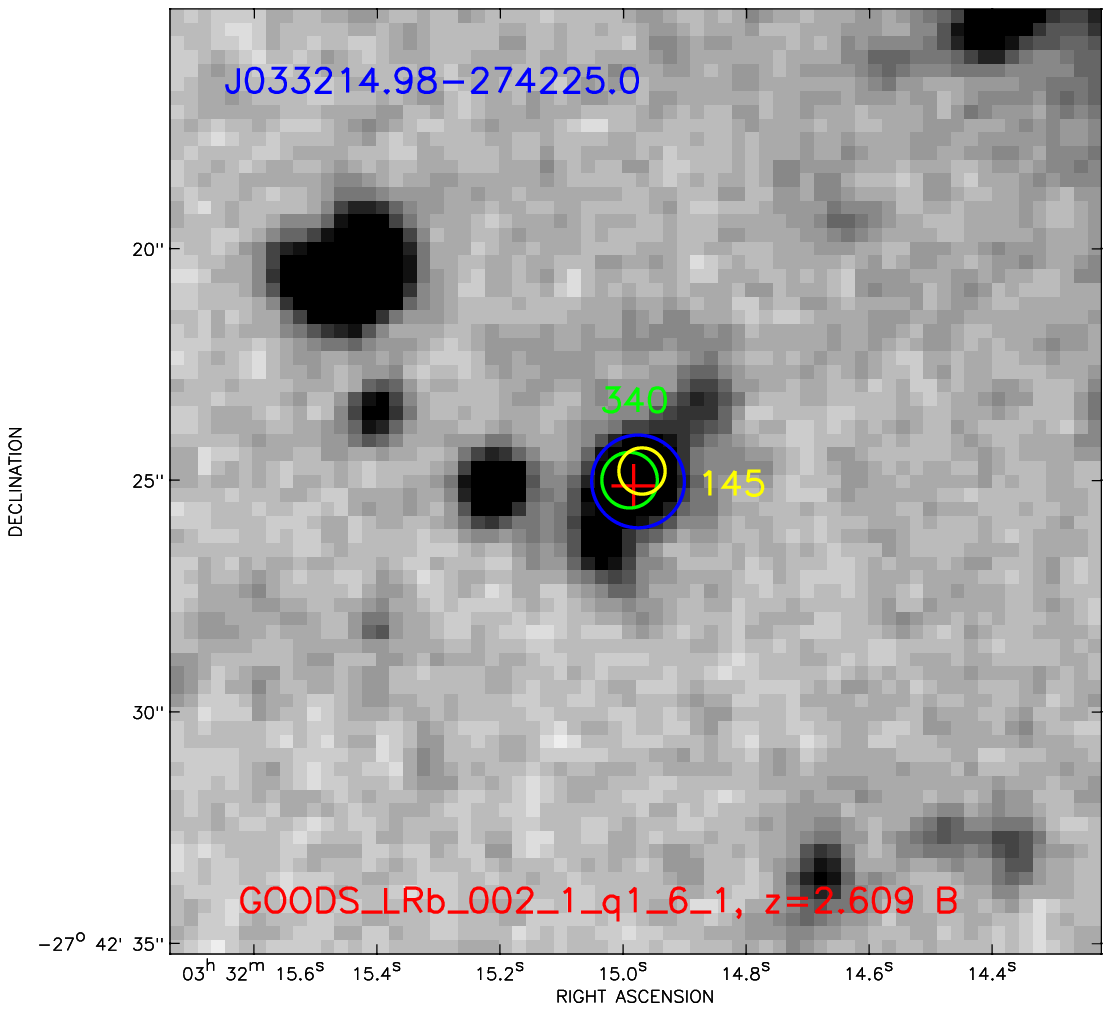}
\caption{New VIMOS spectra of X-ray sources with WFI $R$-band $20''\times20''$ 
cutout centered on the position of the matching WFI source. Red crosses indicate 
the reconstructed VIMOS coordinates. The blue circles show the position of the 
WFI-matched objects. The yellow and green circles display the position of 
X-ray sources from the 2Ms CDFS or from the ECDFS catalog, respectively. Labels
with corresponding colors indicate source identification numbers.}
\label{x1}
\end{figure*}

\begin{figure*}
\centering
\includegraphics[width=8 cm, angle=0]{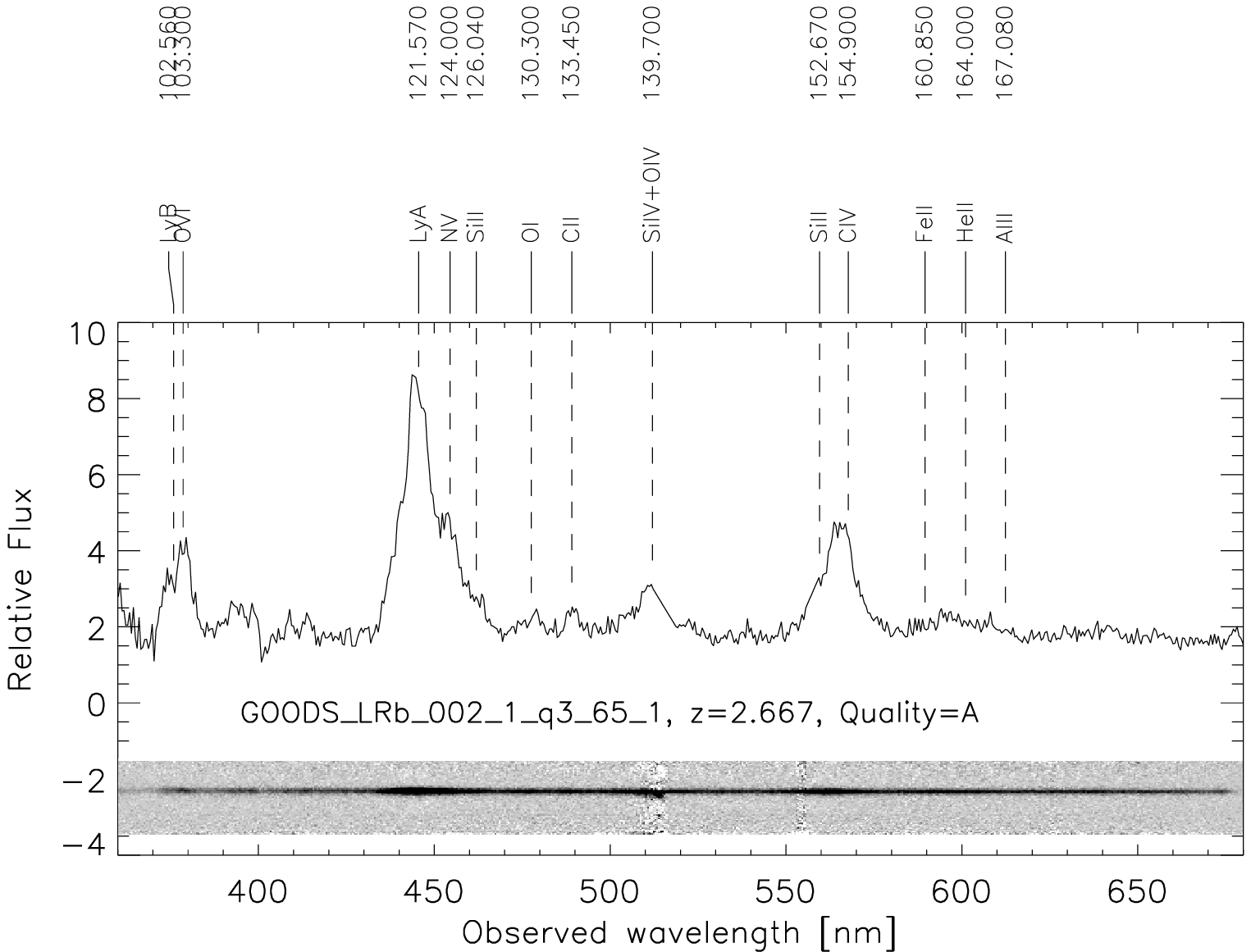}
\includegraphics[width=5 cm, angle=0]{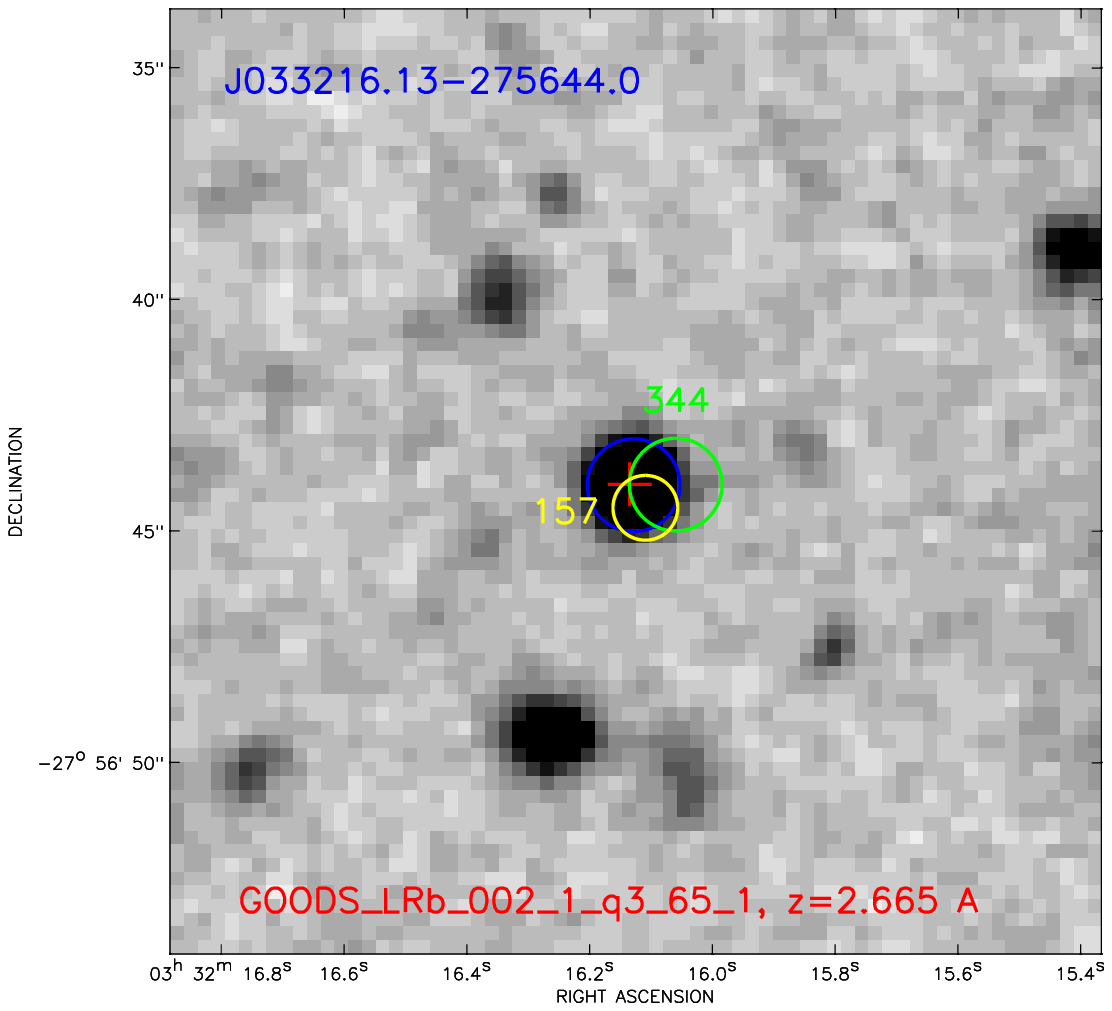}
\includegraphics[width=8 cm, angle=0]{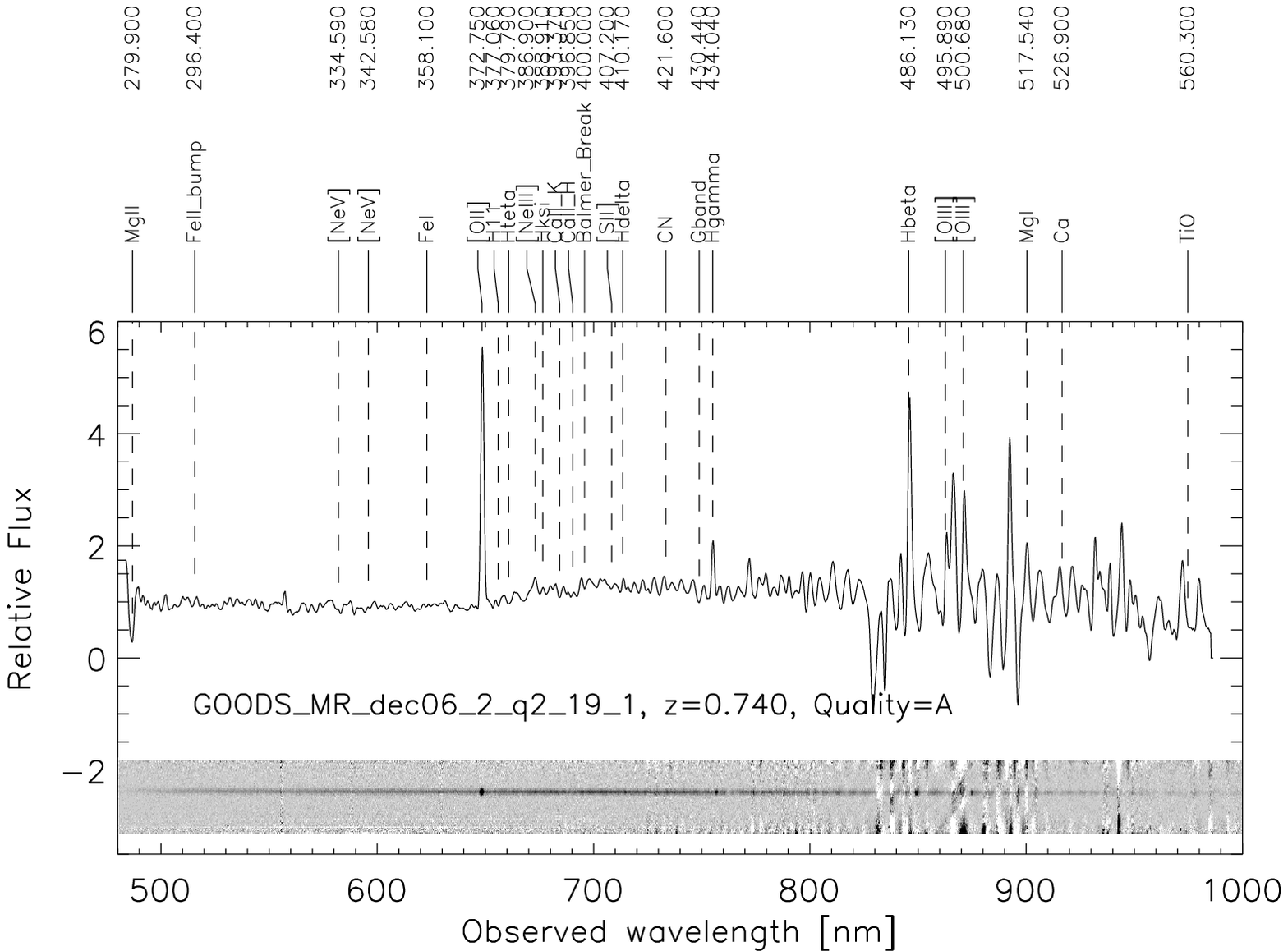}
\includegraphics[width=5 cm, angle=0]{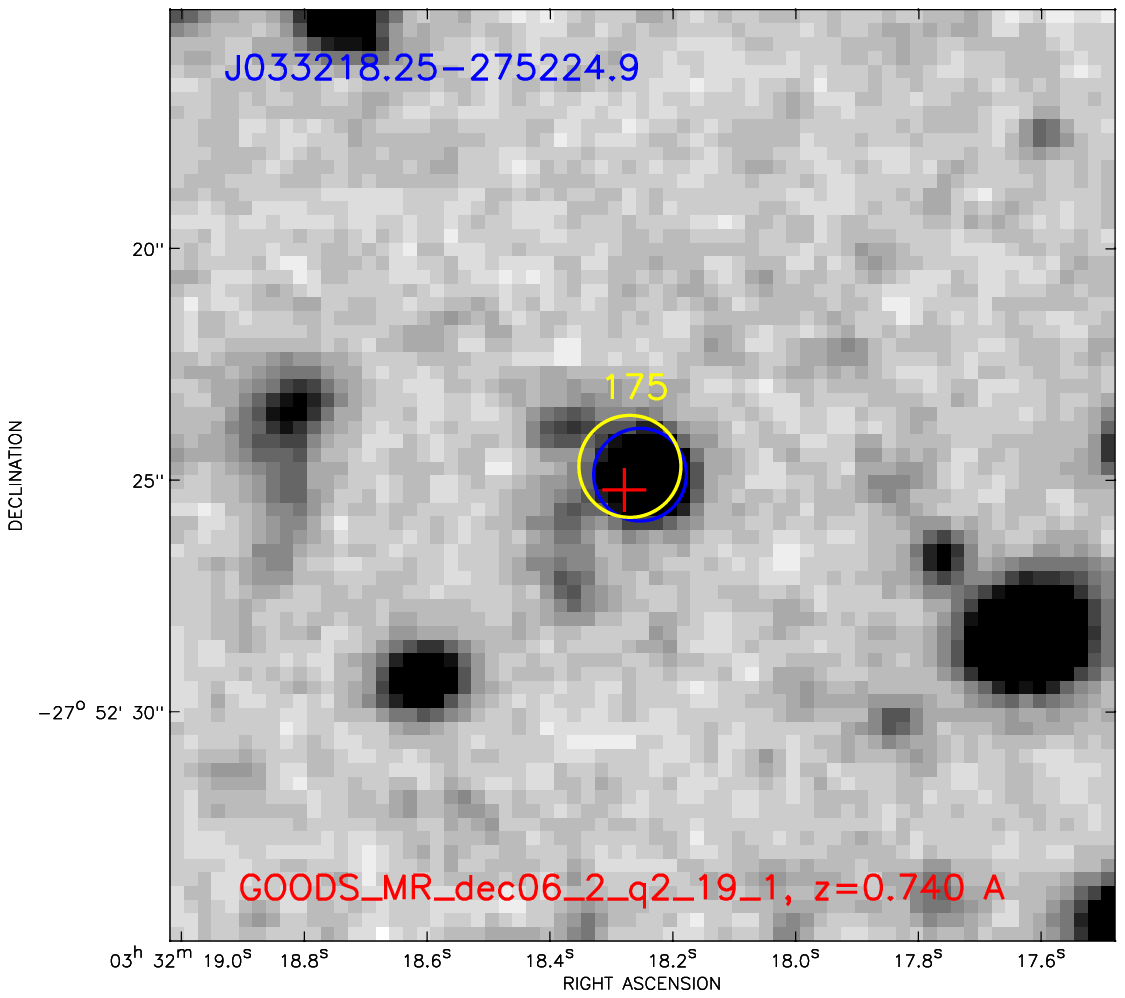}
\includegraphics[width=8 cm, angle=0]{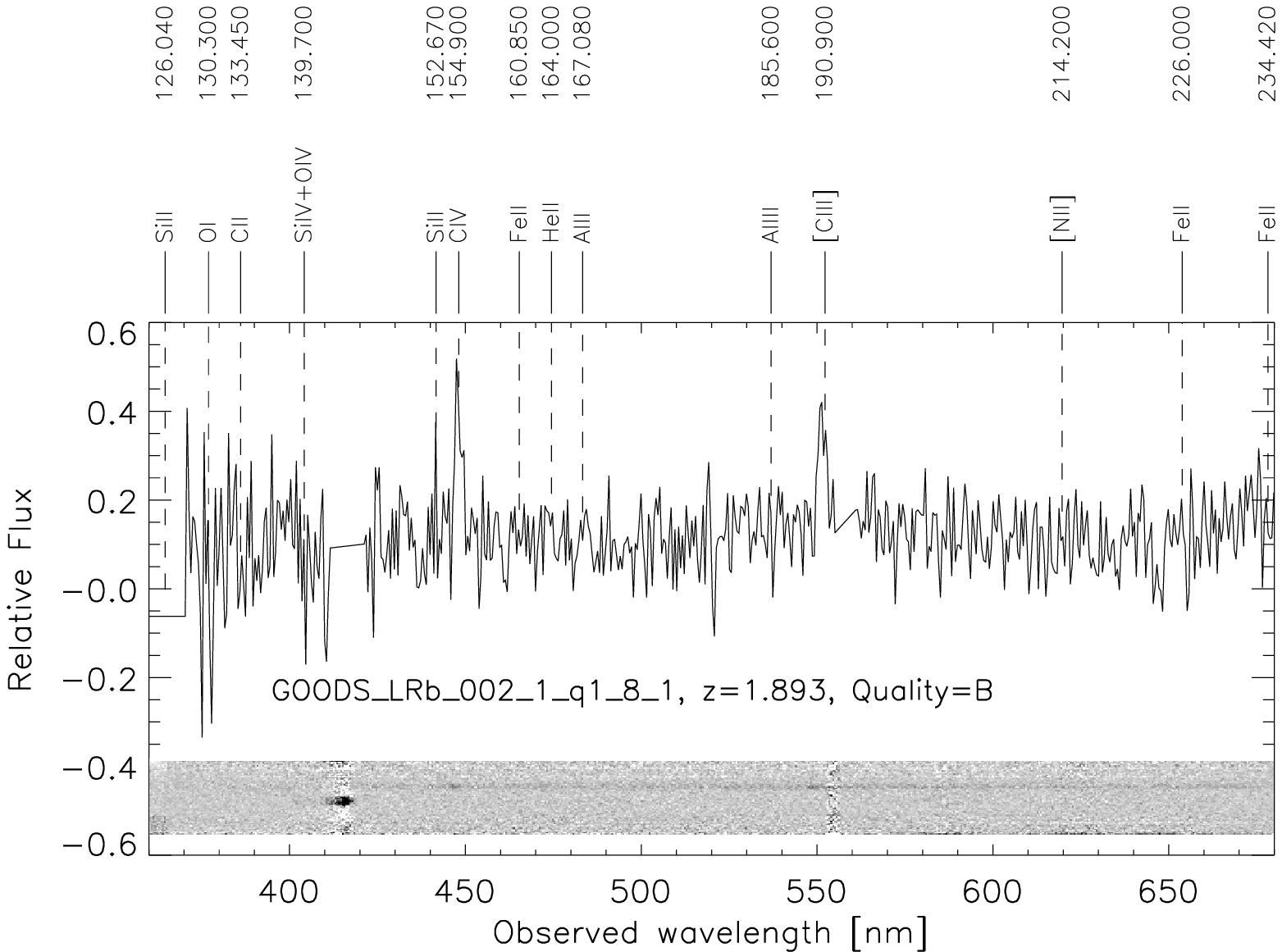}
\includegraphics[width=5 cm, angle=0]{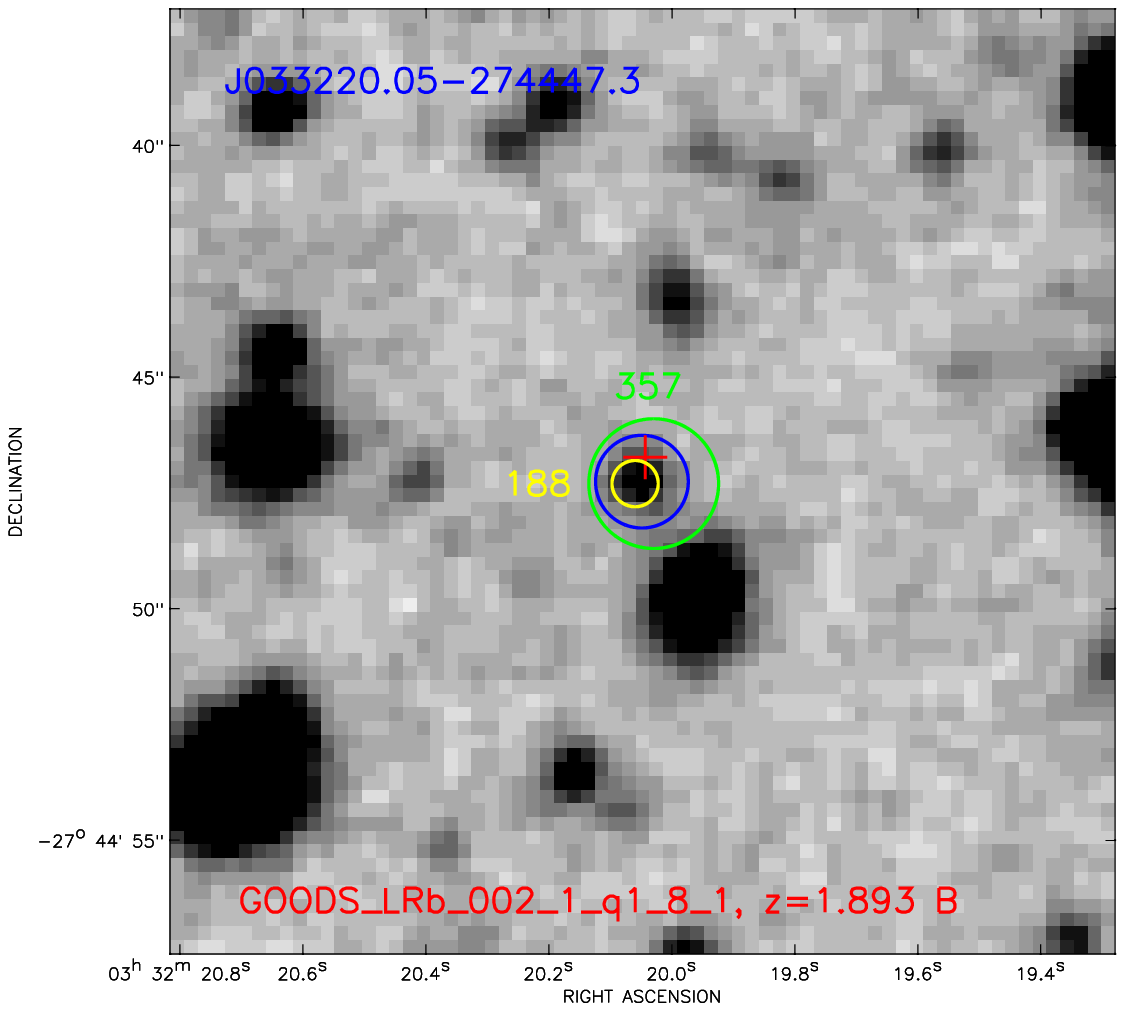}
\includegraphics[width=8 cm, angle=0]{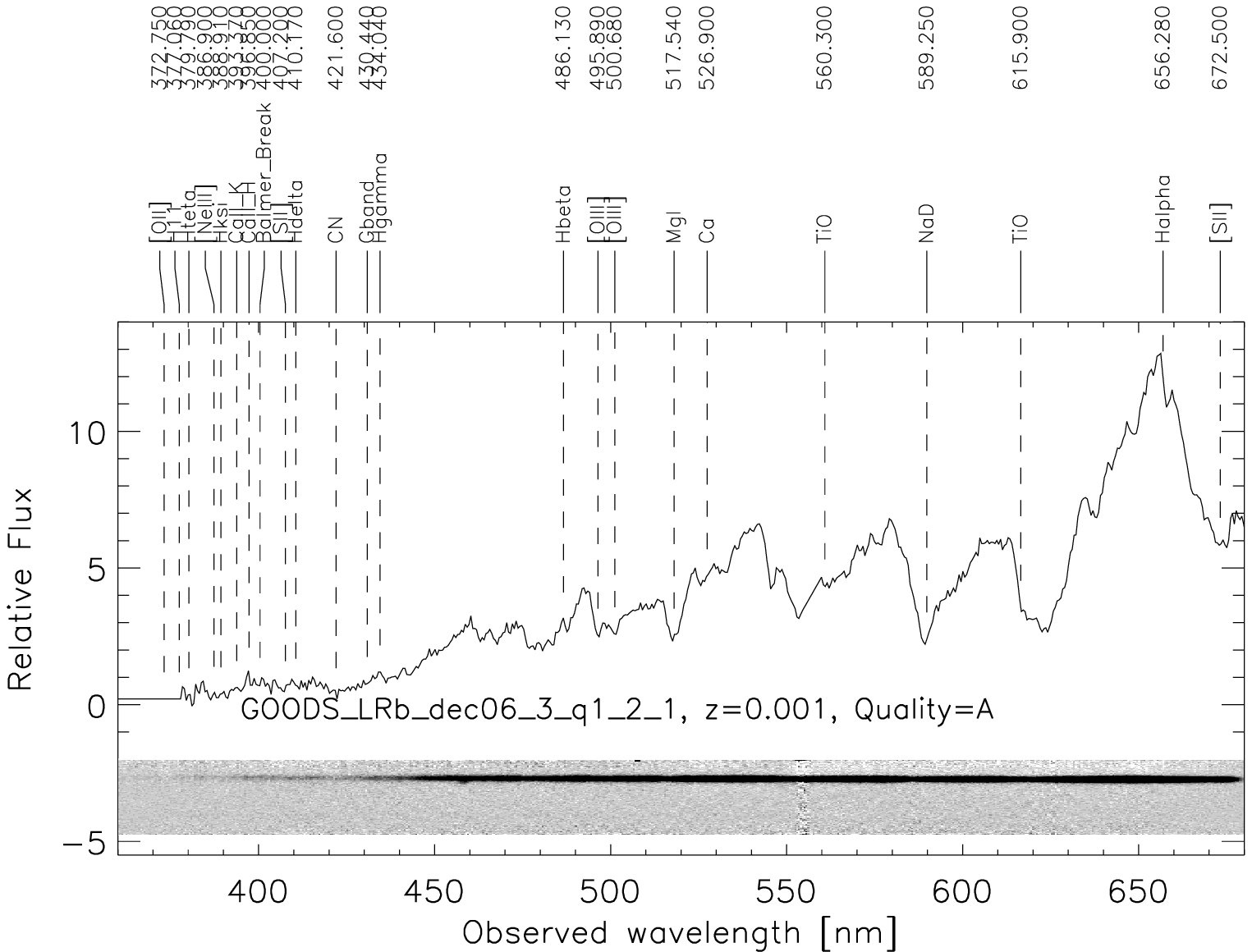}
\includegraphics[width=5 cm, angle=0]{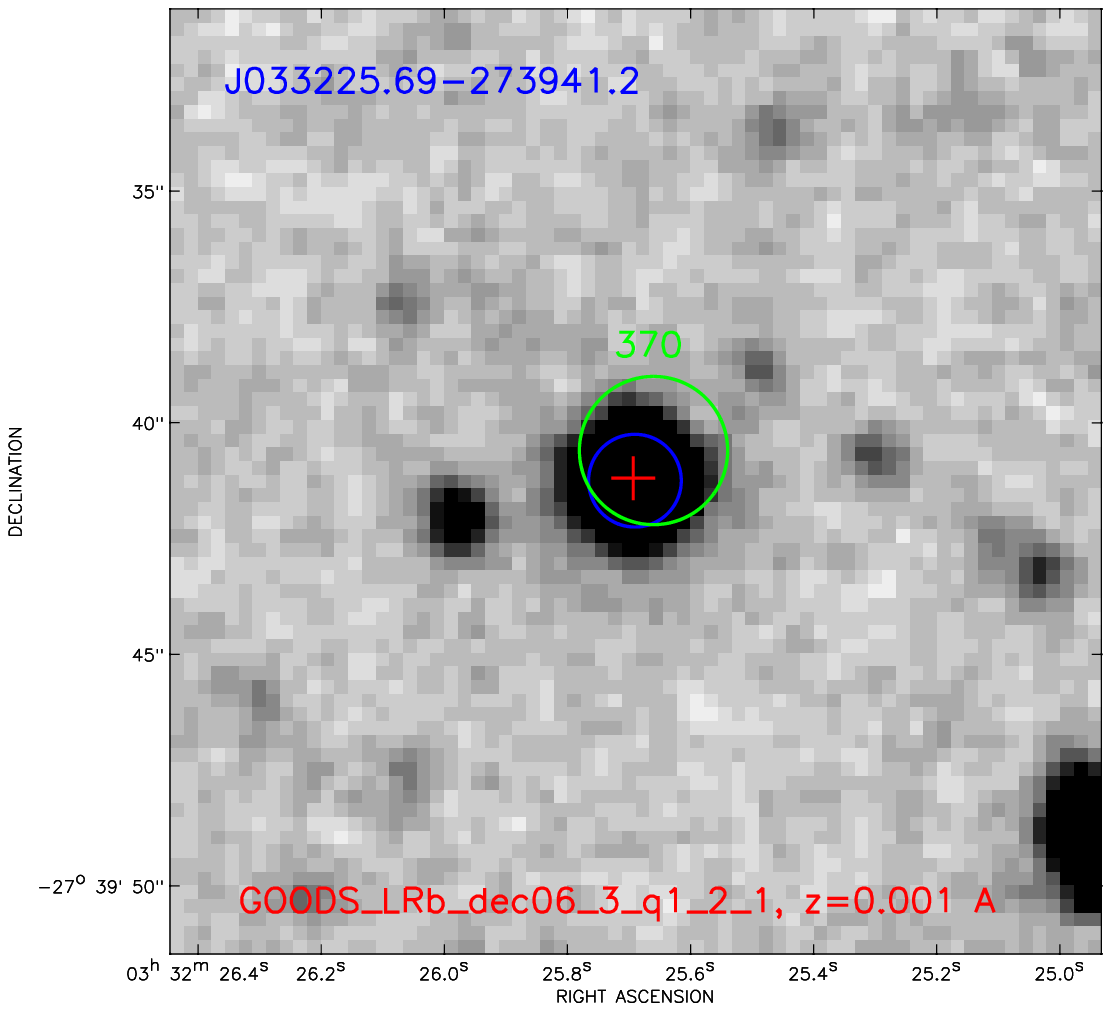}
\caption{New VIMOS spectra of X-ray sources with WFI $R$-band $20''\times20''$ 
cutout centered on the position of the matching WFI source. Red crosses indicate 
the reconstructed VIMOS coordinates. The blue circles show the position of the 
WFI-matched objects. The yellow and green circles display the position of 
X-ray sources from the 2Ms CDFS or from the ECDFS catalog, respectively. Labels
with corresponding colors indicate source identification numbers.}
\label{x2}
\end{figure*}

\begin{figure*}
\centering
\includegraphics[width=8 cm, angle=0]{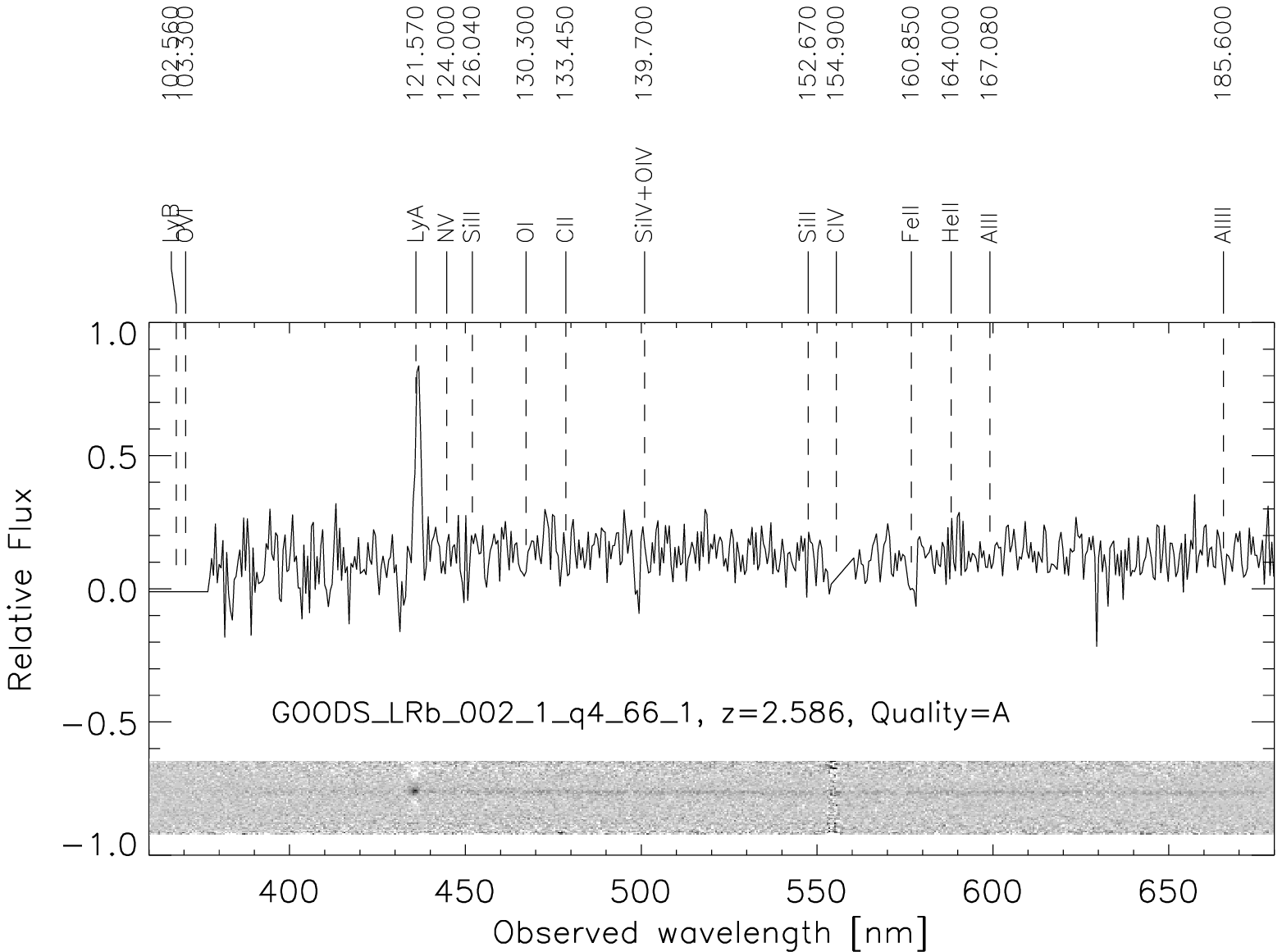}
\includegraphics[width=5 cm, angle=0]{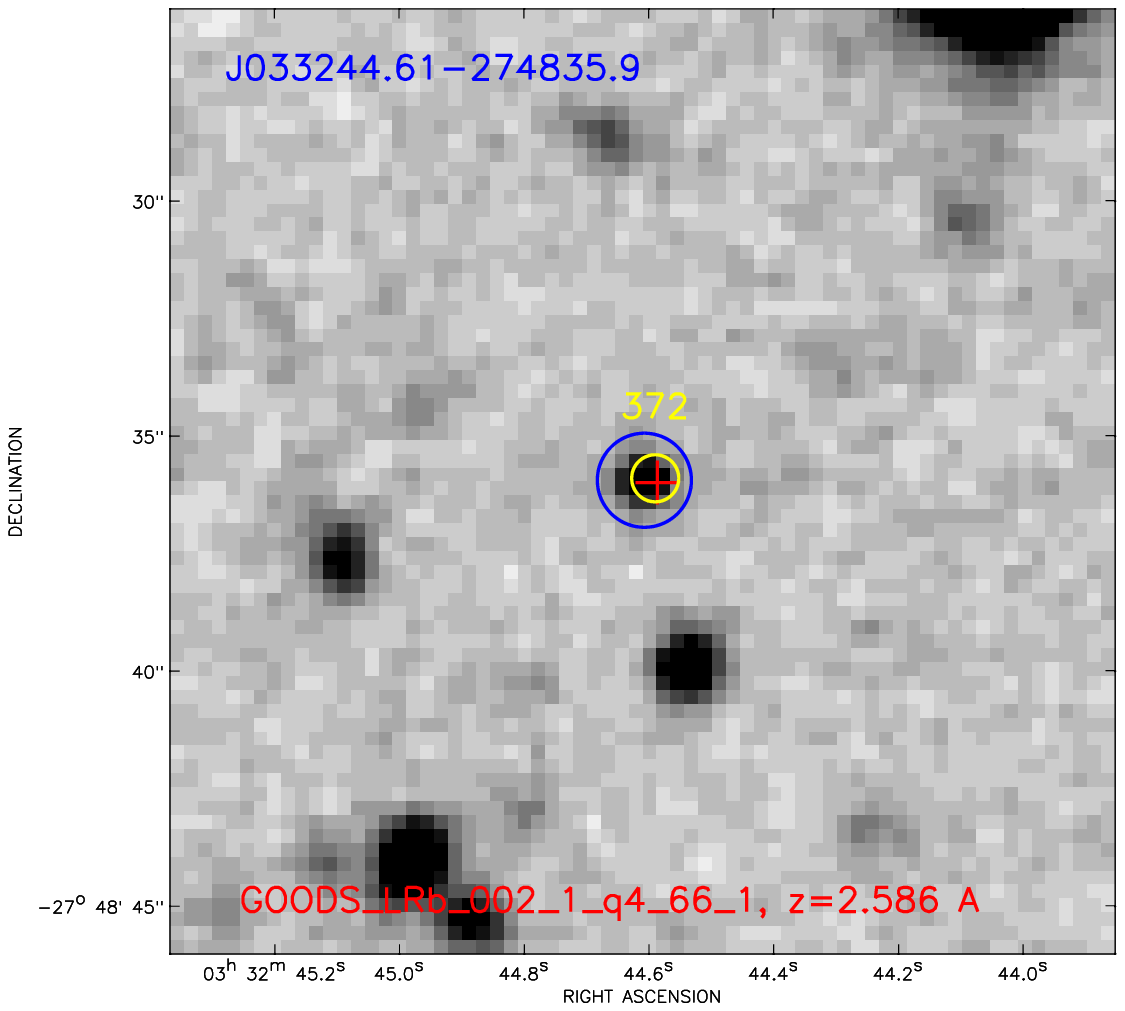}
\includegraphics[width=8 cm, angle=0]{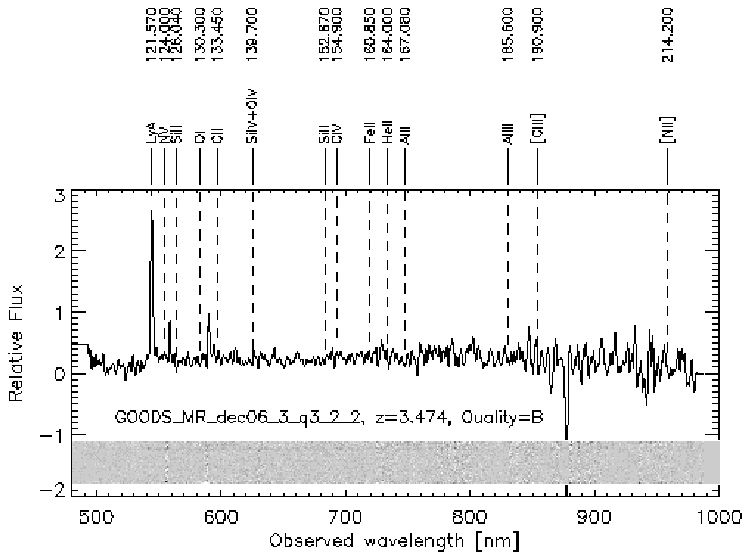}
\includegraphics[width=5 cm, angle=0]{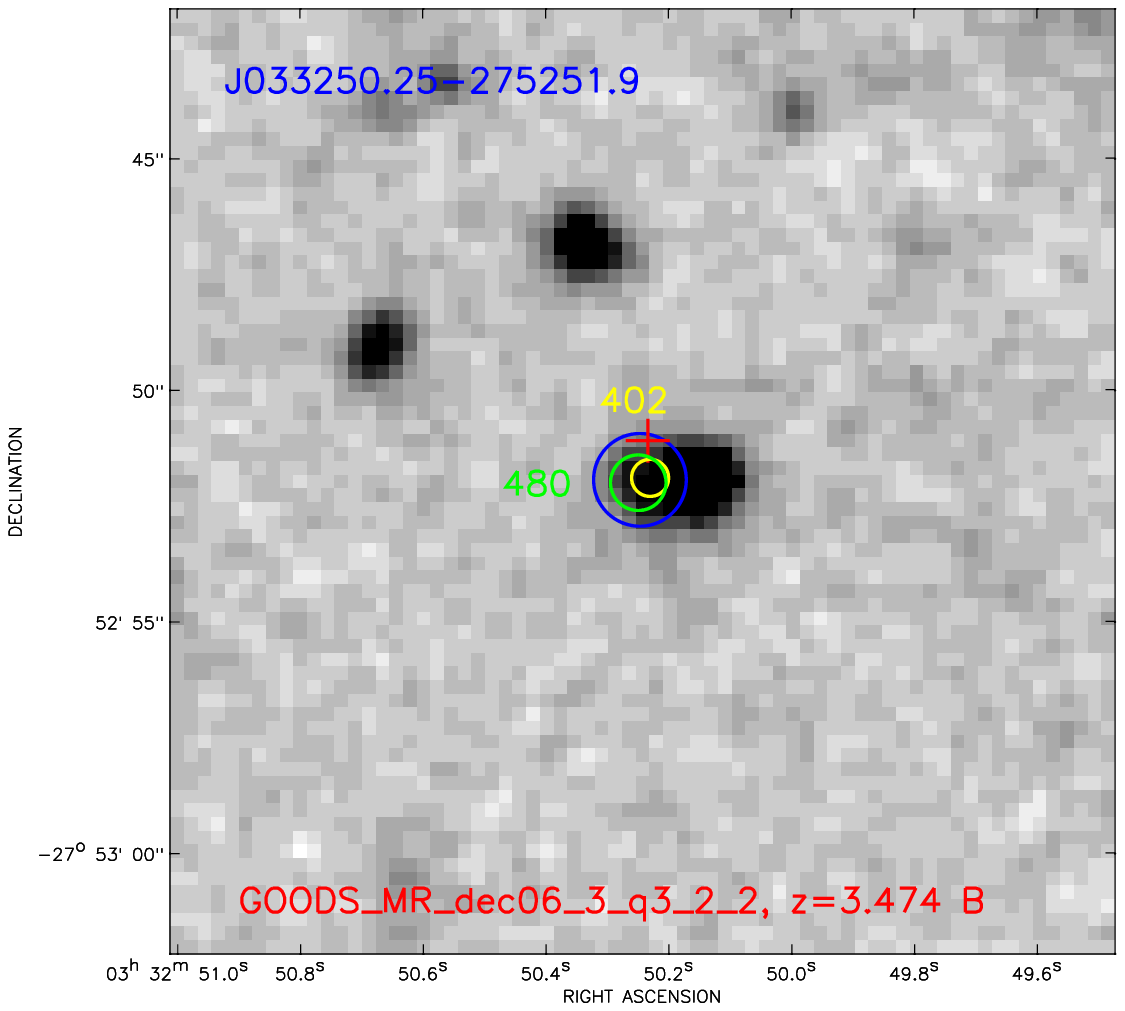}
\includegraphics[width=8 cm, angle=0]{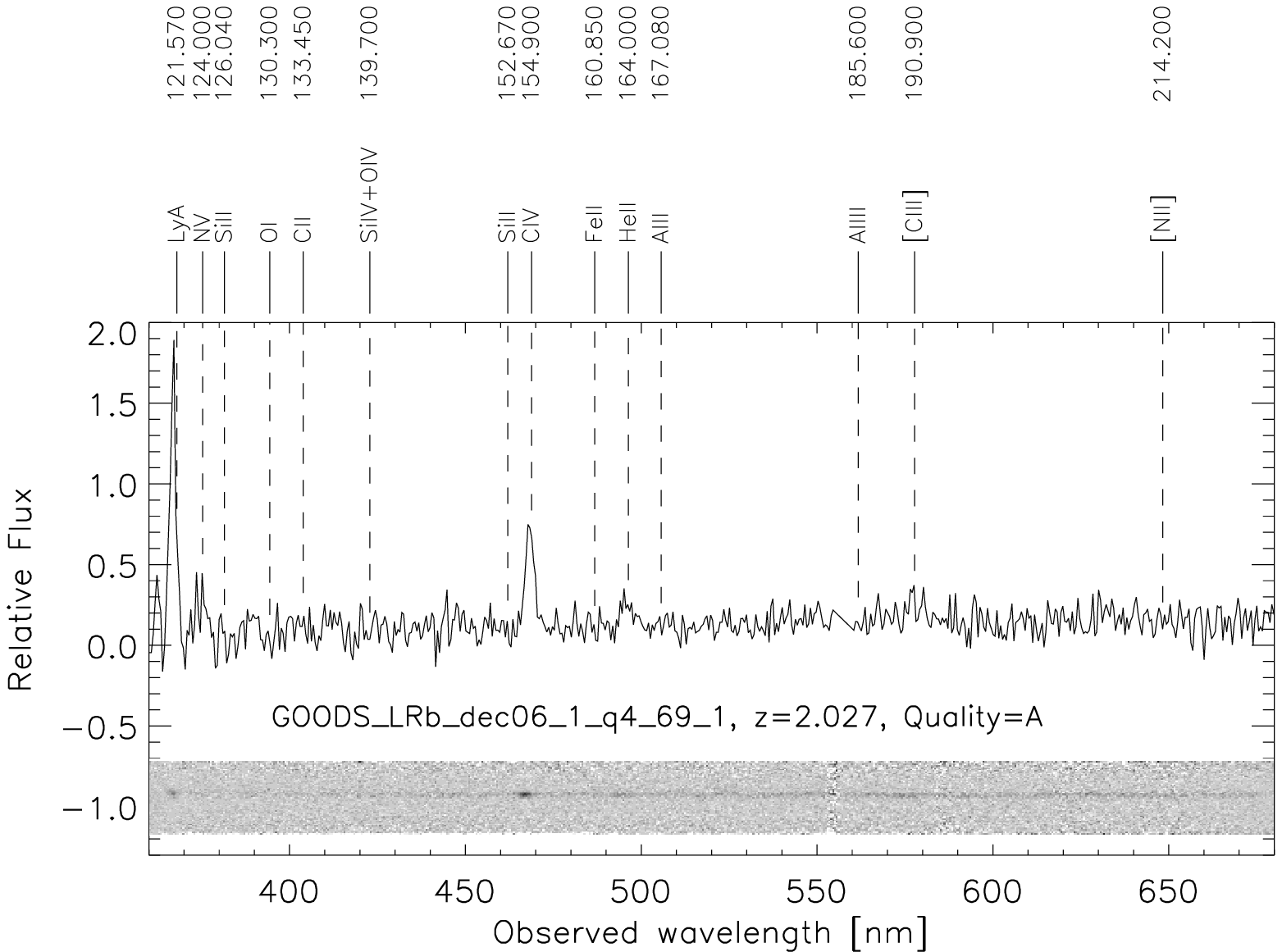}
\includegraphics[width=5 cm, angle=0]{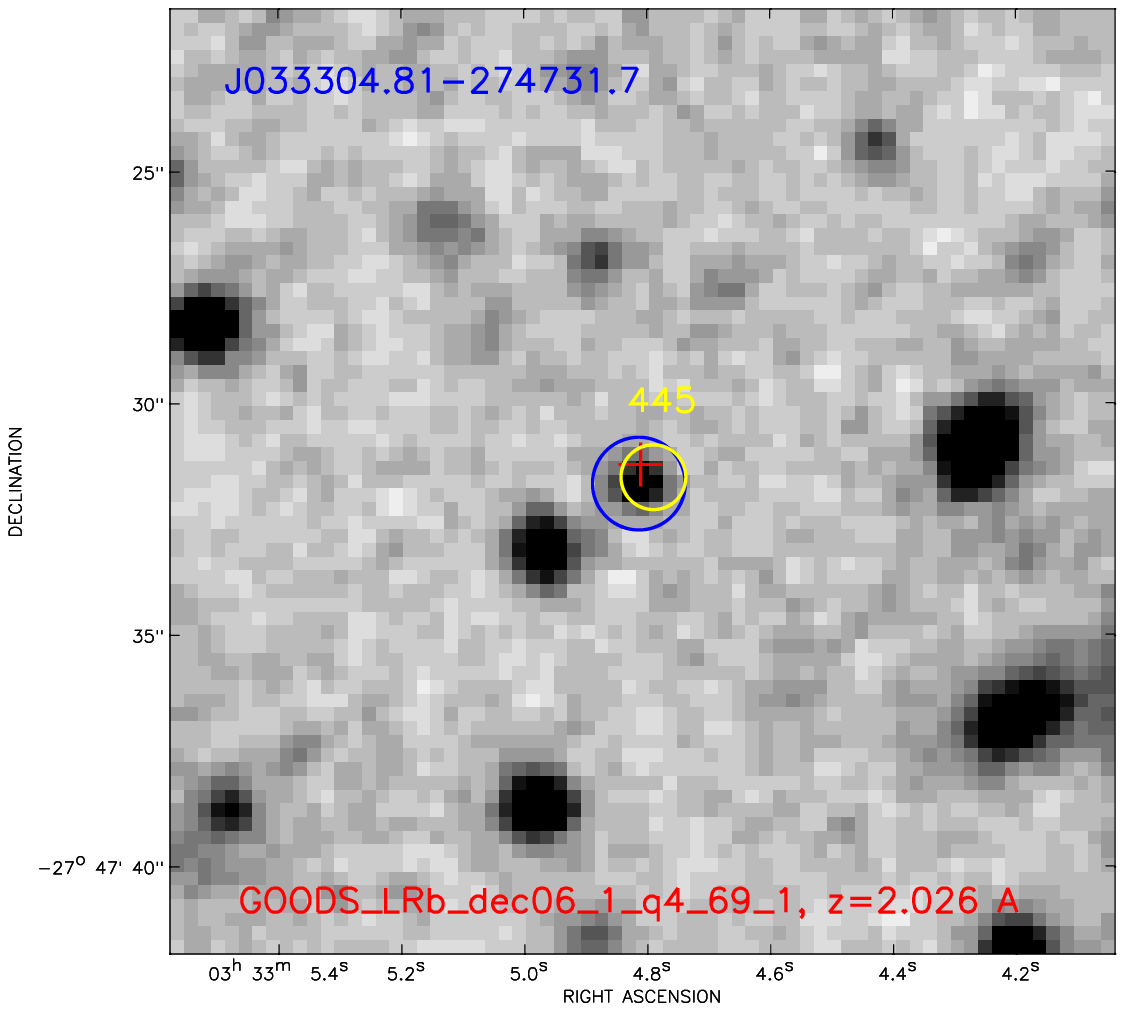}
\includegraphics[width=8 cm, angle=0]{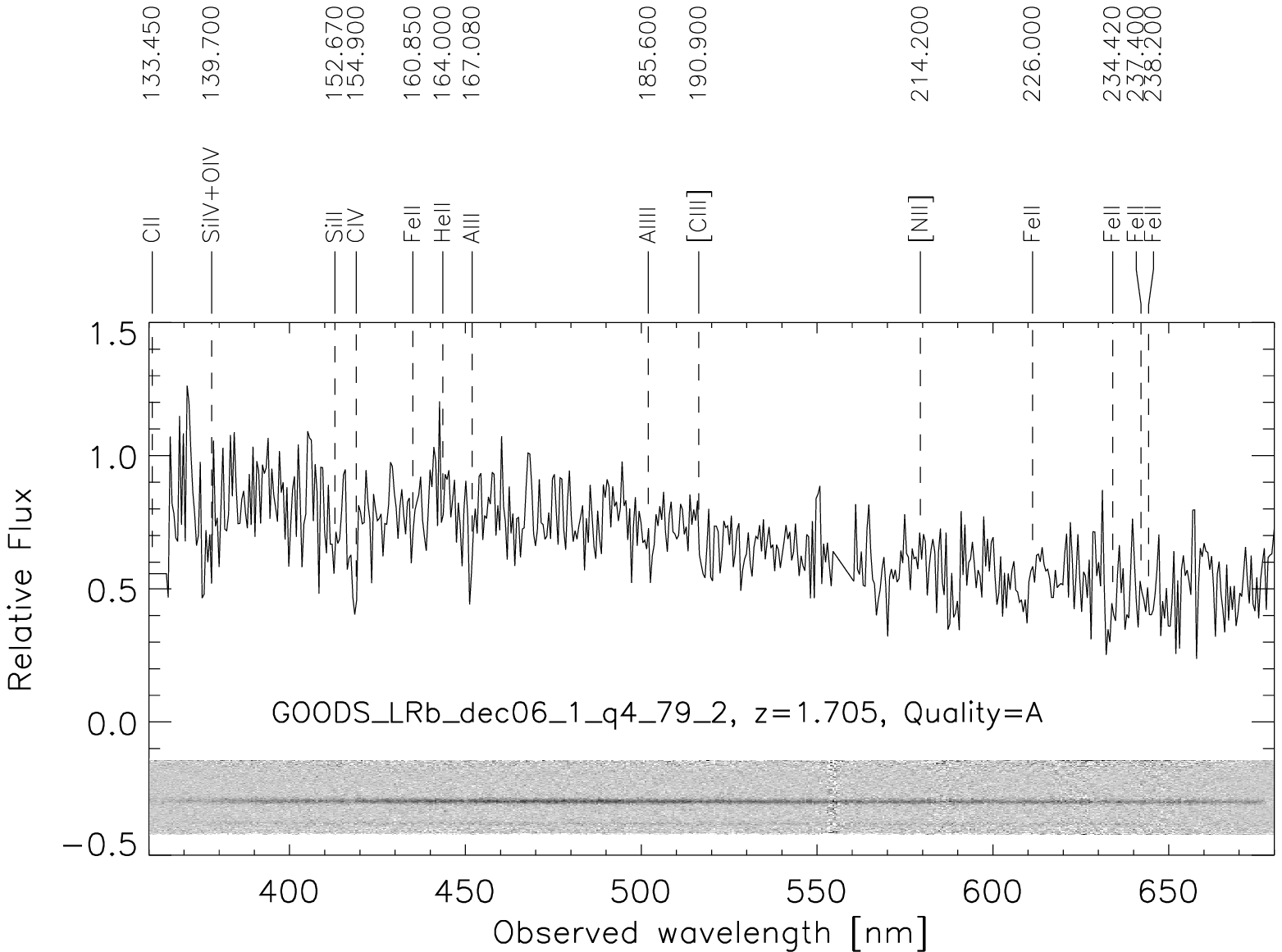}
\includegraphics[width=5 cm, angle=0]{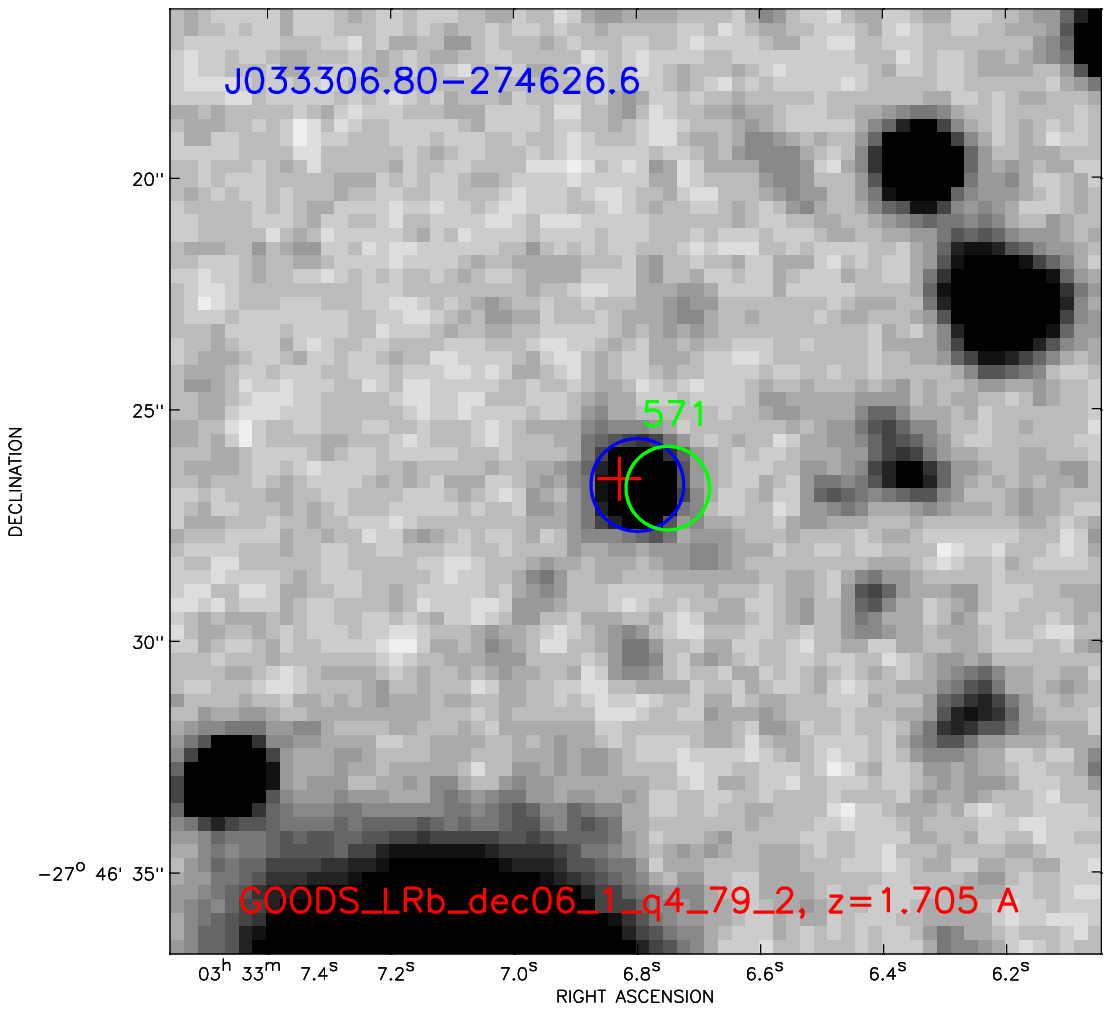}
\caption{New VIMOS spectra of X-ray sources with WFI $R$-band $20''\times20''$ 
cutout centered on the position of the matching WFI source. Red crosses indicate 
the reconstructed VIMOS coordinates. The blue circles show the position of the 
WFI-matched objects. The yellow and green circles display the position of 
X-ray sources from the 2Ms CDFS or from the ECDFS catalog, respectively. Labels
with corresponding colors indicate source identification numbers.}
\label{x3}
\end{figure*}

In Fig.~\ref{x1}, \ref{x2}, and \ref{x3} we show 12 VIMOS spectra of X-ray 
sources with new redshift determinations (i.e. either unknown before or 
improving previous estimates), obtained from the last 8 VIMOS masks
released in this paper. Together with each spectrum we provide a finding chart
with the position of the corresponding X-ray sources. These sources also
appear in the new catalog of X-ray sources detected in the ECDFS, which includes the
identification of optical and near-IR counterparts 
(Silverman et al., in preparation).

\newpage
\newpage

\end{document}